\documentclass[fleqn,usenatbib]{mnras}

\usepackage{newtxtext,newtxmath}
\usepackage{diagbox}
\usepackage[T1]{fontenc}
\usepackage{bm}
\usepackage[dvipsnames]{xcolor} 

\usepackage{graphicx}	
\usepackage{amsmath}	
\usepackage{subcaption} 
\usepackage[normalem]{ulem} 

\title[Black Hole Discs and Spheres in Galactic Nuclei]{Black Hole Discs and Spheres in Galactic Nuclei -- Exploring the Landscape of Vector Resonant Relaxation Equilibria}

\author[Gergely Máthé et al.]{
Gergely Máthé,$^{1}$\thanks{E-mail: gergely.mathe@ttk.elte.hu}
Ákos Szölgyén,$^{1}$
Bence Kocsis$^{2,1}$
\\
$^{1}$Institute of Physics, Eötvös University, Pázmány P. s.  1/A, Budapest, 1117, Hungargary\\
$^{2}$Rudolf Peierls Centre for Theoretical Physics, University of Oxford, Parks Road, Oxford, OX1 3PU, United Kingdom\\
}

\date{Accepted XXX. Received YYY; in original form ZZZ}

\pubyear{2022}

\begin{document}
\label{firstpage}
\pagerange{\pageref{firstpage}--\pageref{lastpage}}
\maketitle

\begin{abstract}
Vector resonant relaxation (VRR) is known to be the fastest gravitational process that shapes the geometry of stellar orbits in nuclear star clusters. This leads to the realignment of the orbital planes on the corresponding VRR time scale $t_{\rm VRR}$ of a few million years, while the eccentricity $e$ and semimajor axis $a$ of the individual orbits are approximately conserved. The distribution of orbital inclinations reaches an internal equilibrium characterised by two conserved quantities, the total potential energy among stellar orbits, $E_{\rm tot}$, and the total angular momentum, $L_{\rm tot}$. On timescales longer than $t_{\rm VRR}$, the eccentricities and semimajor axes change slowly and the distribution of orbital inclinations are expected to evolve through a series of VRR equilibria. Using a Monte Carlo Markov Chain method, we determine the equilibrium distribution of orbital inclinations in the microcanonical ensemble with fixed $E_{\rm tot}$ and $L_{\rm tot}$ for isolated nuclear star clusters with a power-law distribution of $a$, $e$, and $m$, where $m$ is the stellar mass. We explore the possible equilibria for $9$ representative $E_{\rm tot}$--$L_{\rm tot}$ pairs that cover the possible parameter space. For all cases, the equilibria show anisotropic mass segregation where the distribution of more massive objects is more flattened than that for lighter objects. Given that stellar black holes are more massive than the average main sequence stars, these findings suggest that black holes reside in disc-like structures within nuclear star clusters for a wide range of initial conditions. 
\end{abstract}




\section{Introduction}
\label{sec:introduction}

Supermassive black holes (SMBH) are observed in the cores of most nearby galaxies \citep{Genzel2010,Kormendy2013}. In many cases, these SMBHs are surrounded by a dense  population of stars and compact objects which form the nuclear star cluster \citep{Neumayer_Seth_Boker2020}. The nuclear star cluster of the Milky Way exhibits a rich diversity of spatial structures. Orbits of old, low-mass stars follow a spherical distribution, while a young coeval population of massive (mostly O, B, WR type) stars orbits in one or two warped structures, the so-called clockwise and counterclockwise discs \citep{Bartko2009,Lu2009,Yelda2014,Seth2008}. The innermost cluster of stars, also know as the S-cluster, displays anisotropy too; it is also possibly characterised by two discs, the so-called black and red discs \citep{Ali2020,Peissker2020,vonFellenberg2022}. The origin of this complex structure is controversial but two main formation channels were proposed: (i) the in-situ episodic star formation \citep[e.g.][]{Loose+1982,Mihos_Hernquist1994,Mapelli+2012,Mastrobuono2019} and (ii) the episodic migration events of massive stars and globular clusters into the nuclear star cluster \citep{Tremaine1975,Milosavljevic2001,Antonini2012,Antonini2013,Gnedin2014,Antonini2014,Antonini2015,ArcaSedda2015,arca_sedda_2019}. Both channels may contribute to the observed distribution but it is unclear if and how they may give rise to the observed warped structures. 

The gravitational processes shaping the geometry of stellar orbits have a temporal hierarchy in nuclear star clusters \citep{Kocsis2011}. Within 0.001 -- 1 pc, the central SMBH dominates the potential and drives orbital motion on Keplerian ellipses with orbital periods of $t_{\rm orb}=10-10^4$ yr. The spherical component of the stellar distribution and general relativity drive apsidal precession on timescales of $t_{\rm prec}=10^{4-5}$ yr. Mutual gravitational torques between Keplerian orbits accumulate coherently leading to an enhanced rate of relaxation of orbital inclination (vector resonant relaxation, VRR) and eccentricity (scalar resonant relaxation, SRR) \citep{Rauch1996}. The coherence time for non-axisymmetric torques (which drive eccentricity change) is limited by apsidal precession, $t_{\rm prec}$, while the axisymmetric component of the torque continues to accumulate coherently well beyond $t_{\rm prec}$ until the orbital planes reorient. The orbital plane orientation is described by its angular momentum vector direction $\hat{\bm{L}}_i$, which changes on the corresponding VRR timescale of $t_{\rm VRR}=10^{6-7}$\, yr while the eccentricity ($e$) and semimajor axis ($a$) are conserved \citep{Eilon2009,Kocsis2015}. Eccentricity diffusion takes place on the longer scalar resonant relaxation timescale, $t_{\rm SRR}=10^{8-9}$\, yr
\citep{Fouvry_Bar-Or2018,Bar-Or_Fouvry2018,Fouvry2019}. The semimajor axes exhibit Brownian motion with the shortest coherence time, i.e. $t_{\rm orb}$, and the longest diffusion time, $t_{\rm 2b}=10^{8-10}$\,yr, describing two-body relaxation. In summary, phase space mixing unfolds at a different rate in different subspaces according to the following hierarchy: (1) mean anomaly on $t_{\rm orb}$, (2) argument of periapsis on $t_{\rm prec}$, (3--4) argument of ascending node and orbital inclination on $t_{\rm VRR}$, (5) eccentricity on $t_{\rm SRR}$, and (6) semimajor axis on $t_{\rm 2b}$.

\citet{Hopman2006} argued that VRR can randomize orbital inclinations of old low mass stars to produce the observed spherical geometry from an initially flattened distribution. Given that the ages of young O, B, WR stars in the Galactic centre are marginally longer than the VRR timescale, \citet{Kocsis2011} argued that the warped clockwise disc may display a realization of the VRR statistical equilibrium which may be far from isotropic. However their model assumed a thin disc approximation and did not self-consistently account for the backreaction of the disc on the objects in the spherical distribution. This approximation was relaxed in \citet{Szolgyen2018}, where the statistical equilibrium of VRR was computed self-consistently for all stars in the system using a Monte Carlo Markov Chain (MCMC) method. That work showed that more massive objects settle to a more flattened distribution than low mass objects. In a series of papers on direct $N$-body simulations,  \citet{Antonini2012,Perets_Mastro2014,Tsatsi2017} investigated single-mass nuclear structures formed from the merger of 12 star clusters on initially inclined orbits and found that the system remains anisotropic and retains rotation for 12 Gyr \citep[see also][for simulations of the merger of 11 star clusters]{ArcaSedda_Kocsis2018}. Further, \citet{Mastrobuono2019} have simulated the evolution with five inclined single-mass discs and also found that the system remains flattened throughout the simulation. Recently, \citet{Panamarev_Kocsis2022} have run direct $N$-body simulations of nuclear star clusters with initially a disc embedded in a spherical population, and found that anisotropy persists for 10 Myr in systems lacking a massive spherical component, but if the spherical component is massive and nearly isotropic, the disc dissolves in the simulation and becomes spherical. Anisotropy has been shown to persist in simulations of rotating globular clusters too without a central massive object \citep{Einsel1999,Breen2017,Tiongco2017,Tiongco2018} consistently with observed systems \citep{Bianchini2013,Boberg2017,Jeffreson2017,Ferraro2018,Kamann2018,Lanzoni2018}. Direct $N$-body simulations have also shown evidence of anisotropic mass segregation in globular clusters \citep{Szolgyen+2019,Tiongco+2021}, which is in part due to VRR especially in massive clusters \citep{Meiron2019}. In \citet{Szolgyen+2021}, we have confirmed the conclusion of \citet{Szolgyen2018} that massive objects rapidly settle to the midplane of a disc due to VRR  with time-dependent direct $N$-body simulations of nuclear star clusters harboring an SMBH. Recently, \citet{Gruzinov+2020} and \citet{MagnanFouvry2021} also confirmed these results using mean field theory methods. Given that the stars observed in the clockwise disc in the Milky Way's centre are much more massive than average \citep{Bartko2010}, the VRR statistical equilibria have the potential to explain the observed structures. However existing of studies were limited to special initial configurations and did not attempt to explore the total landscape of possible outcomes for a wide range of initial conditions.

In this paper, we generalize the study of \citet{Szolgyen2018} to explore the VRR equilibria of isolated nuclear star clusters for a range of initial conditions. We use the same \textsc{Nring-MCMC} method, and assume a power-law density cusp with a distribution of $m$, $a$, and $e$. We sample the microcanoncial ensemble of the system with fixed total VRR energy and angular momentum, $E_{\rm tot}$ and $L_{\rm tot}$. Here VRR energy refers to the time-averaged potential energy among the stellar orbits over the orbital period and the apsidal precession period. Given that $a$ and $e$ are approximately conserved for each star during VRR, $E_{\rm tot}$ is also (approximately) conserved during VRR. Furthermore, $L_{\rm tot}$ is exactly conserved for an isolated system. These two conserved quantities characterise the possible statistical equilibria of VRR independently of the initial conditions. We explore the possible outcomes in different systems with different $E_{\rm tot}$-$L_{\rm tot}$ pairs through 9 representative cases. For each, we generate $100$ realizations for the initial distributions and evolve the systems with \textsc{Nring-MCMC} to reach the equilibrium. We examine the level of anisotropy as a function of $m$, $a$, and $e$, and investigate under what conditions can the massive objects such as stellar mass black holes form flattened substructures in equilibrium.

The rest of this paper is organised as follows. First, in Section~\ref{sec:methods}, we specify the VRR model and the adopted initial distributions, and introduce the \textsc{Nring-MCMC} method used in this paper. In Section~\ref{sec:results}, we present the resulting statistical equilibrium distributions of orbital inclinations for different masses and semimajor axes for different initial conditions. Finally in Section~\ref{sec:conclusion}, we discuss the implications of our findings.

\section{Methods}\label{sec:methods}

\subsection{Effective Hamiltonian of VRR}

In nuclear star clusters, the reorientation of the orbital planes is accelerated by the coherent accumulation of the axisymmetric torques between stellar orbits. The corresponding effective two-body Hamiltonian of VRR is derived by averaging the Hamiltonian over the orbital and the apsidal precession period. This eliminates the mean anomaly and argument of pericenter from the dynamical variables and leads to the Hamiltonian $H_{\rm VRR}=\sum_{i<j} H^{\rm VRR}_{ij} $ such that
\begin{equation}
H^{\rm VRR}_{ij} = -G \int_{r_{\mathrm{p},i}}^{r_{\mathrm{a},i}} dr \int_{r_{\mathrm{p},j}}^{r_{\mathrm{a},j}} dr' \frac{\sigma_i \left( r \right) \sigma_j \left( r' \right) }{ | r - r' |}
\label{eq:Hvrr}
\end{equation}
where $r$ is the distance from the SMBH, $r_{\mathrm{p}}=a(1-e)$ and $r_{\mathrm{a}}=a(1+e)$ are the periapsis and apoapsis, respectively, $i$ and $j$ denote the $i^{\rm th}$ and $j^{\rm th}$ particle, and 
\begin{equation}
\sigma_i\left(r\right) = \frac{m_i}{2 \pi^2 a_i \sqrt{\left(r_{\mathrm{a},i}-r\right)\left(r-r_{\mathrm{p},i}\right)}}
\end{equation}
is the surface density of the smeared (i.e. time-averaged) stellar orbit for the $i^{\rm th}$ particle \citep{Kocsis2015}. The semimajor axis and eccentricity, or equivalently the Keplerian orbital energy, $H_{{\rm Kep}, i}=-G m_{\rm SMBH}m_i/(2 a_i)$, and the magnitude of the angular momentum, $L_i=m_i [G m_{\rm SMBH} a_i (1-e_i^2)]^{1/2}$, are approximately conserved during VRR given the axisymmetric and stationary nature of the pairwise potentials. The dynamical variables are the $z$ component of the angular momentum and the argument of node in an arbitrary reference frame, or equivalently, the unit vectors $\hat{\bm{L}}_i$ of the angular momentum direction. Eq.~(\ref{eq:Hvrr}) is expressed with these variables explicitly in a multipole expansion as
\begin{equation}\label{eq:HVRRij}
    H^{\rm VRR}_{ij} = -\sum_{\ell=0}^{\infty} \mathcal{J}_{ij\ell}P_{\ell}\left(\hat{\bm{L}}_i\cdot\hat{\bm{L}}_j\right)
\end{equation}
where $\ell$ is the multipole index which is a non-negative integer\footnote{ $\mathcal{J}_{ij\ell}=0$ for $\ell=2n+1$ if $n$ is an integer, and $\ell=0$ does not affect the evolution, leaving $\ell\in\{2,4,\dots\infty\}$. In practice we truncate the Hamiltonian and keep $\ell\leq \ell_{\rm max}=50$ as in \citet{Szolgyen2018}. This effectively amounts to gravitational softening for angular separation $\cos^{-1}(\hat{\bm{L}}_i\cdot\hat{\bm{L}}_j) \leq \pi/(2\ell_{\rm max})$ \citep{Kocsis2015}}, and $\mathcal{J}_{ij\ell}$ are pairwise coupling coefficients which are positive constants given explicitly with the parameters $\{m_i,m_j,a_i,a_{j},e_i,e_{j}\}$ by Eqs.~(7) and (10) in \citet{Kocsis2015}. During VRR $m_i,a_i,e_i$ are constant for all $i$, drawn randomly from their respective distribution functions, implying that $\mathcal{J}_{ij\ell}$ is a quenched random matrix for each $\ell$. 

The orbit- and precession-averaged interactions are modeled with $H_{\rm VRR}$ which is expected to approximately reproduce the correct statistical properties on the corresponding $t_{\rm VRR}$ time scale, i.e. after a timescale much longer than the apsidal precession time but for much less than the scalar resonant relaxation and two-body relaxation time  \citep{Rauch1996,Eilon2009,Kocsis2015,Meiron2019,Szolgyen+2021}. The angular momentum direction unit vectors are expected to settle into statistical equilibrium for the instantaneous values of $(m_i,a_i,e_i)$. As $a,e$ change slowly on longer timescales, the system is expected to evolve through a series of VRR equilibria. In this paper, we determine the distribution of $\hat{\bm{L}}$ for different initial conditions and $(m,a,e)$ distributions.

\subsection{Conserved quantities}

\subsubsection{Semimajor axis, eccentricity, mass parameters}
During VRR, the  masses, semimajor axes, and eccentricities are approximately conserved for each star, respectively. In all our models, we draw these parameters from power-law distributions. In our fiducial model, we adopt distributions $\propto (m^{-2}\rm, a^0\rm, e^1)$ in the ranges $m/m_{\rm min}\in\left[1,100\right]$, $a/a_{\rm min}\in\left[1,100\right]$, and $e\in\left[0,0.3\right]$, respectively. We also explore cases with thermal and superthermal eccentricity distributions. These simple models lack the compositional complexity of the Milky Way's nuclear stellar cluster in order to allow us to clearly identify features generated purely by VRR dynamics.

Note that the mass distribution of young stars in the clockwise disc of the Milky Way is extremely top heavy $dN/dm\propto m^{-0.45\pm 0.3}$ but this may be at least in part due to the preferential anisotropic mass segregation caused by VRR \citep{Szolgyen2018}. The mass function of all components scales with $m^{-1.7 \pm 0.2}$ \citep{Lu+2013}.

The adopted semimajor axis distribution corresponds to a mean 3D number density of $n(r) \propto r^{-2}$ which corresponds to the equilibrium configuration for the massive objects in mass-segregated systems \citep{Bachall1977,OLeary2009} and is also close to the observed number density for the young massive stars in the Galactic centre \citep{Bartko2010}. Note however that the spherical distribution of old stars follows a shallower density slope of $n(r) \propto r^{-1.38}$ \citep{Schodel+2018,Gallego-Cano+2020} and other estimates of the young stars suggest a steeper profile of $r^{-2.9}$ \citep{Bartko2009,Lu2009}. Nuclear star clusters in other galaxies show a deprojected 3D number density profile scaling between $r^{-1}$ and $r^{-3}$ \citep{Neumayer_Seth_Boker2020}.

The adopted fiducial eccentricity distribution, truncated at $e=0.3$, is representative of the massive young O-type stars in the Galactic centre \citep{Yelda2014}. To explore the effects of the eccentricity distribution we run additional models with a thermal $f(e)\propto e$  and a superthermal $f(e)\propto e/(1-e^2)^{1/2}$ eccentricity distribution for $0\leq e \leq 0.95$. Here ``thermal'' refers to the fact that this distribution represents a uniform phase-space distribution for an isotropic system, which represents statistical equilibrium for isolated single-component systems, while this particular ``superthermal'' eccentricity distribution has more high eccentricity orbits than the isotropic thermal distribution, and this represents statistical equilibrium for razor thin disks \citep{Valtonen2006,Samsing2020}. The observed distribution of S-stars in the innermost region of the Galactic centre is super-thermal, see Figure 21 in \citet{Gillessen+2009}. In these additional models, we explore cases with a high level of initial anisotropy as described in Section~\ref{sec:conserved:E-L}.

\subsubsection{Total VRR energy and total angular momentum}
\label{sec:conserved:E-L}

For an isolated nuclear star cluster undergoing VRR, the statistical equilibrium is specified by maximising the entropy for fixed total angular momentum and VRR energy, ${L}_{\rm tot}=|\sum_i \bm{L}_{i}|$ and $E_{\rm tot}=H_{\rm VRR}$. In the remainder of this paper, we use the normalized dimensionless $L_{\rm tot}$ and $E_{\rm  tot}$ quantities defined as
\begin{align}\label{eq:Etotnorm}
    E_{\rm tot} &= -\dfrac{\sum_{ij}\sum_{\ell=2}^{\ell_{\rm max}} \mathcal{J}_{ij\ell}P_{\ell}\left(\hat{\bm{L}}_i\cdot\hat{\bm{L}}_j\right)}{\sum_{ij}\sum_{\ell=2}^{\ell_{\rm max}} \mathcal{J}_{ij\ell}}\,,\\
    L_{\rm tot} &= \dfrac{\left|\sum_{i=1}^{N} \bm{L}_i\right|}{\sum_{i=1}^{N}|\bm{L}_i|}\,,
    \label{eq:Ltotnorm}
\end{align}
which are bounded between 
$-1\leq E_{\rm tot} \leq 0.5$ and $0\leq L_{\rm tot}\leq 1$ since $\mathcal{J}_{ij\ell}>0$ and $-0.5\leq P_{\ell}(x)\leq 1$ for all even $\ell$ and $|x|\leq 1$.\footnote{Here $\mathcal{J}_{ij\ell}=0$ for odd $\ell$ \citep{Kocsis2015}.} The case of $E_{\rm tot}=-1$ and $L_{\rm tot}=1$ corresponds to a razor thin disc in physical space in which all angular momentum vectors are parallel so that $\hat{\bm{L}}_i\cdot\hat{\bm{L}}_j=1$. Further, $(E_{\rm tot}, L_{\rm tot})=(-1,0)$ is a razor thin counter-rotating disc where half of the stars orbit in the opposite sense in the same plane. The limit $(E_{\rm tot}, L_{\rm tot})=(0,0)$ represents an isotropic distribution, and $(0,0.5)$ is the spherical distribution with the largest possible net rotation where there are no retrograde orbits in projection.\footnote{Note that $(E_{\rm tot},L_{\rm tot})$ with $E_{\rm tot}=0$ and $L_{\rm tot}>0.5$ cannot be attained by any system.} An example with maximum positive $E_{\rm tot}=0.5$ is obtained in the case with $N=2$, two orthogonal orbits having a semimajor axis ratio approaching infinity.\footnote{Note that the $E_{\rm tot}$ may be positive for the adopted definition (Eq.~\ref{eq:HVRRij}) which omits the unimportant $\ell=0$ monopole term as that term does not directly affect the time-evolution of VRR.} 

\begin{table}
\centering
\begin{tabular}{|c|c|c|c|c|c|c|}
\hline 
\diagbox{$E_{\rm tot}$}{$L_{\rm tot}$} & $L_{\rm low}$ & $L_{\rm med}$ & $L_{\rm high}$\\
\hline 
\hline 
$E_{\rm high}$         & $(-0.03,0.16)$ & $(-0.03,0.20)$ & $(-0.03,0.38)$ \\
\hline 
$E_{\rm med}$    & $(-0.33,0.15)$ & $(-0.33,0.35)$ &  $(-0.32,0.82)$\\
\hline 
$E_{\rm low}$        & $(-0.58,0.15)$ & $(-0.58,0.40)$  & $(-0.58,0.88)$\\
\hline 
\end{tabular}
\caption{Representative systems in dimensionless VRR energy and total angular momentum $(E_{\rm tot}, L_{\rm tot})$, Eqs.~(\ref{eq:Etotnorm}--\ref{eq:Ltotnorm}).  
\label{tab:regions} }
\end{table}

\begin{figure}
	\includegraphics[width=1.\columnwidth]{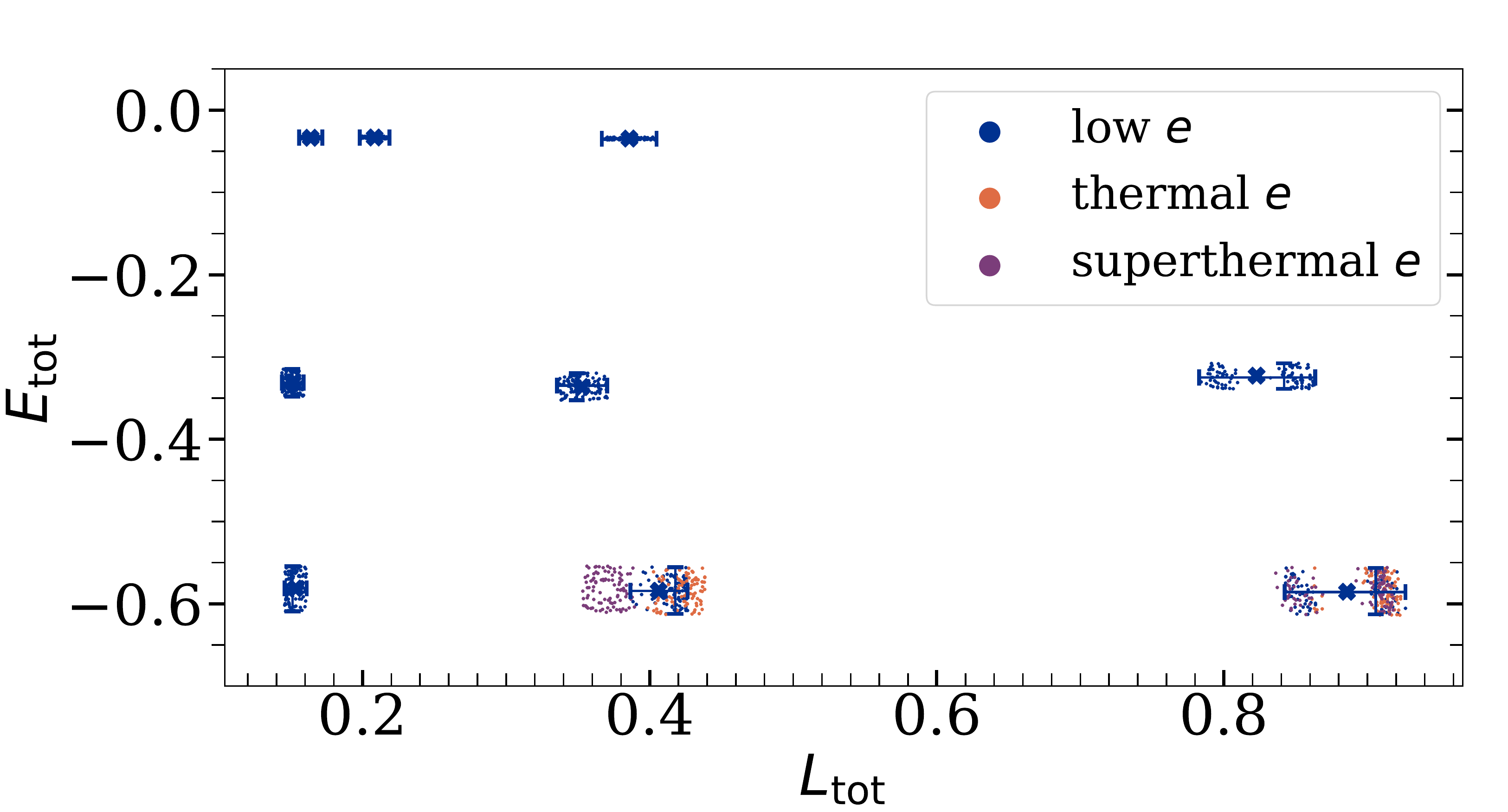}
    \caption{Scatter plot of the normalized total energy and angular momentum $\left(E_{\rm tot}, L_{\rm tot}\right)$ (Eqs.~\ref{eq:Etotnorm}--\ref{eq:Ltotnorm}) adopted in the MCMC simulations.  We generate 100 realisations around the 9 target values shown in Table~\ref{tab:regions} for the fiducial choice of eccentricity distribution truncated at a maximum eccentricity of 0.3. We run two additional models with low energy for a thermal and a superthermal disrtibution truncated at $e=0.95$ (see legend). Small blue X-s show the target values, blue dots show a scatter plot of the actual individual realisations and the horizontal blue error bars show the full range of $E_{\rm tot}$ for the median $L_{\rm tot}$ and similarly for the vertical error bars for $L_{\rm tot}$. The dots with different colors represent different assumptions for the eccentricity distribution.
    }
    \label{fig:normedIntialEL}
\end{figure}

We explore the behavior of the system for 9 representative $(E_{\rm tot},L_{\rm tot})$ pairs given in Table~\ref{tab:regions} and also shown in Figure~\ref{fig:normedIntialEL}. These values were chosen to cover the whole parameter space from nearly spherical to highly anisotropic configurations and with small to high levels of net rotation. Note that we do not include cases with $E_{\rm tot}>0$ as these cases are rather contrived like the one noted above. Similarly, evolving a razor-thin disc ($E_{\rm tot}=-1$) with VRR would be highly unrealistic as in this case two-body relaxation makes the disc puff up, rapidly increasing $E_{\rm tot}$ \citep{Cuadra+2008}. VRR may dominate once the disc is not razor thin. For this reason, we restrict attention to the range $-0.6 \leq E_{\rm tot} \leq 0$.

Given that $(E_{\rm tot},L_{\rm tot})$ are conserved during VRR, and that the statistical equilibrium maximises entropy for fixed $(E_{\rm tot},L_{\rm tot})$, the equilibrium distribution is not expected to depend on the details of the initial condition other than setting the values of $(E_{\rm tot},L_{\rm tot})$.

\subsection{Initial conditions}
\label{sec:initialConditions}

\begin{figure}
    \centering
    \includegraphics[width=1.\columnwidth]{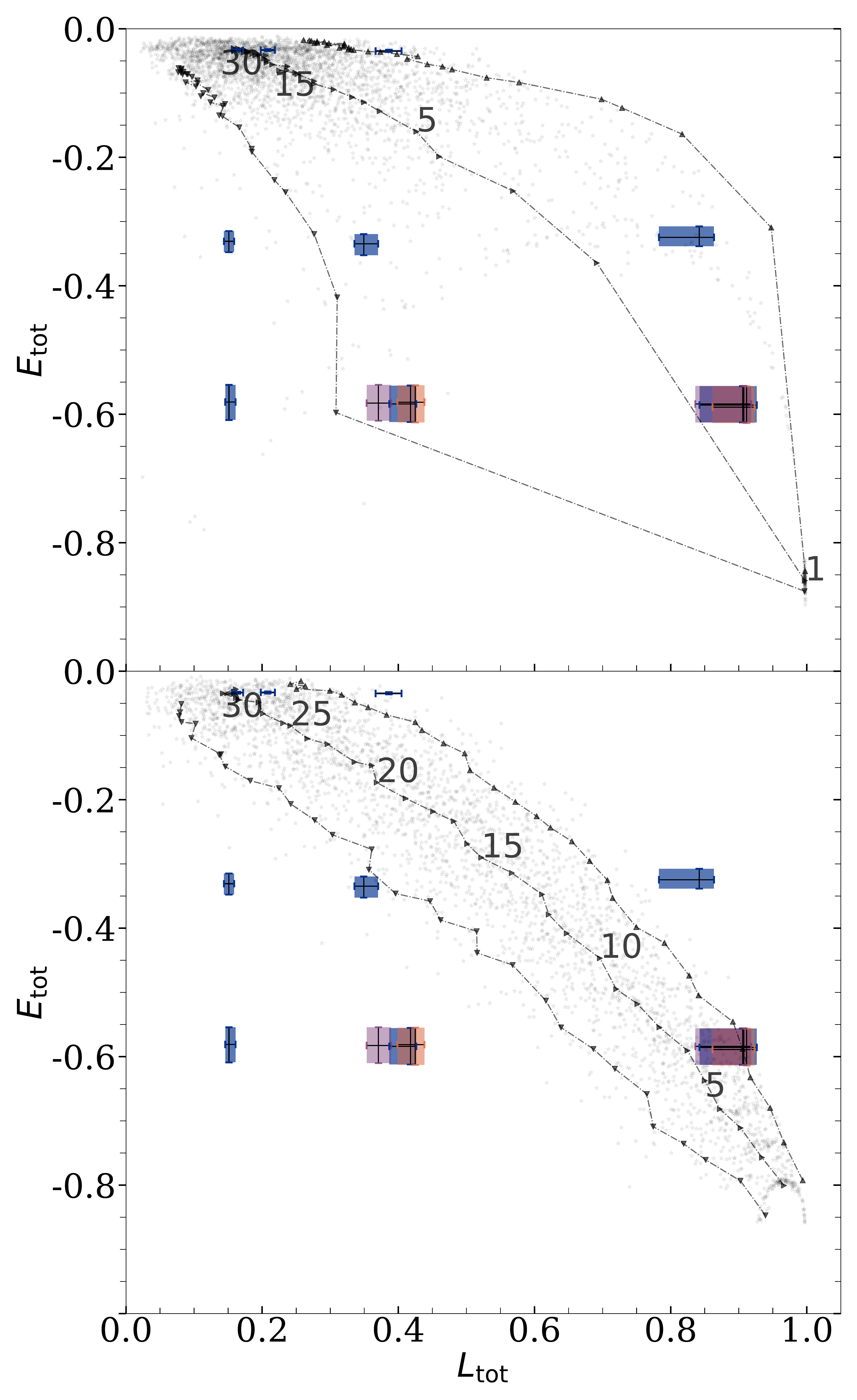}
    \caption{The dimensionless total energy-angular momentum, $(E_{\rm tot},L_{\rm tot})$ pairs of hypothetical astrophysical systems of (i) cluster infall and (ii) in-situ star formation scenarios. Faint black dots show the individual realisations for the ensemble of $32\times100$ different realisations (32 different number of subclusters ($N_{\rm sub}$) with 100 different realisation for each). The three lines (with triangle symbols) correspond to the $25\%,50\%,75\%$ percentile of the cumulative distribution, respectively. The parameter regions covered by our actual MCMC simulations are shown with shaded rectangles for comparison, with the same coloring scheme as in Figure~\ref{fig:normedIntialEL}. In the top panel we test scenario (i) where we assume that the system is built up of $N_{\rm sub}$ number of counter-rotating, thin discs of 128 stars each with randomly chosen rotational axes. By varying $N_{\rm sub}$ all the different $(E_{\rm tot},L_{\rm tot})$ cases can be reproduced except for low $L_{\rm tot}$-low to moderate $E_{\rm tot}$ configurations which correspond to thin discs with almost the same fraction of co- and counter-rotating particles.  The bottom panel tests scenario (ii), where we assume that a fraction of the $4096$ stars are contained in $N_{\rm sub}$ number of thin, counter-rotating discs of $128$ stars with randomly oriented rotational axis just as in the top panel, but the rest of the stars are assumed to reside in a massive counter-rotating thin disc. The number of subsystems (i.e. randomly oriented discs) decreases from the upper left to lower right corner as labelled with numbers. The bottom panel shows that the low $E_{\rm tot}$-high $L_{\rm tot}$, moderate $E_{\rm tot}$-moderate $L_{\rm tot}$ and high $E_{\rm tot}$-low to moderate $L_{\rm tot}$ cases are best represented with $N_{\rm sub}=(6,17,31)$, respectively. The three cases correspond to a disc/(disc+sphere) mass fraction of $81.25\%,46.87\%,3.12\%$, respectively.
    }
    \label{fig:elReprFormationChanelAnalysis}
\end{figure}

To understand the astrophysically relevant range of initial conditions, we identify the energy-angular momentum pairs of (i) the cluster infall scenario and (ii) the in-situ episodic star formation scenario, respectively. Figure~\ref{fig:elReprFormationChanelAnalysis}, shows the potential energy-angular momentum pairs of the 9 combinations corresponding to Table~\ref{tab:regions} simulated in this work (colored regions, same as in Figure~\ref{fig:normedIntialEL}) along with the astrophysically relevant range (black lines). The top panel shows scenario (i) with different numbers of in-falling episodes $N_{\rm sub}$, assuming that each episode delivers a thin  counter-rotating subcluster of $128$ stars with randomly oriented rotational axis. Clearly, the $E_{\rm low}$ and $E_{\rm med}$ cases of Table~\ref{tab:regions} correspond to 3 or less infall episodes, while the 
$E_{\rm high}$ case corresponds to approximately 30 infall episodes. Note that the $(L_{\rm low},E_{\rm low})$ and $(L_{\rm low},E_{\rm med})$ of Table~\ref{tab:regions} do not arise naturally in this formation scenario, these cases represent thin disc configurations with almost equal numbers of co- and counter-rotating particles. The bottom panel shows scenario (ii) for which, in an initial in-situ star formation period, we vary the relative fraction of stars in thin, randomly oriented, counter-rotating discs relative to a massive pre-existing thin-disc population of stars. Here the total number of stars in the cluster is set to be $4096$ while the number $N_{\rm sub}$ of thin discs of $128$ stars decreasing form the to left to the bottom right corner as indicated by the numbers. Scenario (ii) leads to a narrower region in the $(L_{\rm tot},E_{\rm tot})$ space, compared to the selected values in Table~\ref{tab:regions}. Cases of $(E_{\rm low},L_{\rm high})$, $(E_{\rm med}, L_{\rm med})$ and $(E_{\rm high},L_{\rm med})$ can be reproduced with particular numbers of subclusters, $N_{\rm sub}=(6,17,31)$ which correspond to a disc/(disc+sphere) mass fractions of $81.25\%,46.87\%,3.12\%$, respectively.

To obtain an initial realisation of the selected values of $(E_{\rm tot},L_{\rm tot})$ shown in Table~\ref{tab:regions}, we generate the initial conditions for the angular momentum vectors of different stars as follows. We generate the superposition of $N_{\rm sub}=32$ thin discs of 128 stars each. In particular, for $(E_{\rm low},L_{\rm high})$ the normal vectors of the 32 thin discs are nearly parallel, while for $(E_{\rm high},L_{\rm low})$ their normal vectors are drawn from an isotropic distribution. More specifically, the angular momentum vectors of each subsystem were first drawn from a polar cap region with opening angle $\cos\theta^{\rm sub}_{\rm max}=0.994$ so that the spread of orbital inclinations is $\theta^{\rm sub}_{\rm max}=0.11\,\rm rad$. Then the subsystems were rotated respectively as solid bodies by an angle which was again drawn from a polar cap distribution with a different opening angle $\theta^{\rm sys}_{\rm max}$. Furthermore since the rotation by $180^{\circ}$ does not change $E_{\rm tot}$, but reduces $L_{\rm tot}$, a $\kappa_{\rm sys}$ fraction of the subsystems was flipped by $180^{\circ}$ to explore different values of $L_{\rm tot}$ for fixed $E_{\rm tot}$.\footnote{Note that $\mathcal{J}_{ij \ell}\neq 0$ in Eq.~(\ref{eq:Hvrr}) only for even $\ell$ for which $P_{\ell}(x)=P_{\ell}(-x)$.} We repeat this procedure by adjusting $\theta^{\rm sys}_{\rm max}$ and $\kappa_{\rm sys}$ to obtain systems with $(E_{\rm tot}, L_{\rm tot})$ within a $5\%$ tolerance around the target values shown in Table~\ref{tab:regions}. 

\subsection{Monte Carlo Markov Chain simulator}
For each initial distribution, we use the \textsc{Nring-MCMC} method \citep{Szolgyen2018} to find the equilibrium distribution. This algorithm perturbs the system randomly in steps to reach the maximal entropy state in phase space while keeping $(E_{\rm tot},L_{\rm tot})$ constant to approach the microcanonical ensemble. In particular in each step, a randomly selected pair of angular momentum vectors are rotated by a random angle around their pairwise-total angular momentum axis. This procedure preserves the total angular momentum of the system and conserves $H_{ij}^{\rm VRR}$ for the pair exactly while the total VRR energy changes by $\Delta E_{\rm tot}$ due to the interaction energy between the pair and the rest of the system. The proposed step is accepted if the cumulative change $\Delta E_{\rm tot}$ is lower than a predefined energy tolerance. The energy tolerance is specified such that it is smaller than $10/N$ times the smallest one particle energy of any object within the system $E_{\rm 1p,min}=-\min_{i} \sum_{j} H^{\rm VRR}_{ij}$. Here, $N=4096$ is the number of objects in the system. Note that since the minimum one-particle energy is much lower than the mean one-particle energy, thus $E_{\rm tot}$ does not change more than $0.007\%$ during the simulation. Furthermore, for this choice we find that the acceptance rate for the perturbation proposals is around $\sim 24\%$ which facilitates the rapid convergence of the simulations to maximize entropy. We run the algorithm for $8.39\times 10^7$ number of simulation steps, i.e. $4915$ average number of accepted perturbing steps per star. Note that since the pairwise energy is proportional to the product of the masses, the perturbation acceptance rate is lower for the more massive objects in the cluster. The mean acceptance rate is $(23.25,0.63,0.31)\%$ for the lightest, intermediate, and heaviest stars in linear mass bins of $1\leq m/m_{\rm min}\leq 32.83$, $32.83\leq m/m_{\rm min}\leq 65.66$, and $65.66\leq m/m_{\rm min}\leq 100$, respectively.\footnote{Note that the mass bins are chosen differently here than in Figure~\ref{fig:mSelectedTimeEnsembleAver}.}

The simulations were run in parallel with the openMP version of \textsc{Nring-MCMC} (see Sec.~\ref{sec:results} for a full list of simulations). While one simulation takes about 5 days to run on a single machine with 8 CPU cores, all 900 simulations (i.e. 800k--900k CPU hours) were run within a month using parallel execution on the high performance computing clusters of the National Informatics Infrastructure Development Institute (NIIF) in Hungary. In comparison a direct N-body simulation requires to simulate a much larger number of particles by a factor of $10-100\times$, to avoid excessive two-body relaxation artifacts, i.e. $N\sim$ $10^5-10^6$ particles. This on average takes at least $\approx 2000$ GPU hours or $\approx 90$ days to carry out for an individual simulation. Given that a large sample of simulations with different initial realizations are required to map out the typical behavior in a reliable way, direct $N$-body simulations are not used in this paper. However, this method which relies on double orbit-averaging neglects the possible changes of eccentricity and semimajor axes, which may affect the solution \citep{Panamarev_Kocsis2022}.

\begin{figure}
\centering
    \begin{subfigure}[b]{1.\columnwidth}
    \centering
        	\includegraphics[width=1.\textwidth]{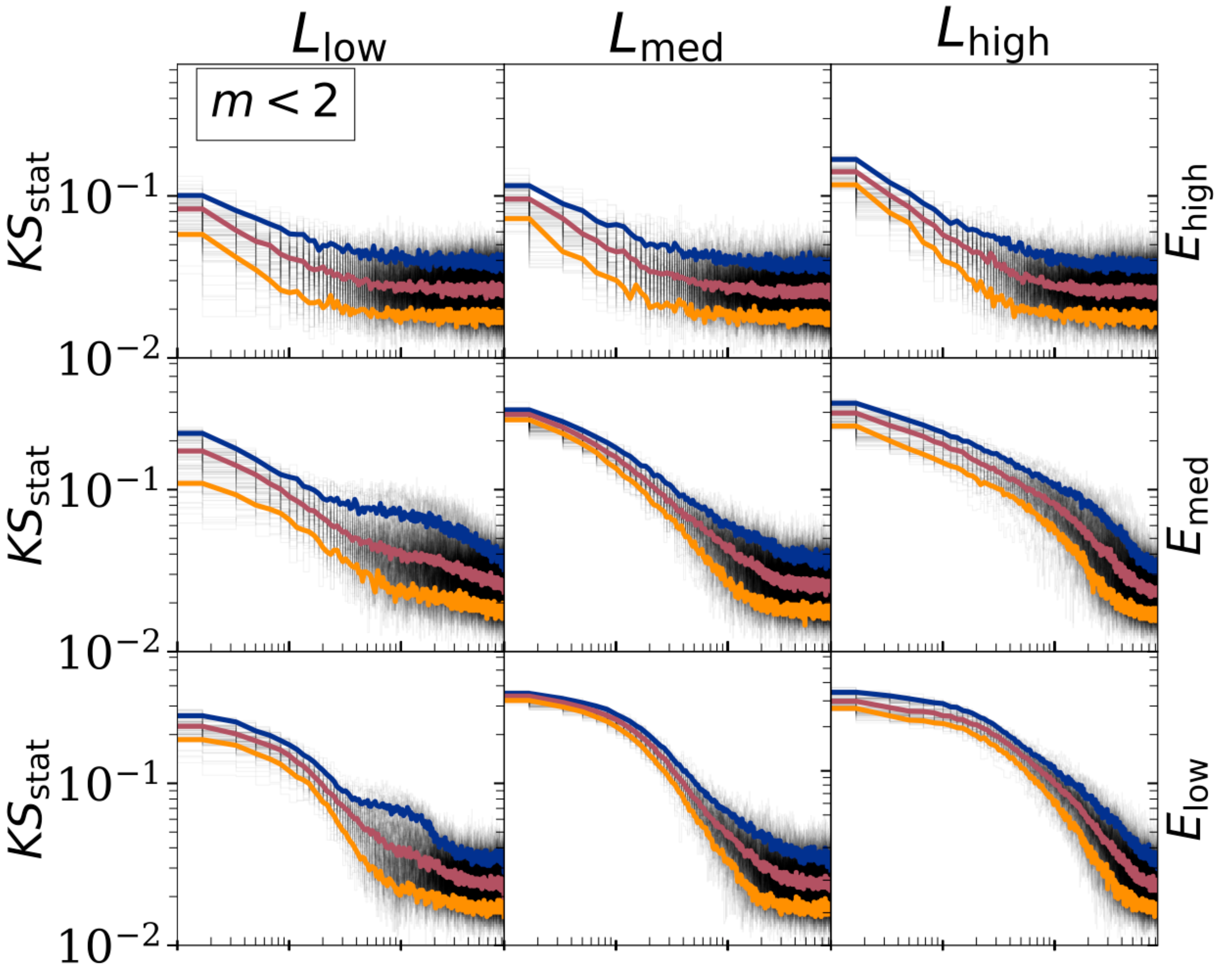}
    \end{subfigure}
    \begin{subfigure}[b]{1.\columnwidth}
    \centering
        	\includegraphics[width=1.\textwidth]{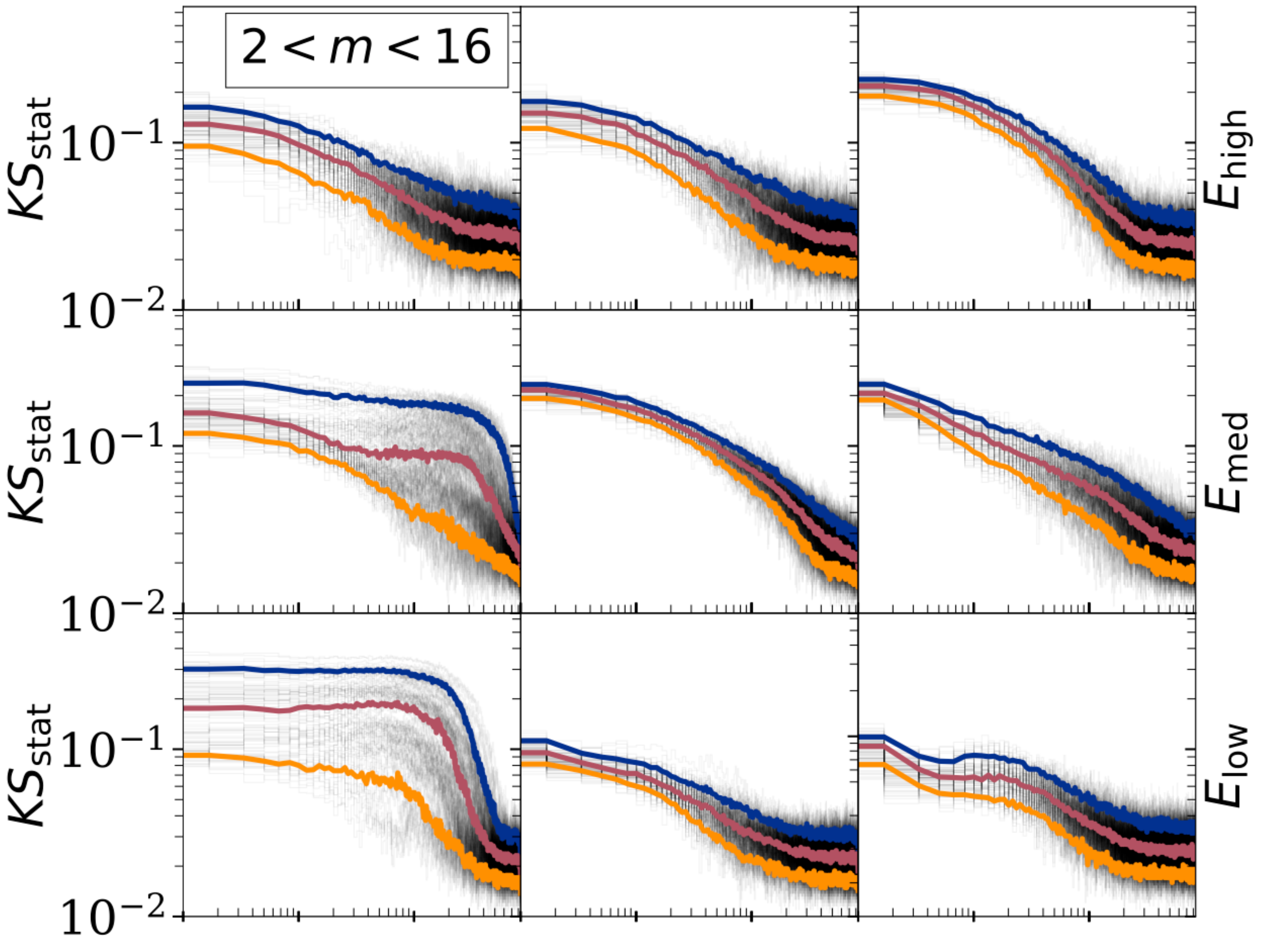}
    \end{subfigure}
    \begin{subfigure}[b]{1.\columnwidth}
    \centering
        	\includegraphics[width=1.\textwidth]{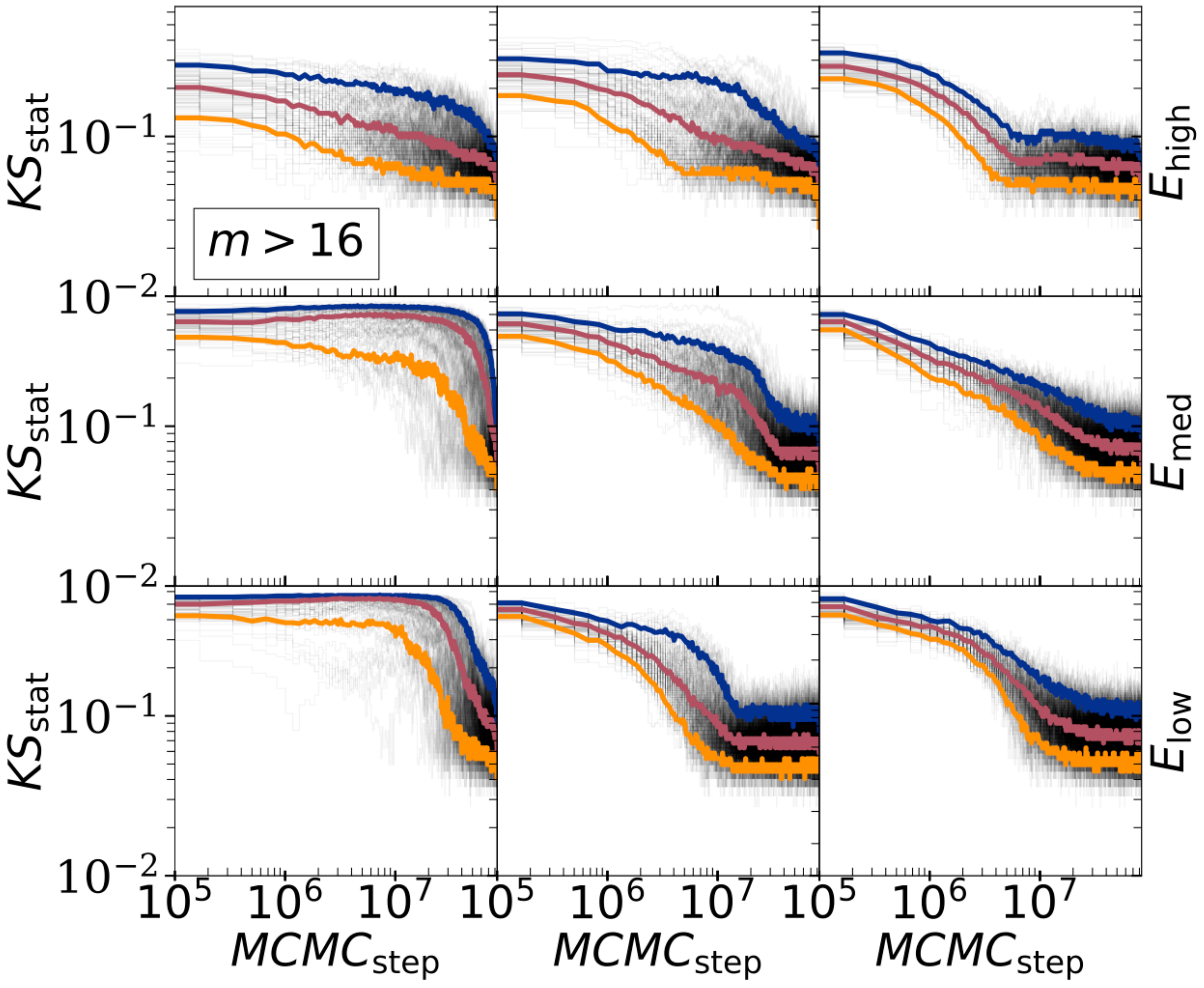}
    \end{subfigure}
    \caption{Kolmogorov-Smirnov test to probe the convergence of the MCMC method for the $(E_{\rm tot},L_{\rm tot})$ of Table~\ref{tab:regions} for 3 different mass ranges increasing from top to bottom as labelled. For each realisation the test is performed with respect to the last MCMC step and the result is shown as a function of the MCMC step number $n_{\rm step}$. The distributions tend to converge with large $n_{\rm step}$. The convergence is best for $m/m_{\rm min}<2$ for all cases particularly $(E_{\rm high},L_{\rm low})$ (spherical system). Convergence is slowest for $m/m_{\rm min}>16$ but even in this case there is clear convergence for $(E_{\rm low},L_{\rm med})$ (flat disc with a limited counter-rotating component) while the convergence may not be fully establised for $m/m_{\rm min}\geq 16$, $(E_{\rm low},L_{\rm low})$ (counter-rotating flat disc). 
    }
    \label{fig:ksTest}
\end{figure}

\begin{figure*}
\centering
    \begin{subfigure}[b]{.33\textwidth}
      \centering
      \includegraphics[width=1.\textwidth]{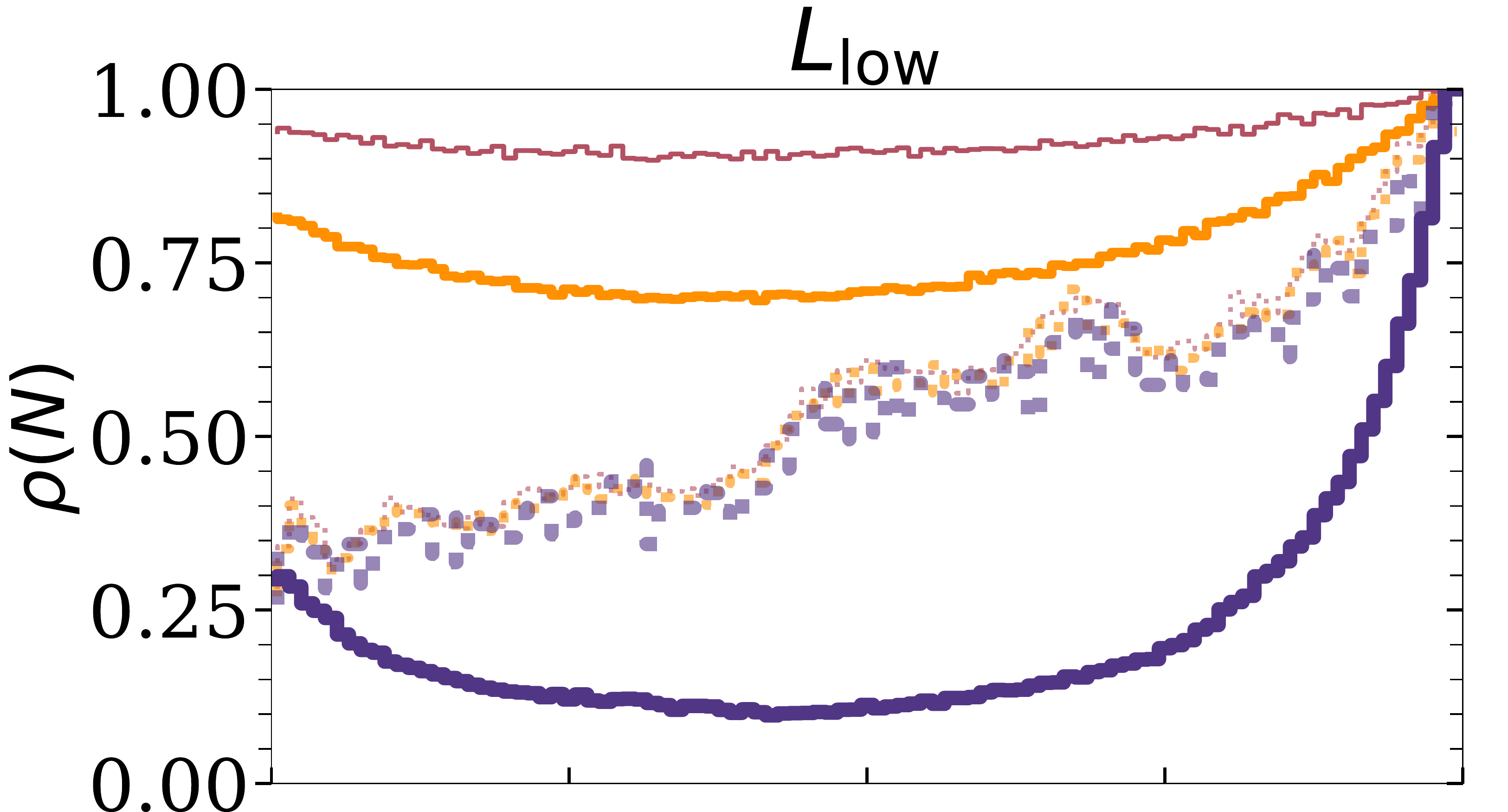}
    \end{subfigure}\hfil
    \begin{subfigure}[b]{.33\textwidth}
      \centering
      \includegraphics[width=1.\textwidth]{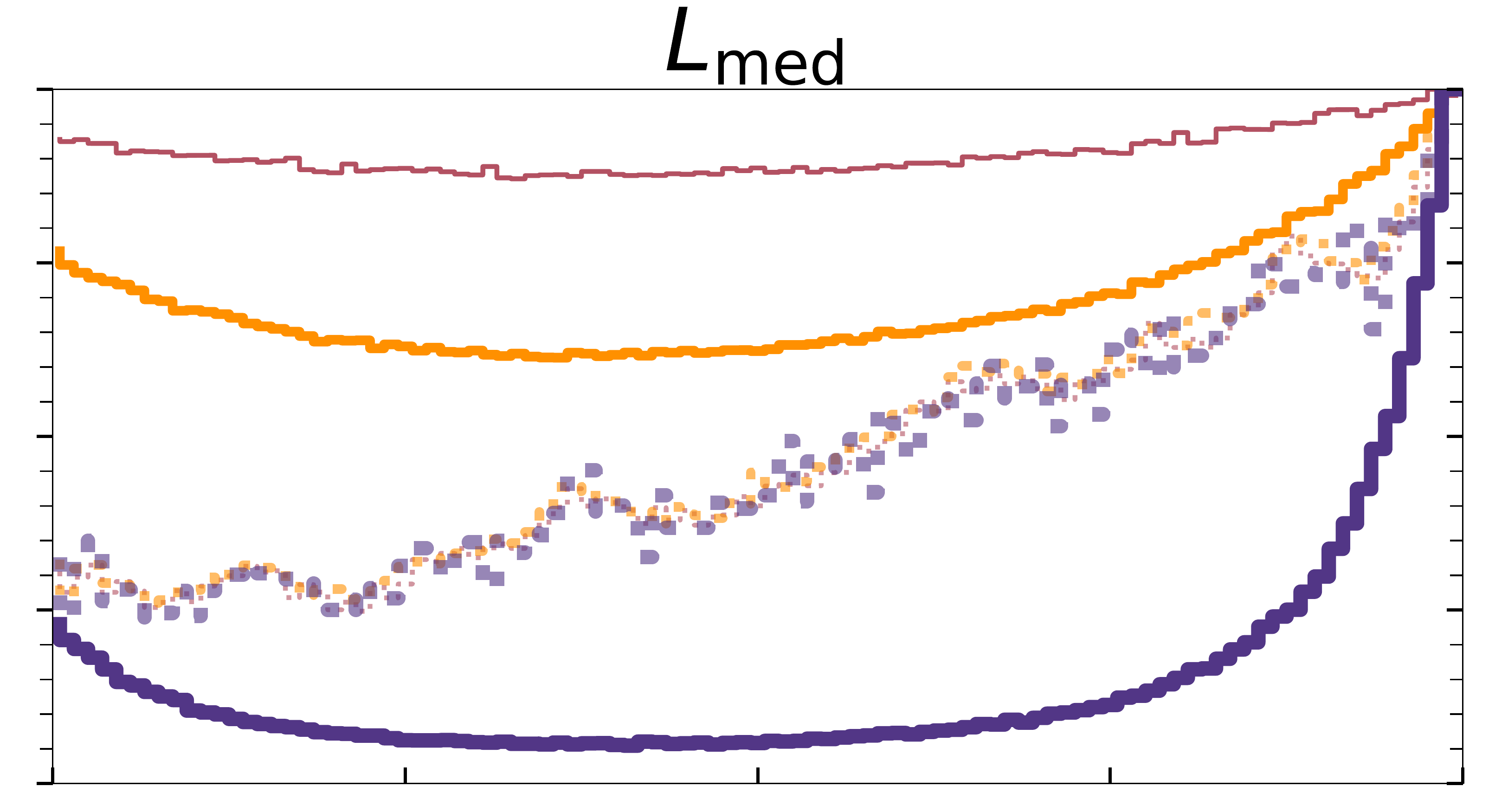}
    \end{subfigure}\hfil
    \begin{subfigure}[b]{.33\textwidth}
      \centering
      \includegraphics[width=1.\textwidth]{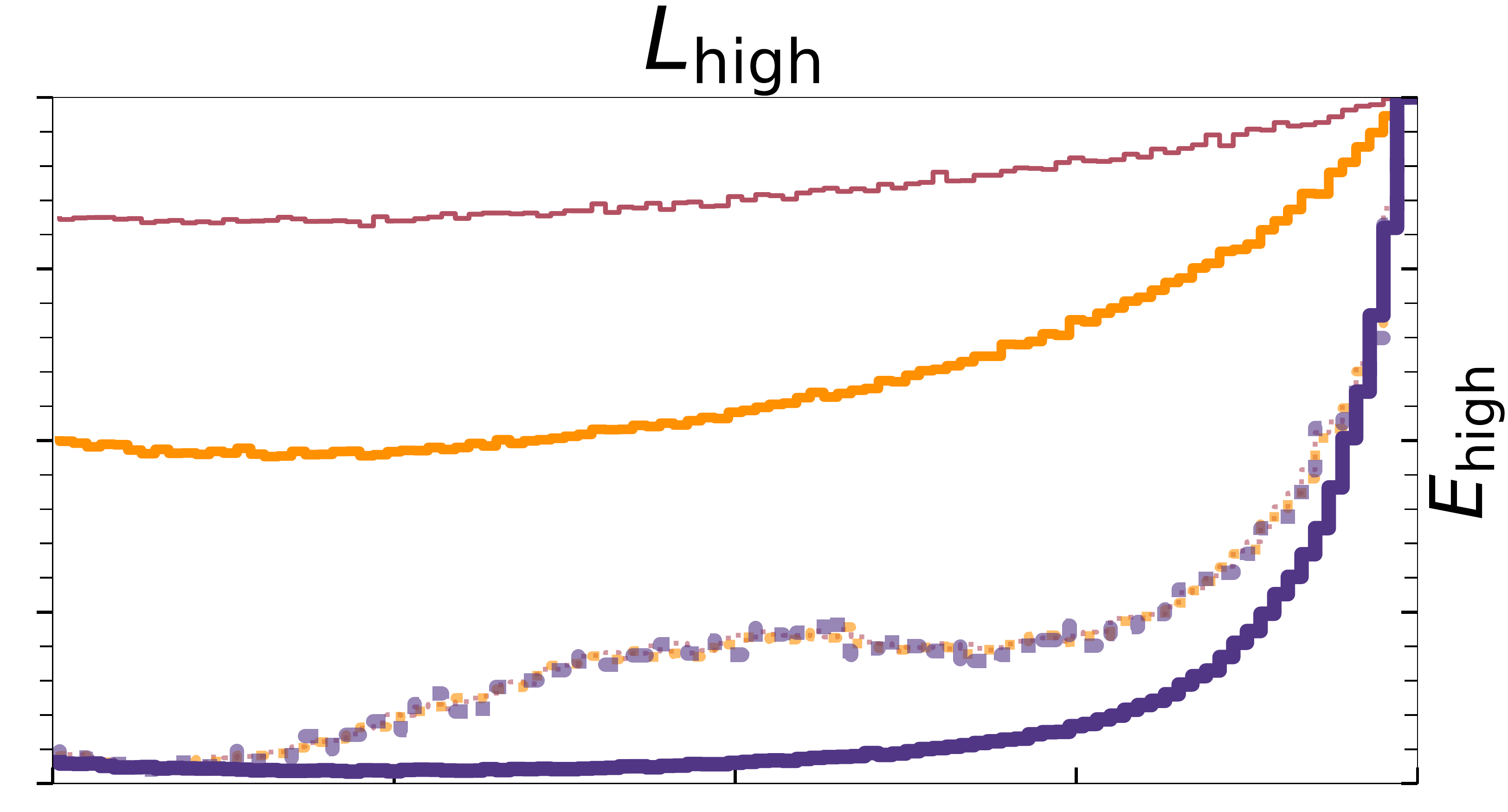}
    \end{subfigure}

    \medskip
    \begin{subfigure}[b]{0.33\textwidth}
      \centering
      \includegraphics[width=1.\textwidth]{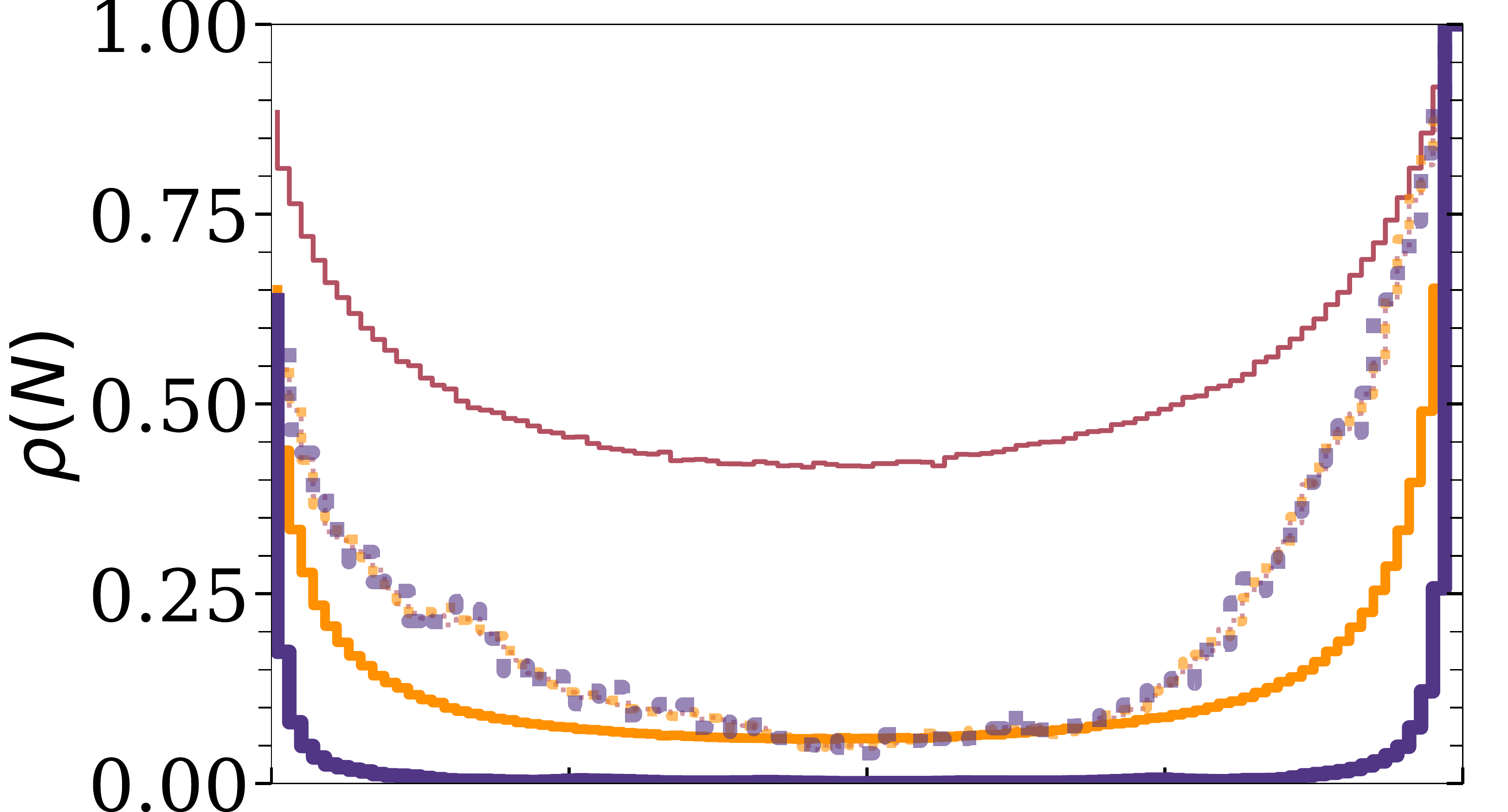}
    \end{subfigure}\hfil
    \begin{subfigure}[b]{0.33\textwidth}
      \centering
      \includegraphics[width=1.\textwidth]{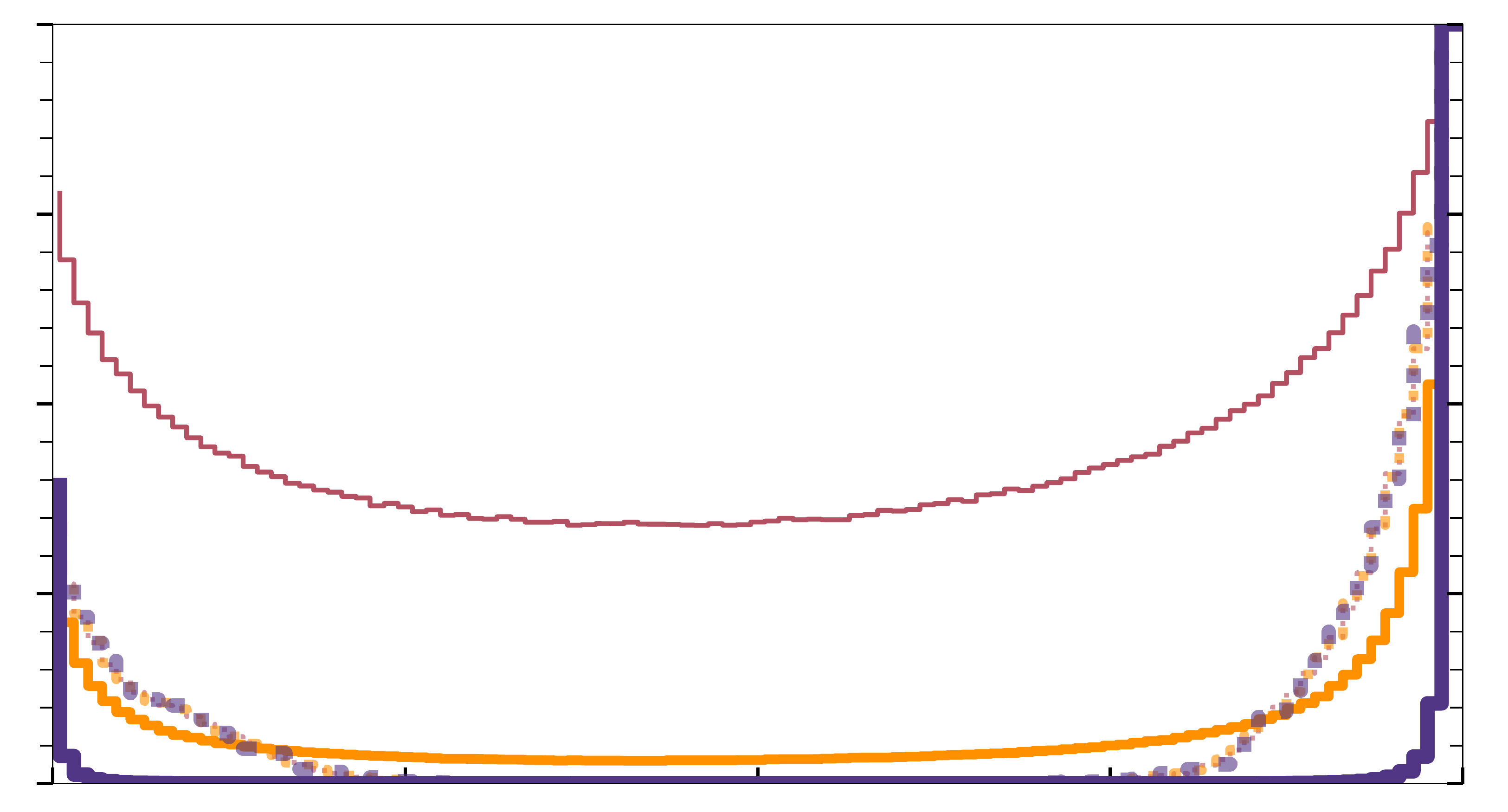}
    \end{subfigure}\hfil
    \begin{subfigure}[b]{0.33\textwidth}
      \centering
      \includegraphics[width=1.\textwidth]{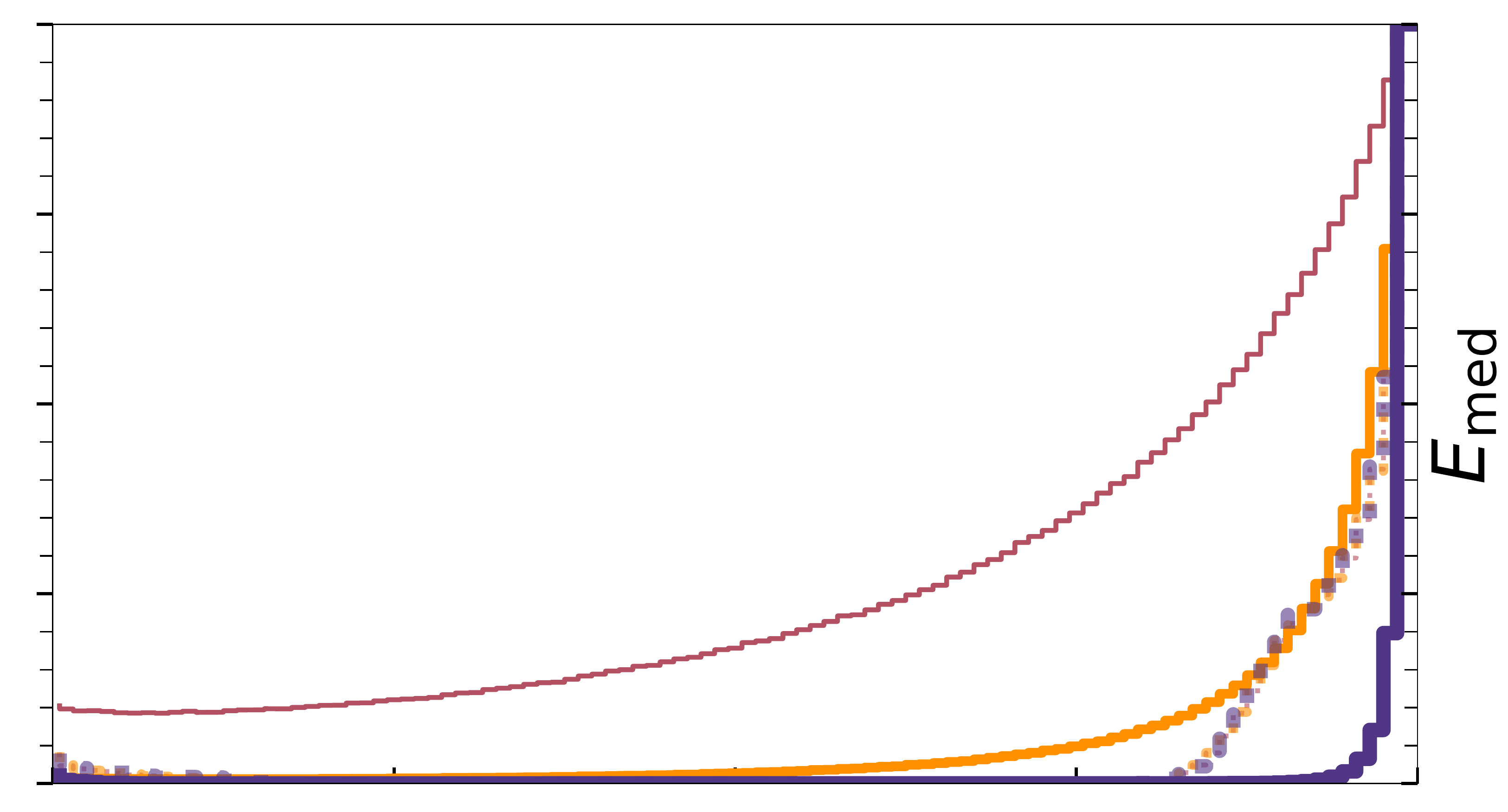}
    \end{subfigure}
    
    \medskip
    \begin{subfigure}[b]{0.33\textwidth}
      \centering
      \includegraphics[width=1.\textwidth]{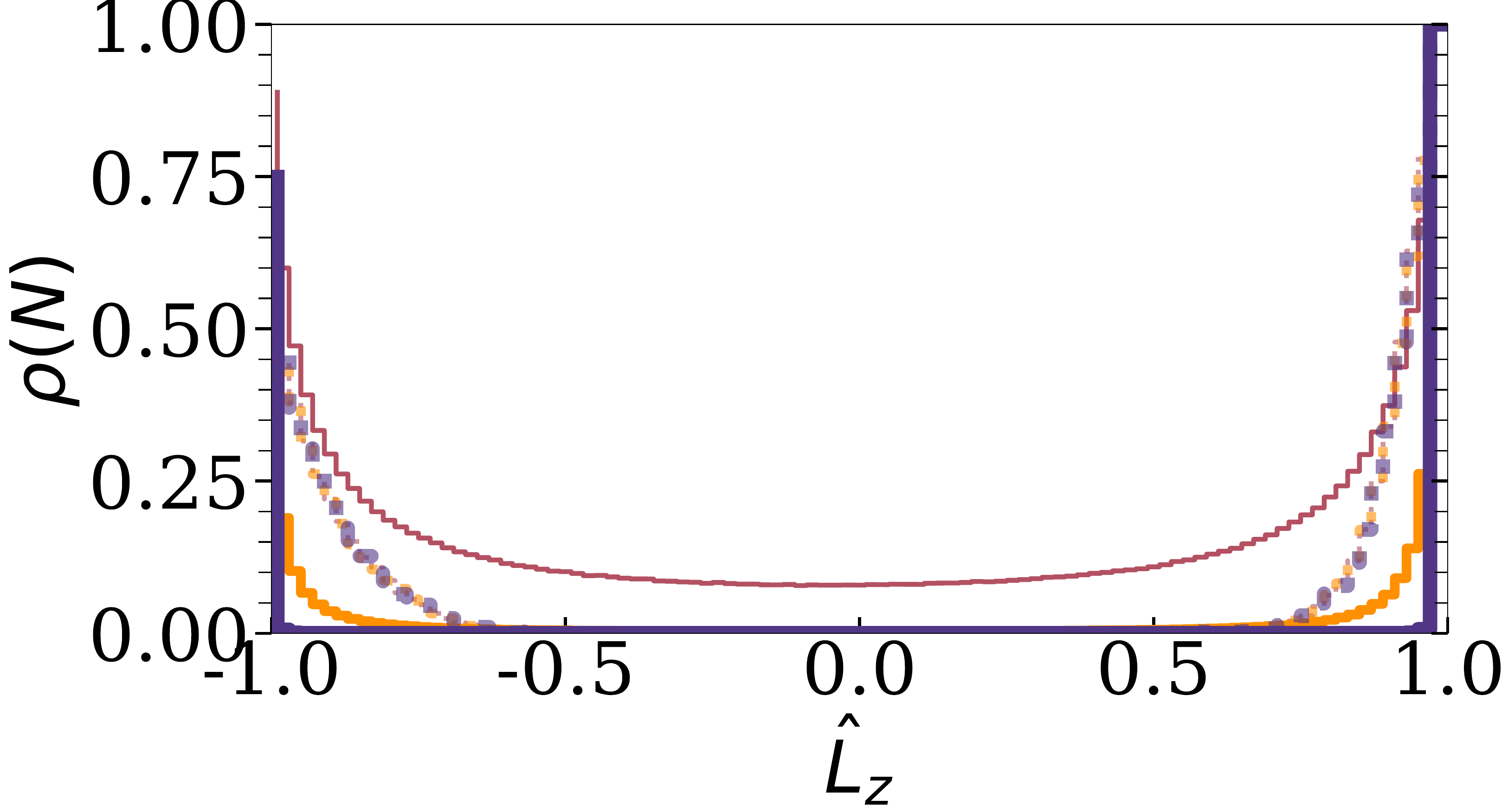}
    \end{subfigure}\hfil
    \begin{subfigure}[b]{0.33\textwidth}
      \centering
      \includegraphics[width=1.\textwidth]{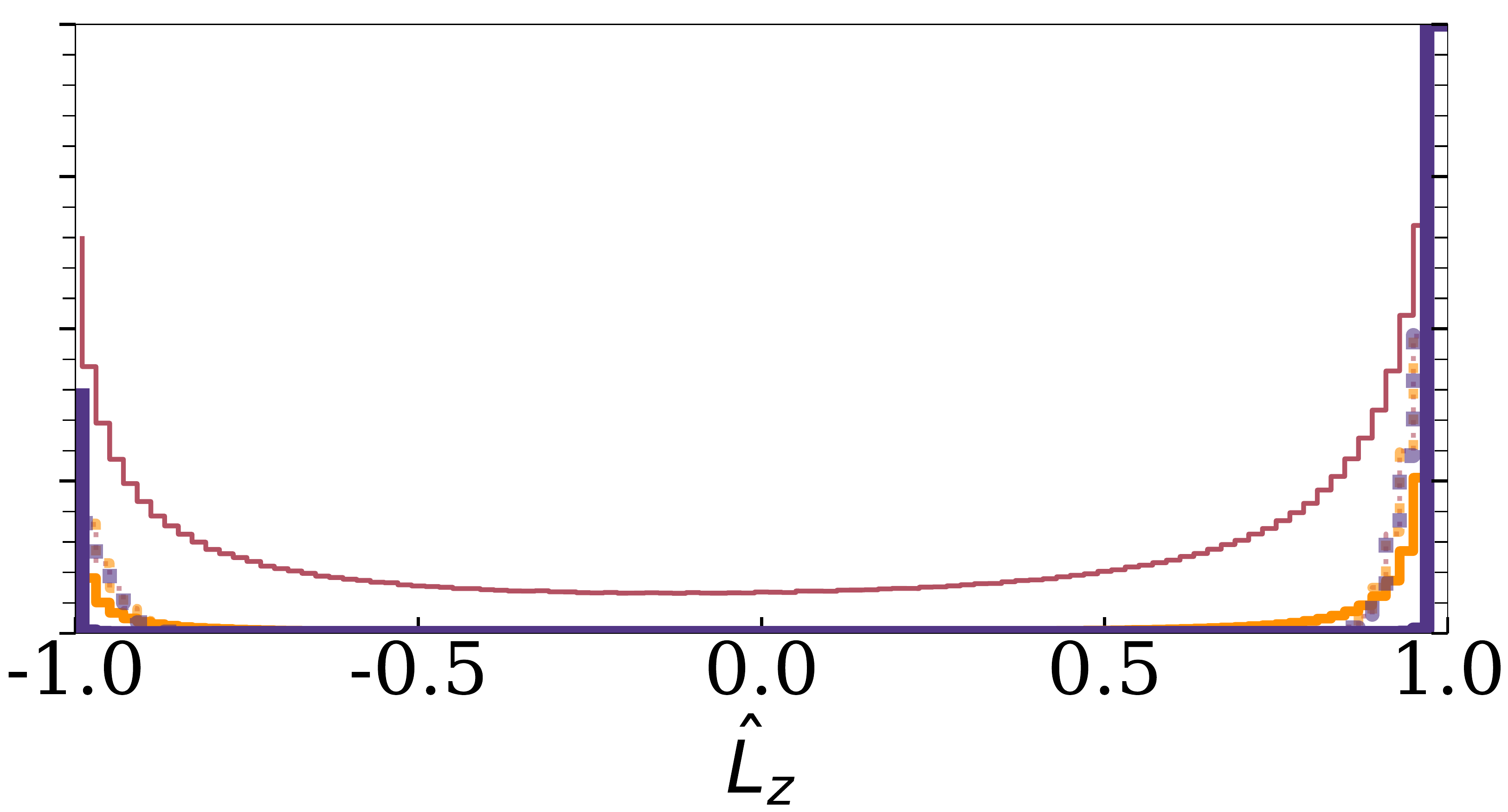}
    \end{subfigure}\hfil
    \begin{subfigure}[b]{0.33\textwidth}
      \centering
      \includegraphics[width=1.\textwidth]{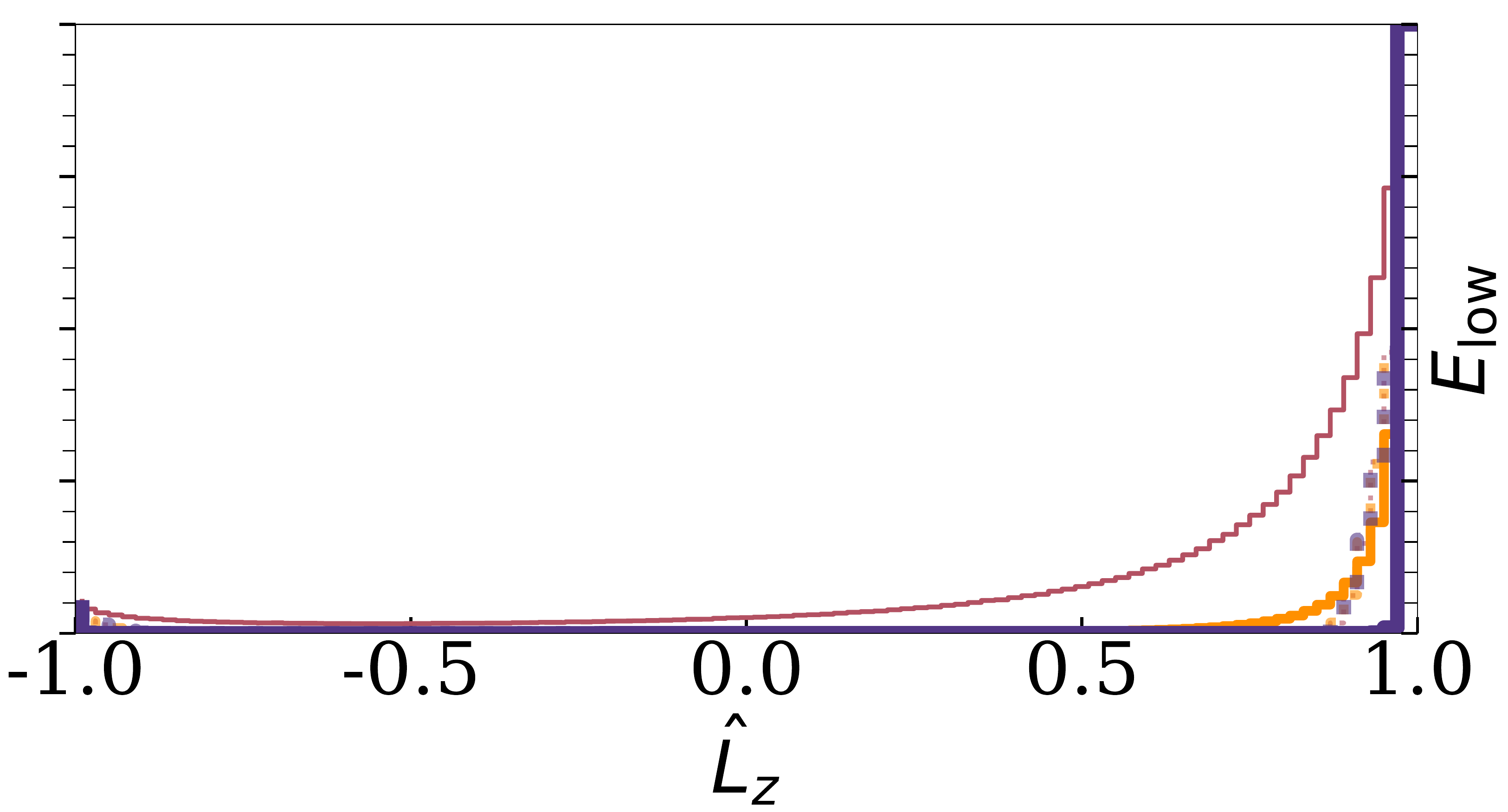}
    \end{subfigure}
\caption{The distributions of the angular momentum unit vector $z$ components, $\hat{L}_z$, (i.e. the cosine of the inclination angle) for different mass populations for systems with different $(E_{\rm tot},L_{\rm tot})$ in different panels. The initial and the final distributions are shown with faint dash-dotted and solid lines, respectively. Burgundy, orange and violet lines show the respective distributions for the mass populations ranges $1\leq m/m_{\rm min}<2$, $2\leq m/m_{\rm min}<16$, $m/m_{\rm min}\geq16$. The distribution functions are ensemble-averaged over 100 different realisations and also averaged over the last $3.35\times 10^6$ MCMC steps of the simulations. The distributions are normalised by the sample size and the maximum bin value for the different mass populations, separately. Different panels show systems with different total VRR energy increasing from top row to bottom, and different net rotation increasing from left column to the right column as specified in Table~\ref{tab:regions}. In all cases, the distributions show evidence of anistropic mass segregation. In comparison to the initial distributions, the equilibrium distributions are more spherical for the low mass components and more flattened for the high mass components. 
The heavy particles settle into a disc-like equilibrium distribution even for the high $E_{\rm tot}$ cases. 
}
\label{fig:mSelectedTimeEnsembleAver}
\end{figure*}

We test the convergence of the Monte Carlo method using the Kolmogorov-Smirnov (KS) test with respect to the last simulation step shown in Figure~\ref{fig:ksTest}. The figure shows the progression of the KS-test at the $n^{\rm th}$ simulation step as a function of $n$. There are 3 groups of panels showing the respective outcomes for low, intermediate, and heavy objects with $1\leq m/m_{\rm min}< 2$, $2\leq m/m_{\rm min}< 16$, and $16\leq m/m_{\rm min}\leq 100$ from top to bottom as labelled. The number of stars in the three mass bins are $(N_{\rm light},N_{\rm intermediate},N_{\rm heavy})=(2058,1816,222)$ representing $(50.3\%, 44.3\%, 5.4\%)$ of all stars. The $3\times 3$ different subpanels correspond to simulations with different $(E_{\rm tot},L_{\rm tot})$ as labelled for the cases shown in Table~\ref{tab:regions}. For example the top left subpanels have $(E_{\rm tot},L_{\rm tot})=(-0.03,0.16)$, labelled $E_{\rm high}-L_{\rm low}$ (these are nearly isotropic configurations), the bottom right subpanels have $(-0.58,0.88)$  labelled $E_{\rm low}-L_{\rm high}$ (these are flattened unidirectional discs). The shaded region in each subpanel shows the range of possible outcomes among the 100 different realisations with the given $(E_{\rm tot},L_{\rm tot})$. The colored lines (orange, burgundy, blue) show the $10\%,50\%,90\%$ levels of the cumulative distribution between the 100 realisations. The output of the KS-test displays a mean convergence within a factor of $\sim 5.6\%$ during the final $3.3\times 10^6$ number of the proposed MCMC steps. The highest median convergence rate is for the light particles with $3.9\%$ and the lowest for the heavy particles with $8.5\%$. On average, all the convergence rates vary in this interval. The level of convergence is attained in a larger number of steps for higher mass objects, which is expected as the high mass stars are moved less in angular momentum direction space if paired with a low mass star and the acceptance rate is also lower. For low and intermediate mass stars, the distributions are clearly well-converged as they show no systematic change in the corresponding panels. However, convergence may not be fully complete for the highest mass stars in the most anisotropic lowest energy configurations with minimal total angular momentum. The statistical fluctuations are also higher for the heavy stars, as there are almost a factor 10 less of them in the cluster than light or intermediate mass stars, respectively.

\begin{figure*}
    \centering
    \begin{subfigure}[b]{.33\textwidth}
      \centering
      \includegraphics[width=1.\textwidth]{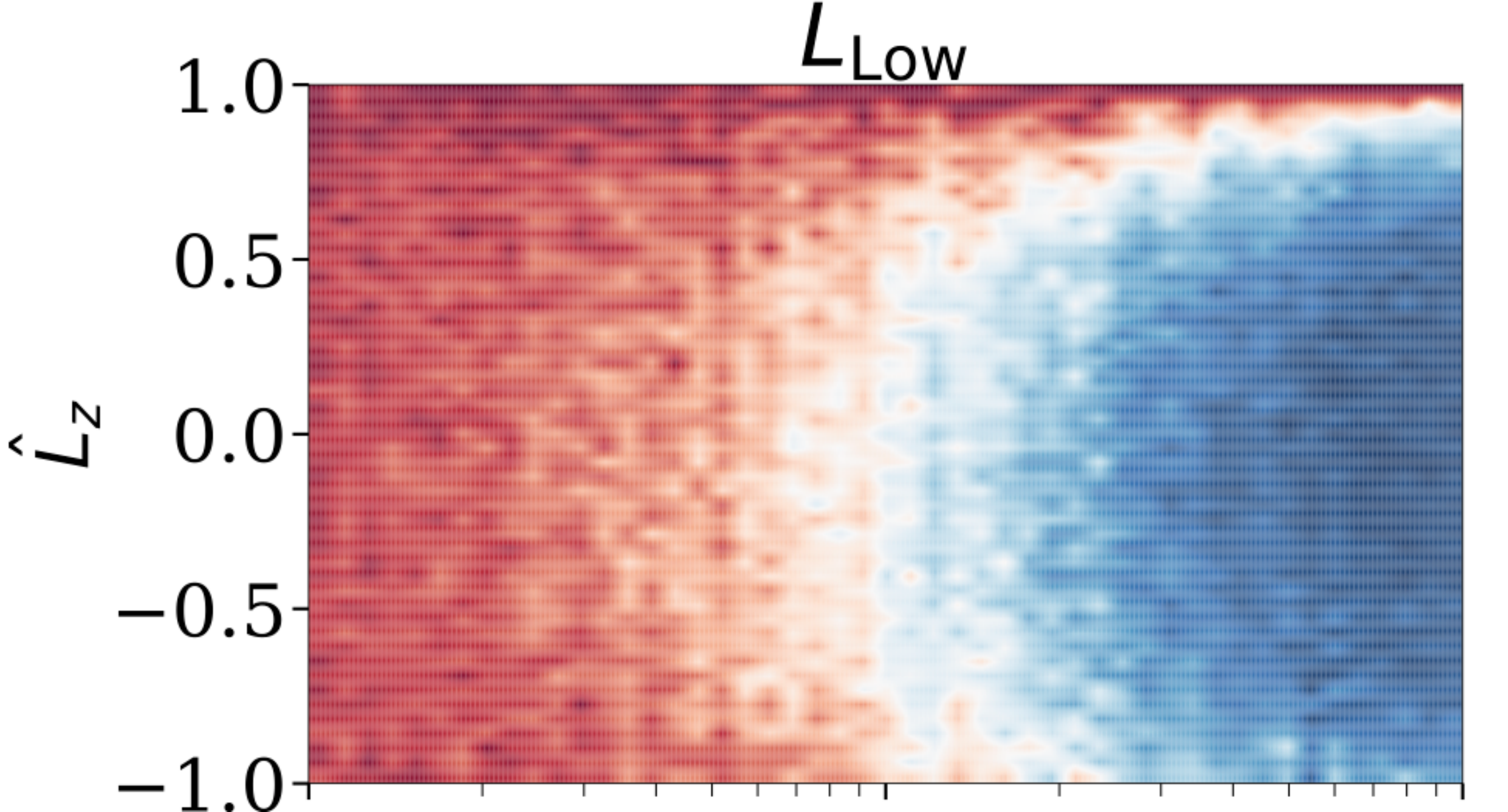}
    \end{subfigure}\hfil
    \begin{subfigure}[b]{.33\textwidth}
      \centering
      \includegraphics[width=1.\textwidth]{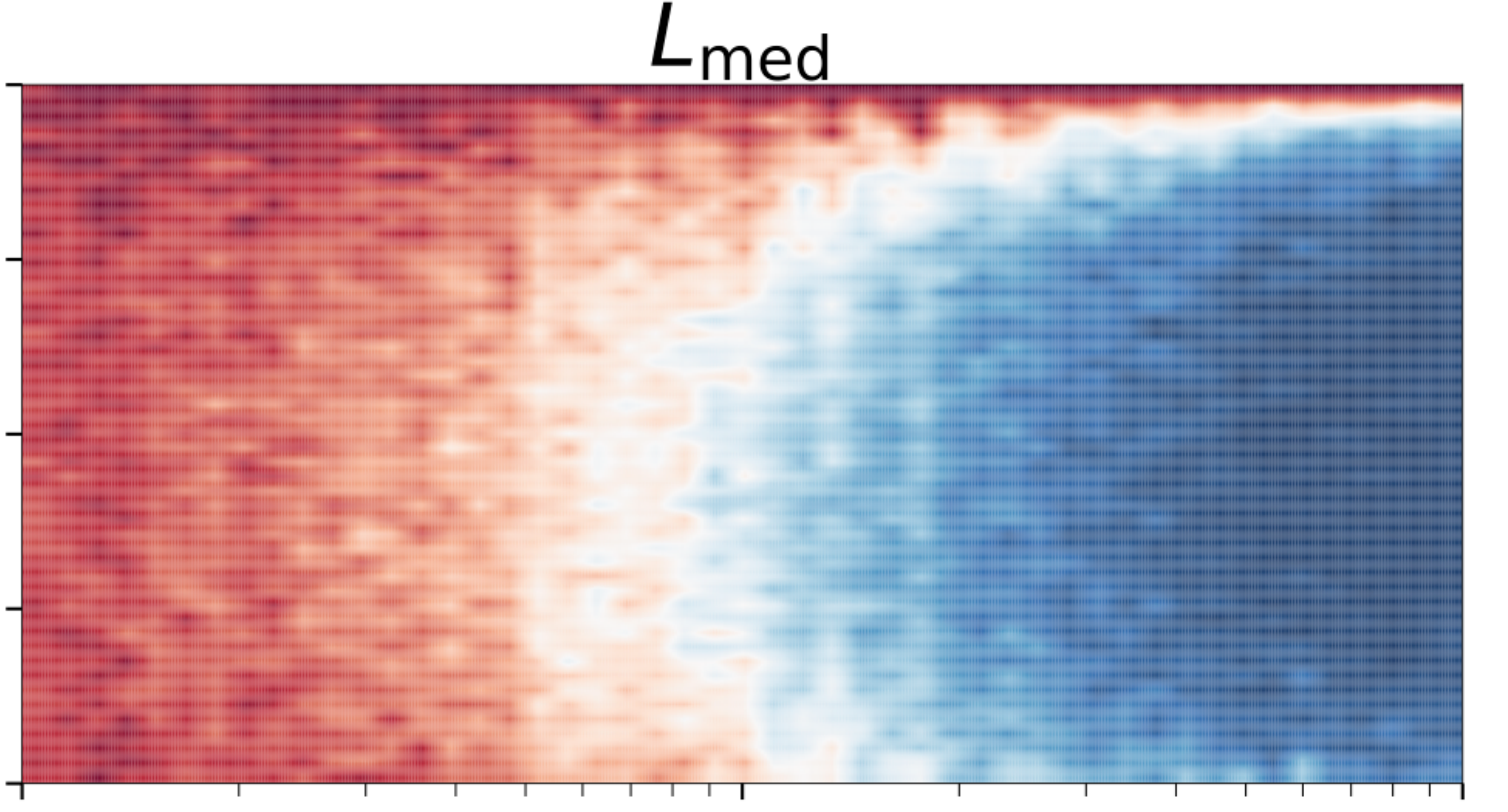}
    \end{subfigure}\hfil
    \begin{subfigure}[b]{.33\textwidth}
      \centering
      \includegraphics[width=1.\textwidth]{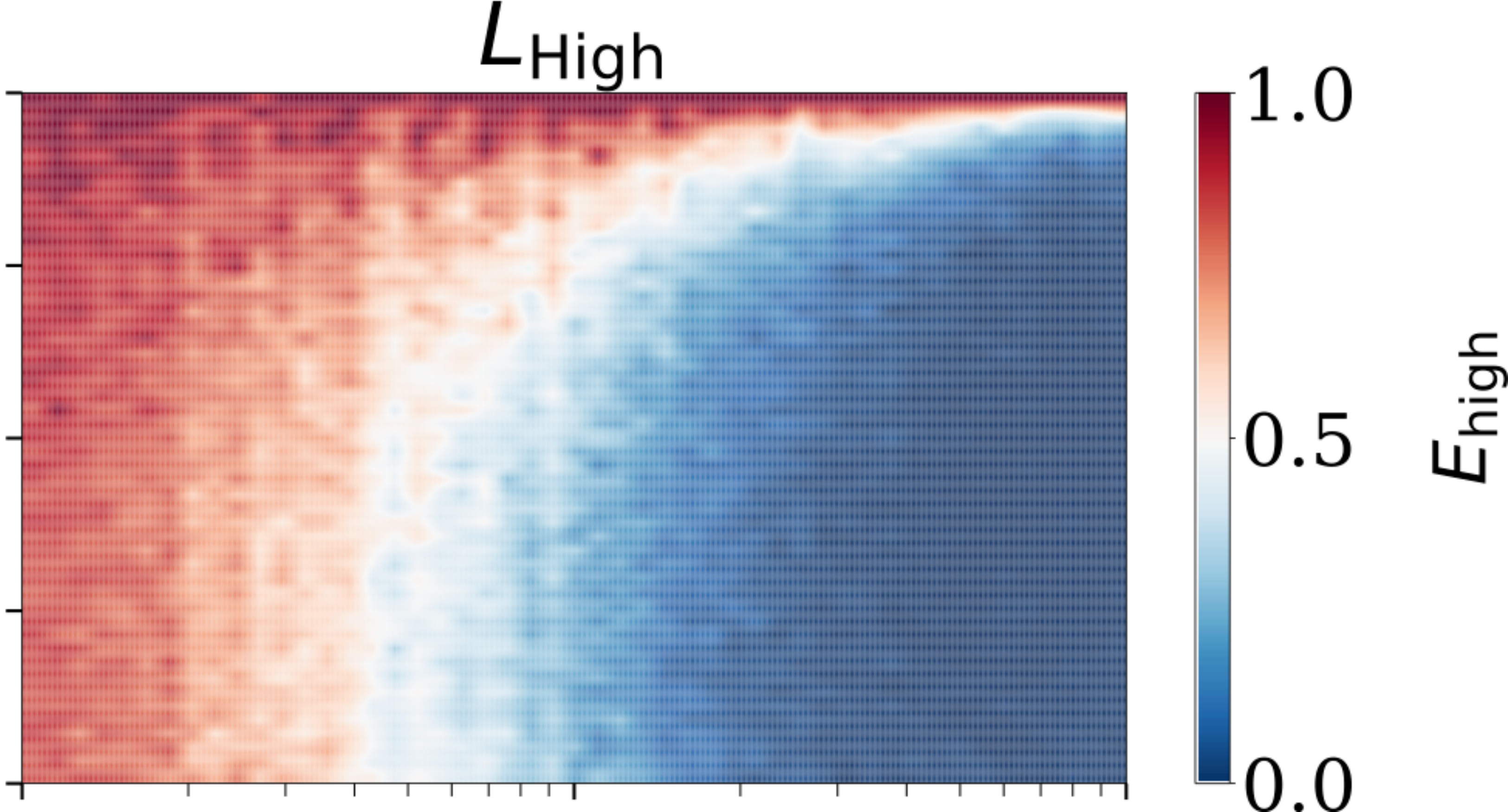}
    \end{subfigure}

    \medskip
    \begin{subfigure}[b]{0.33\textwidth}
      \centering
      \includegraphics[width=1.\textwidth]{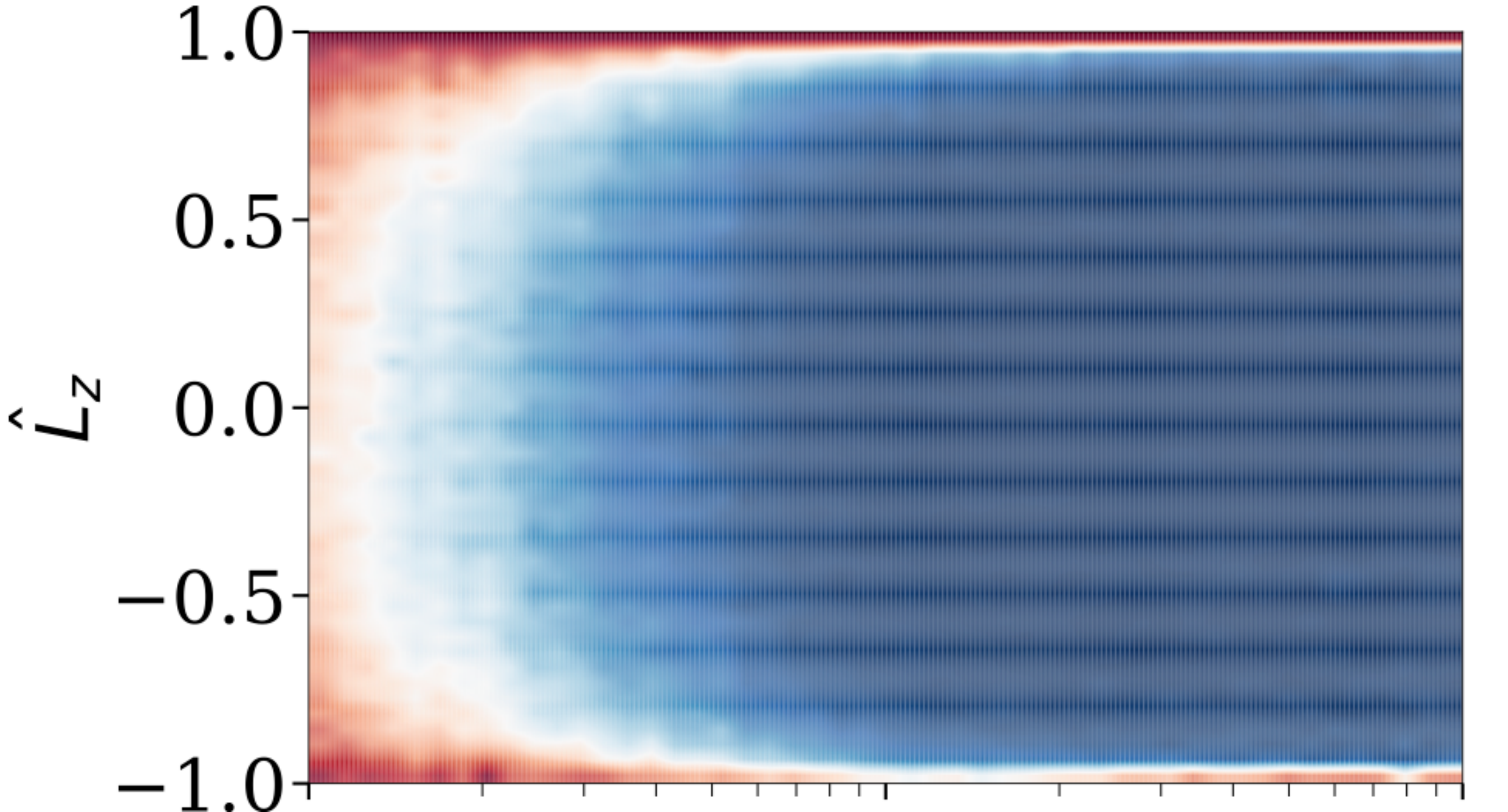}
    \end{subfigure}\hfil
    \begin{subfigure}[b]{0.33\textwidth}
      \centering
      \includegraphics[width=1.\textwidth]{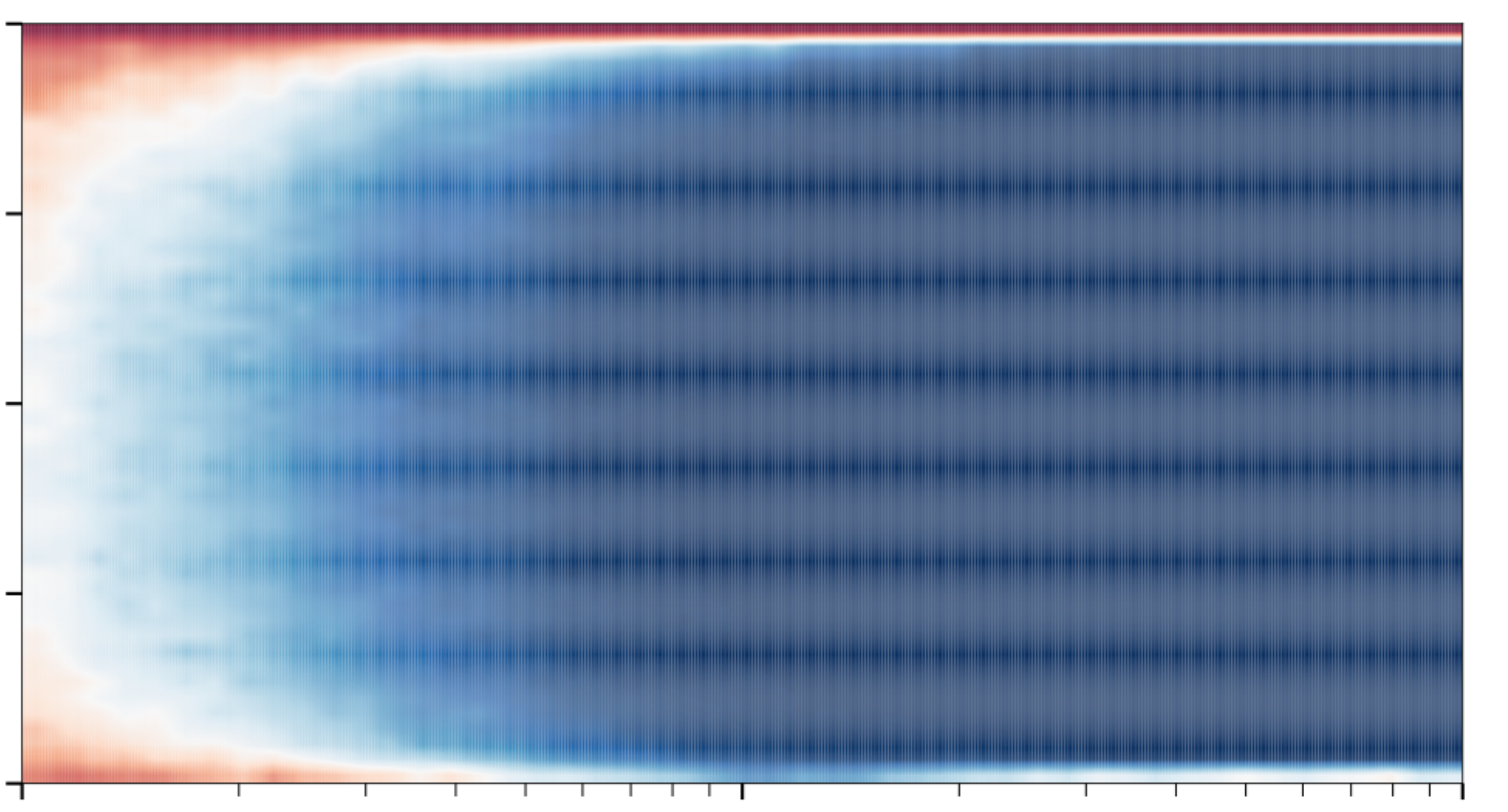}
    \end{subfigure}\hfil
    \begin{subfigure}[b]{0.33\textwidth}
      \centering
      \includegraphics[width=1.\textwidth]{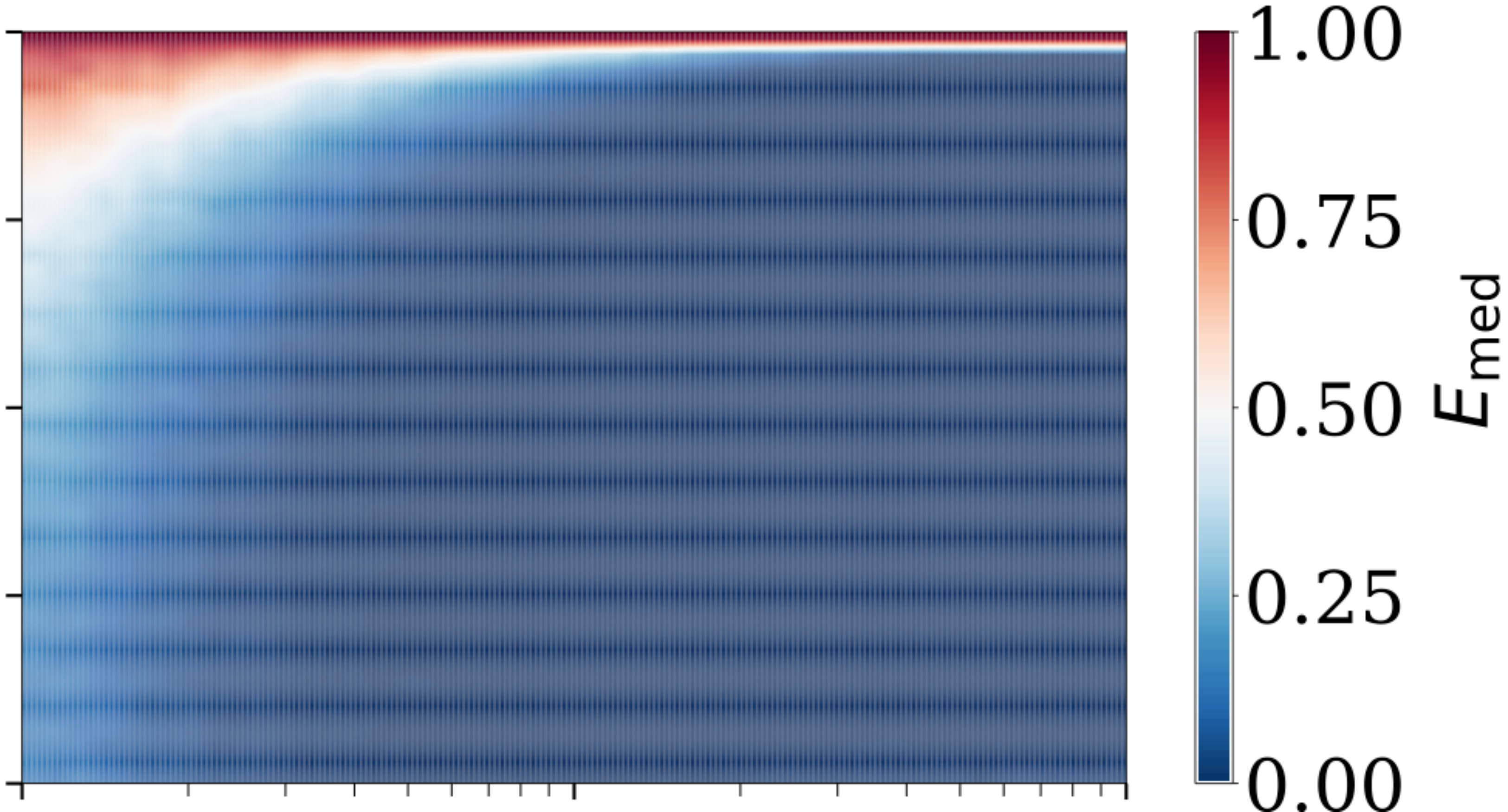}
    \end{subfigure}
    
    \medskip
    \begin{subfigure}[b]{0.33\textwidth}
      \centering
      \includegraphics[width=1.\textwidth]{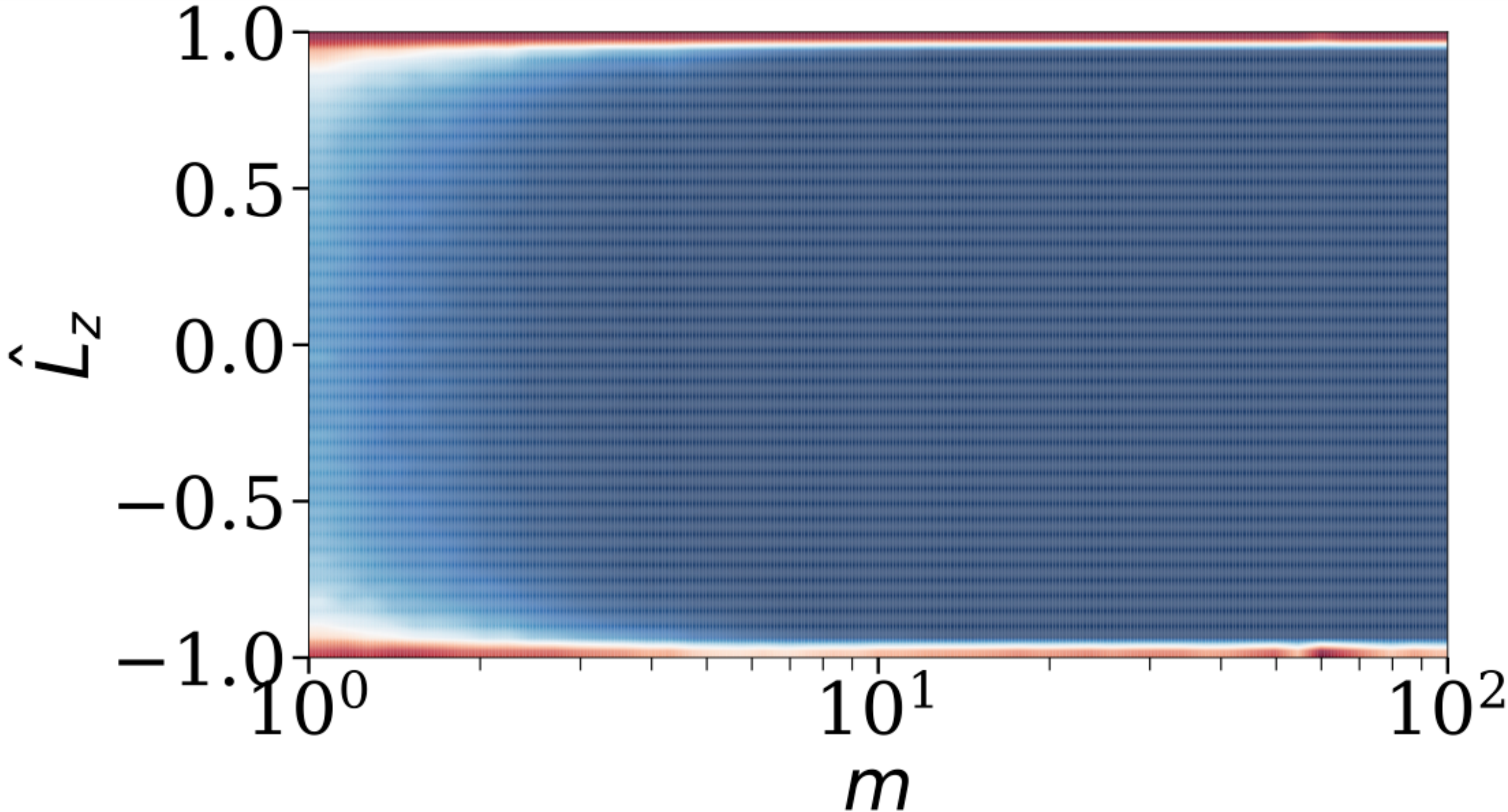}
    \end{subfigure}\hfil
    \begin{subfigure}[b]{0.33\textwidth}
      \centering
      \includegraphics[width=1.\textwidth]{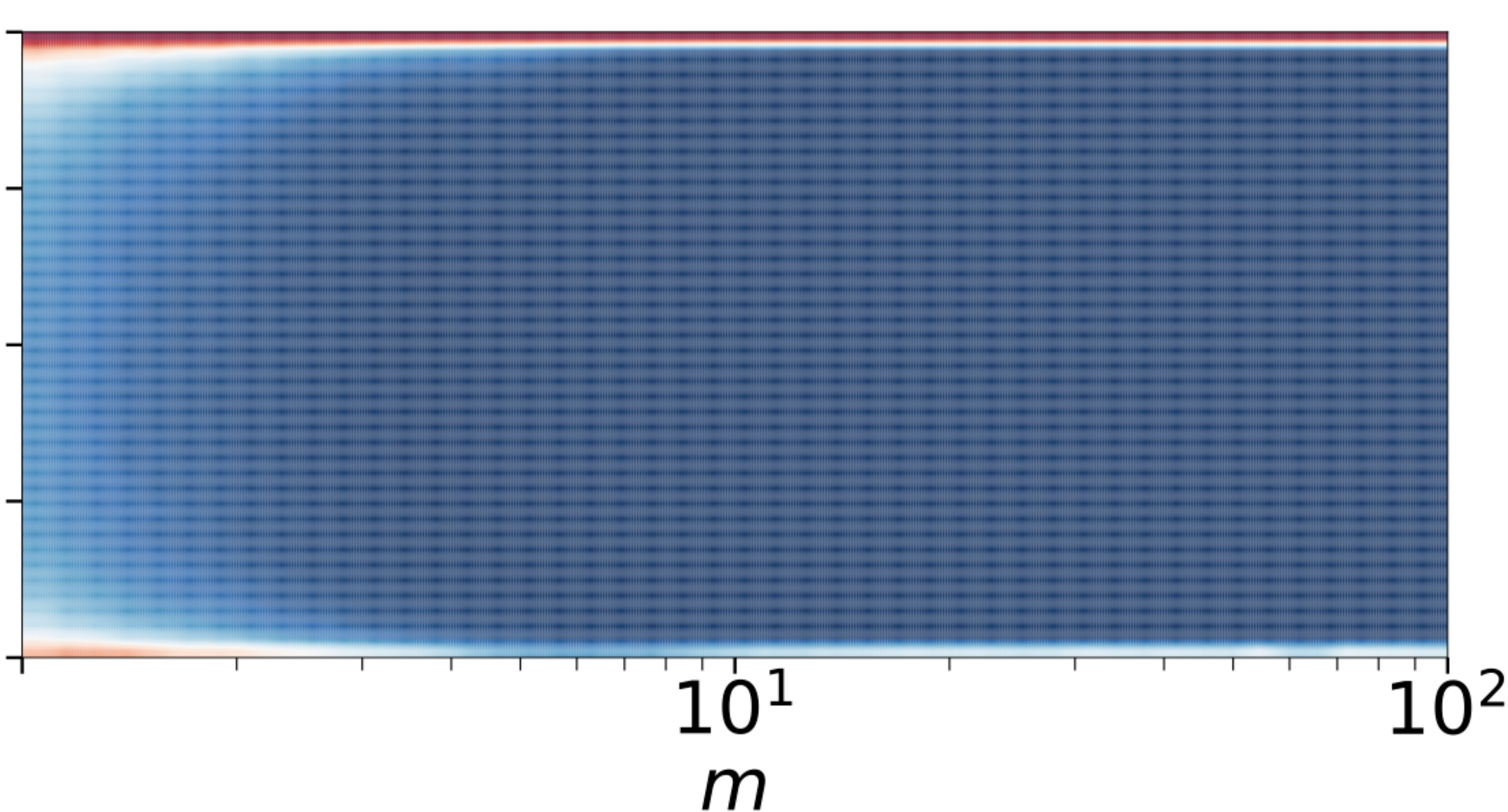}
    \end{subfigure}\hfil
    \begin{subfigure}[b]{0.33\textwidth}
      \centering
      \includegraphics[width=1.\textwidth]{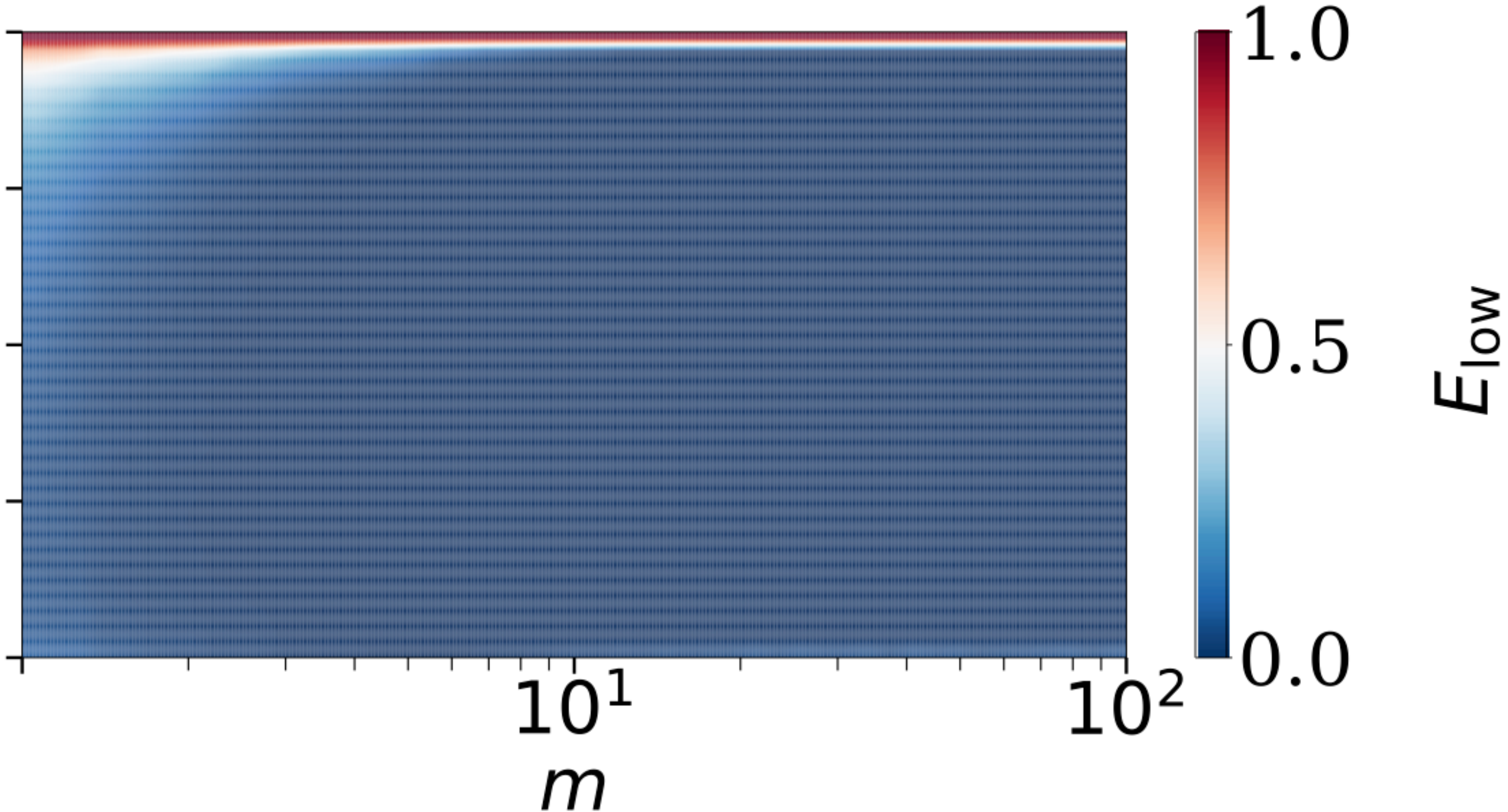}
    \end{subfigure}
\caption{
The two-dimensional equilibrium distribution of VRR as a function of $\hat{L}_z$ and $m$ (i.e. $\cos$ inclination and mass).
We stack the different realisations and the last simulation steps as in Figure~\ref{fig:mSelectedTimeEnsembleAver}, and normalise the histogram with the maximum value for each mass bin. Different panels show systems with different $(E_{\rm tot}, L_{\rm tot})$ as in Table~\ref{tab:regions}. The distribution is clearly more anisotropic for increasing mass decreasing $E_{\rm tot}$, as the angular momentum vector directions are more and more confined to the regions close to $\hat{L}_z=\pm 1$. This resembles a corotating and a counter-rotating disc in physical space. Note that the counter-rotating structure (region near $\hat{L}_z=- 1$) is most prominent for $L_{\rm low}$ and for low-mass stars.
}
\label{fig:cosThetaMassMaxBinNormed}
\end{figure*}

\begin{figure*}
\centering
\centering
    \begin{subfigure}[b]{.33\textwidth}
      \centering
      \includegraphics[width=1.\textwidth]{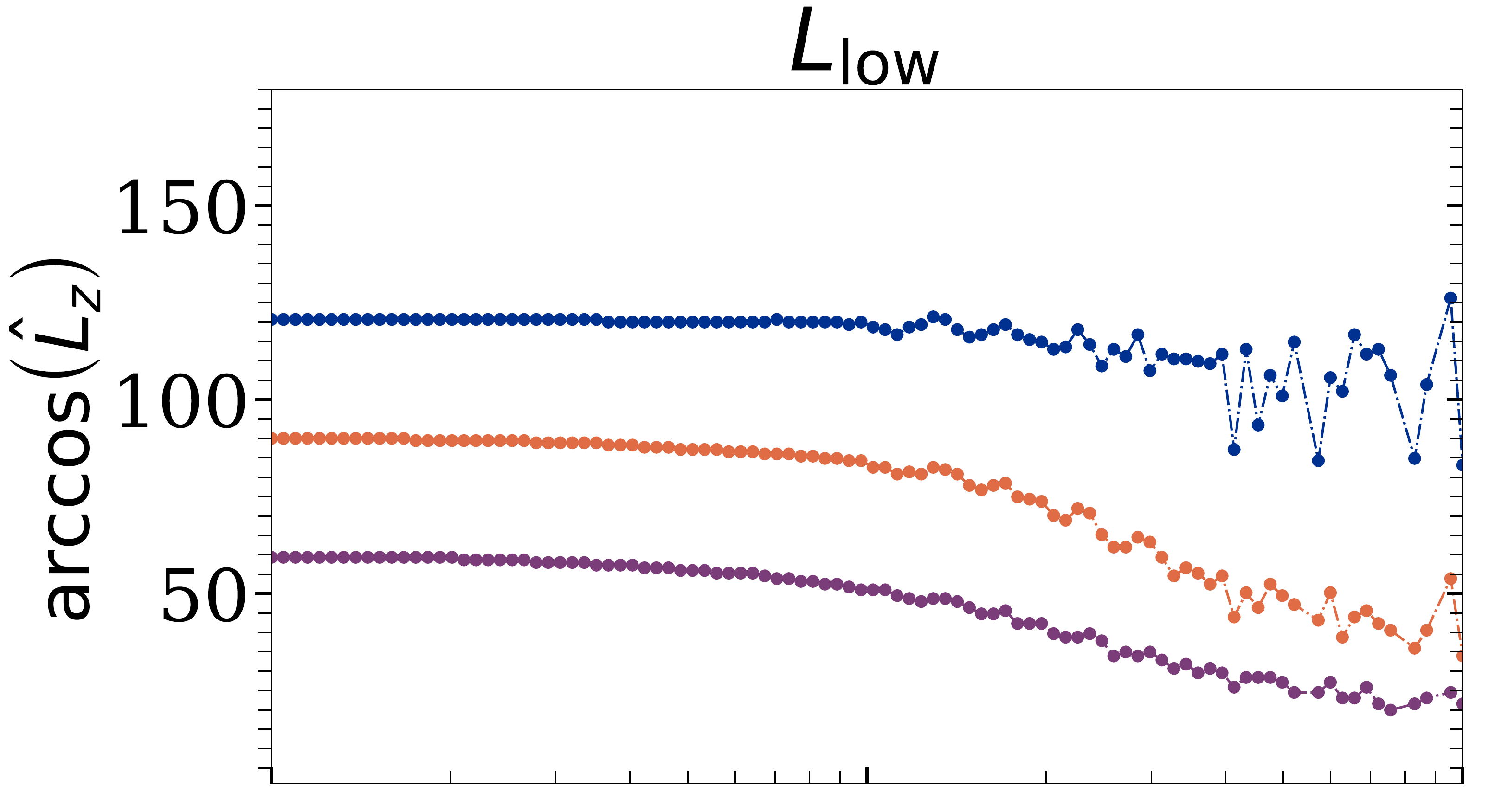}
    \end{subfigure}\hfil
    \begin{subfigure}[b]{.33\textwidth}
      \centering
      \includegraphics[width=1.\textwidth]{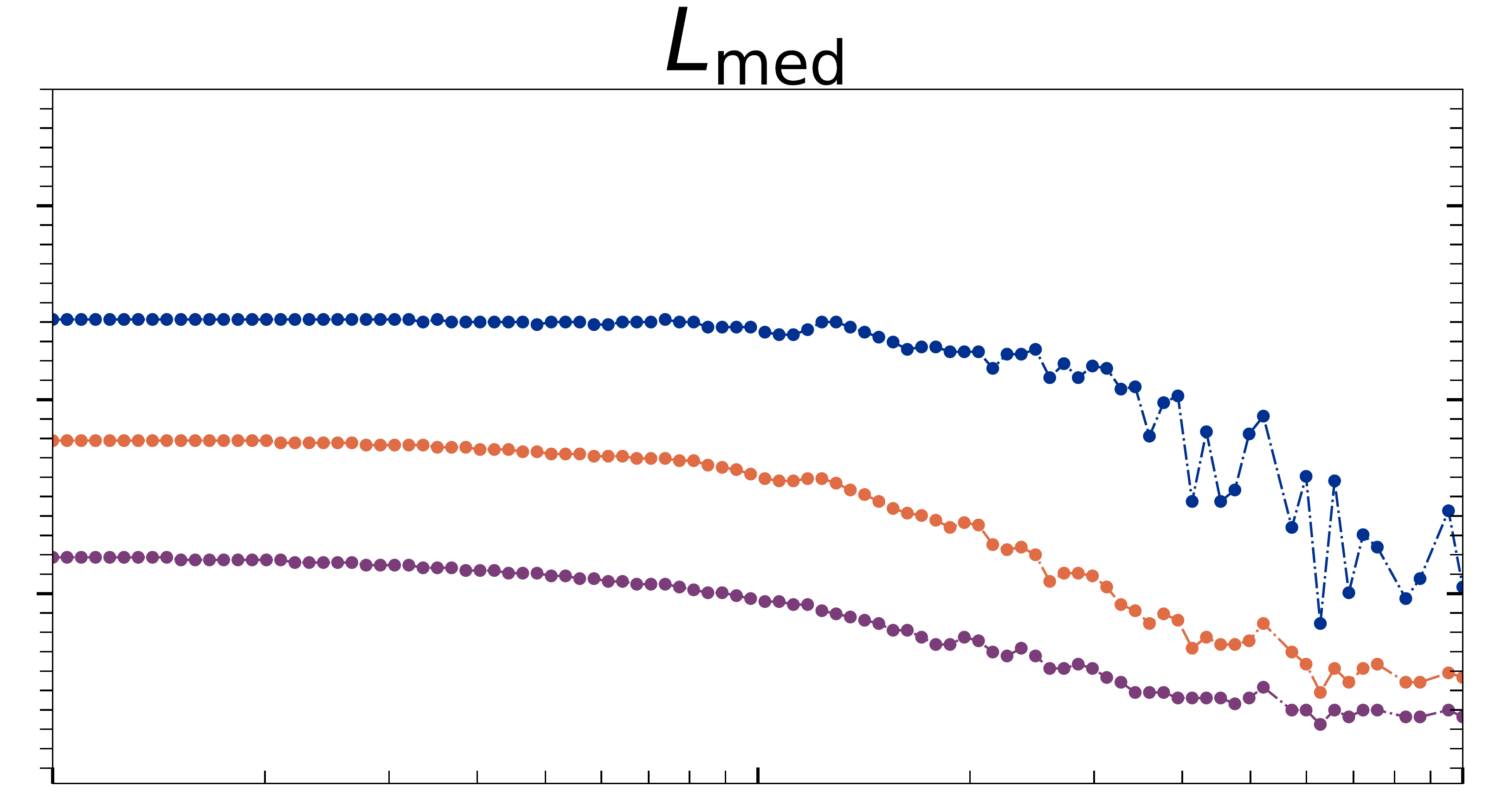}
    \end{subfigure}\hfil
    \begin{subfigure}[b]{.33\textwidth}
      \centering
      \includegraphics[width=1.\textwidth]{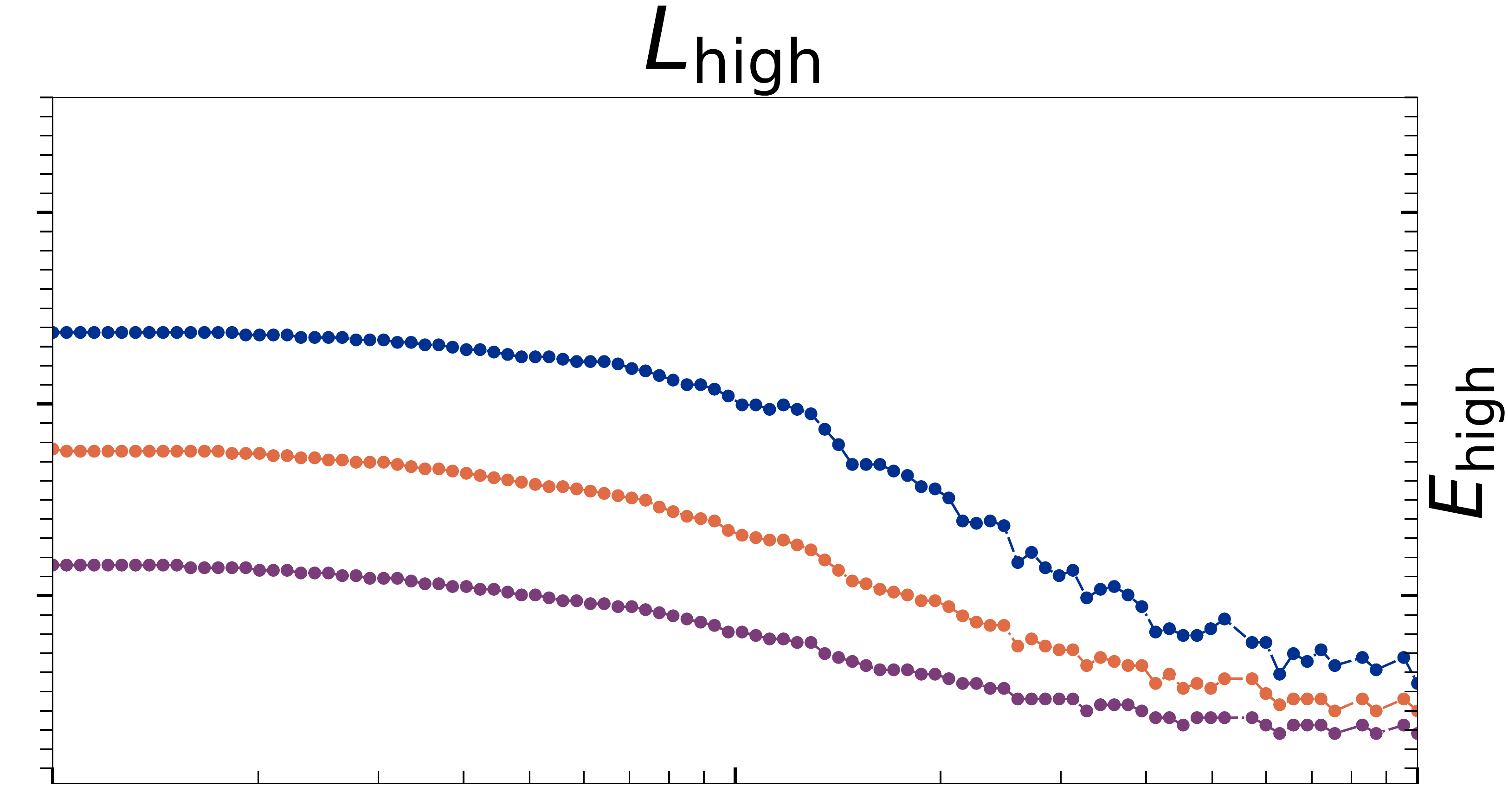}
    \end{subfigure}

    \medskip
    \begin{subfigure}[b]{0.33\textwidth}
      \centering
      \includegraphics[width=1.\textwidth]{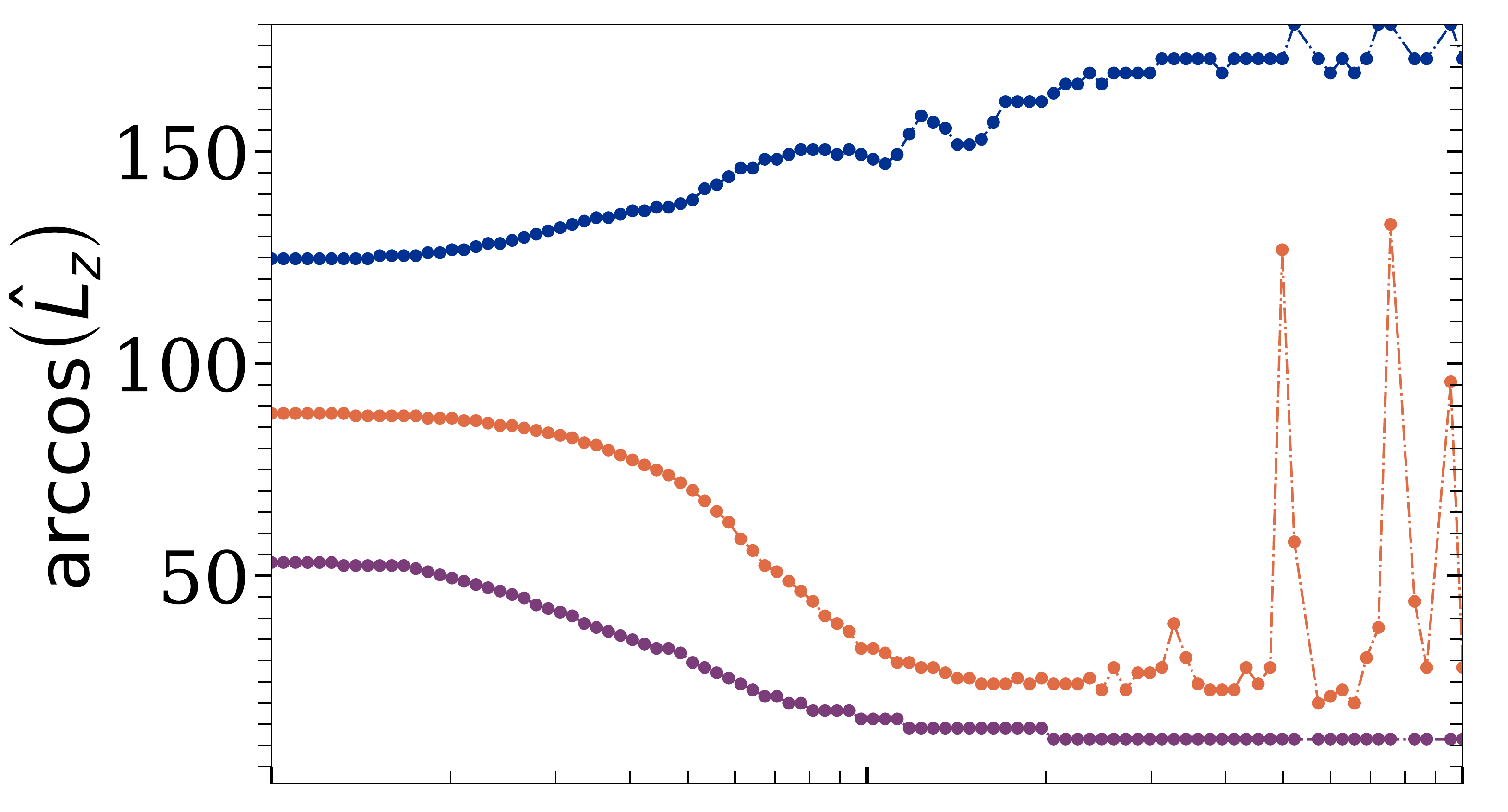}
    \end{subfigure}\hfil
    \begin{subfigure}[b]{0.33\textwidth}
      \centering
      \includegraphics[width=1.\textwidth]{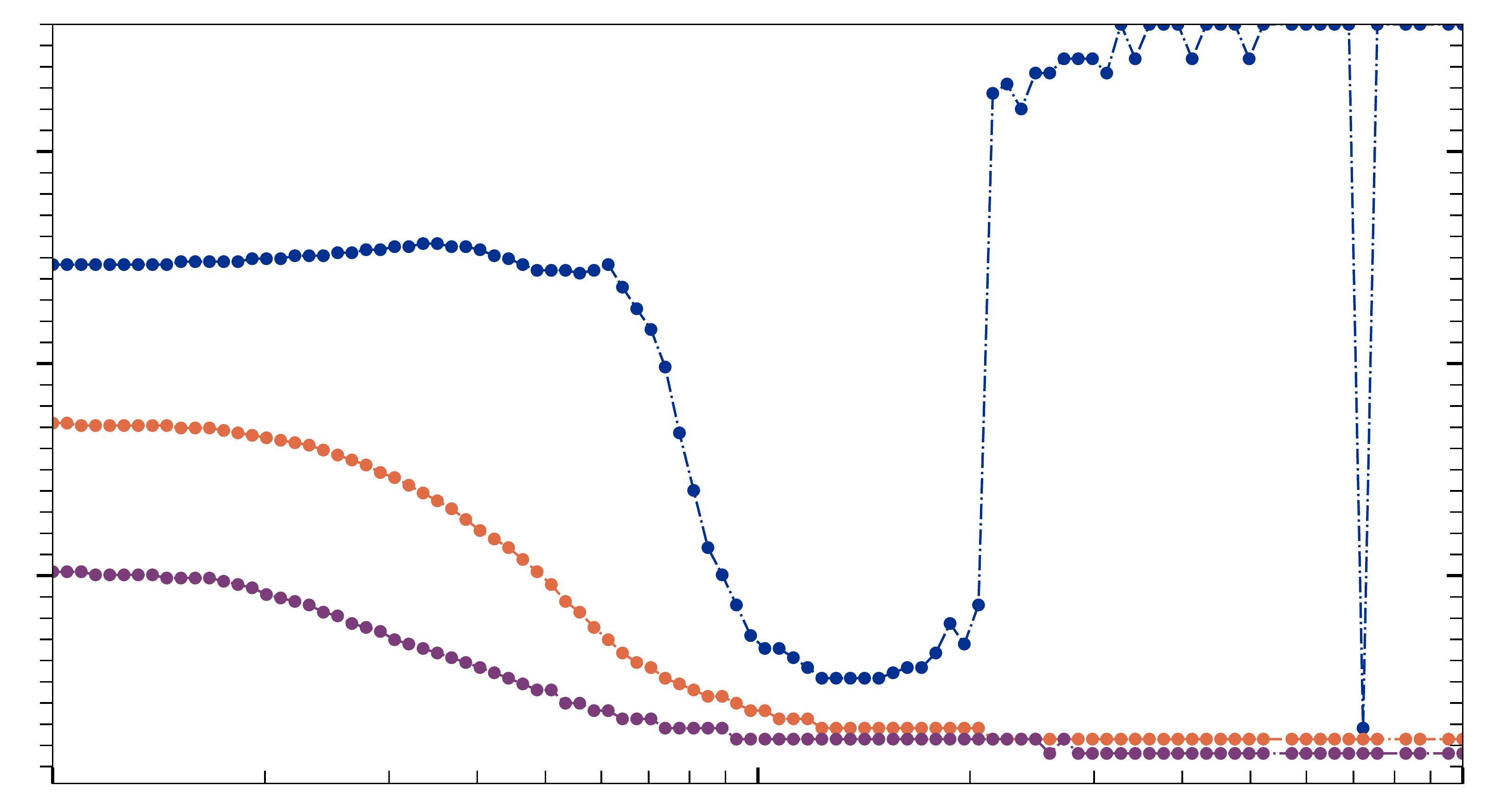}
    \end{subfigure}\hfil
    \begin{subfigure}[b]{0.33\textwidth}
      \centering
      \includegraphics[width=1.\textwidth]{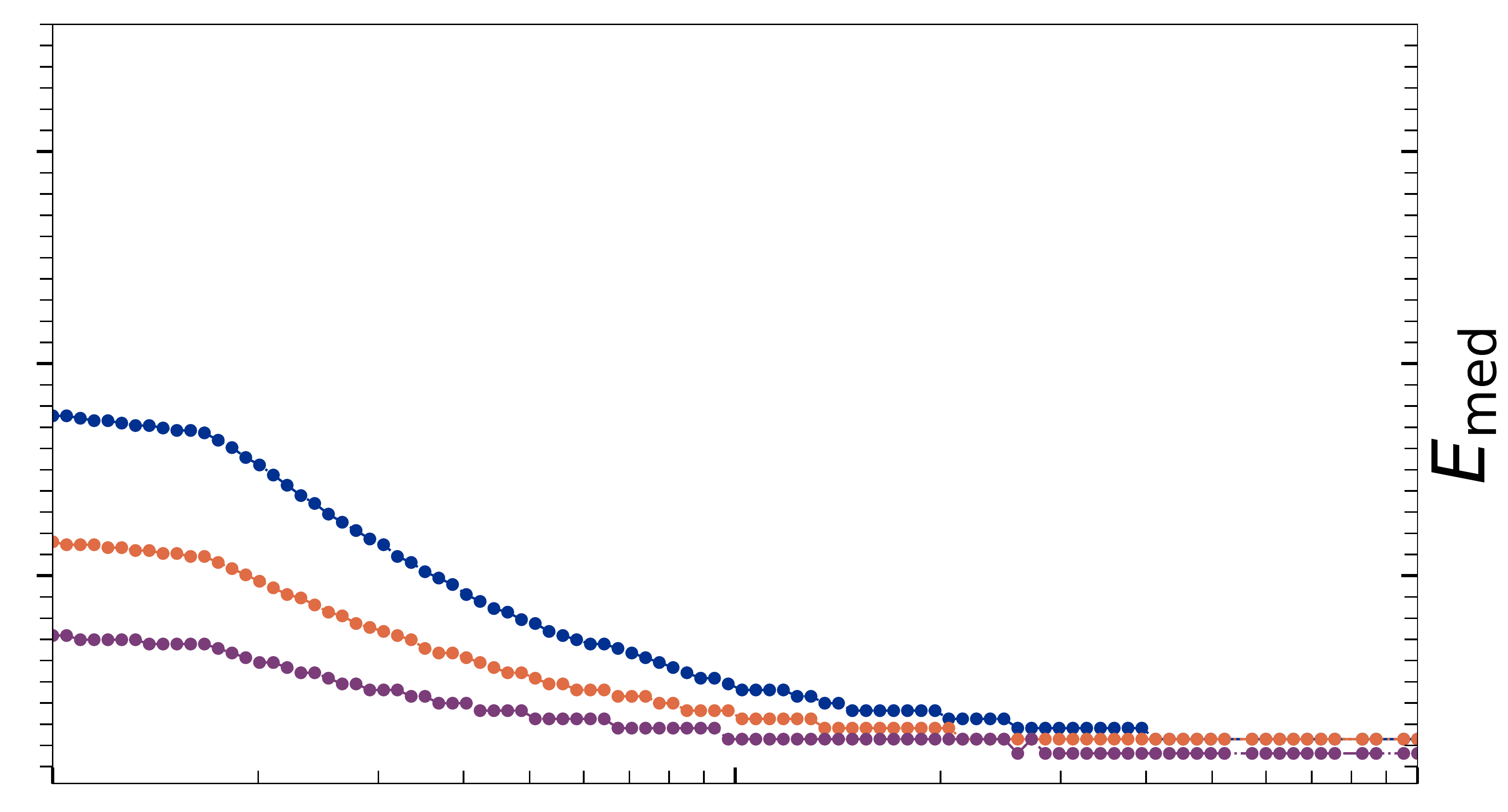}
    \end{subfigure}
    
    \medskip
    \begin{subfigure}[b]{0.33\textwidth}
      \centering
      \includegraphics[width=1.\textwidth]{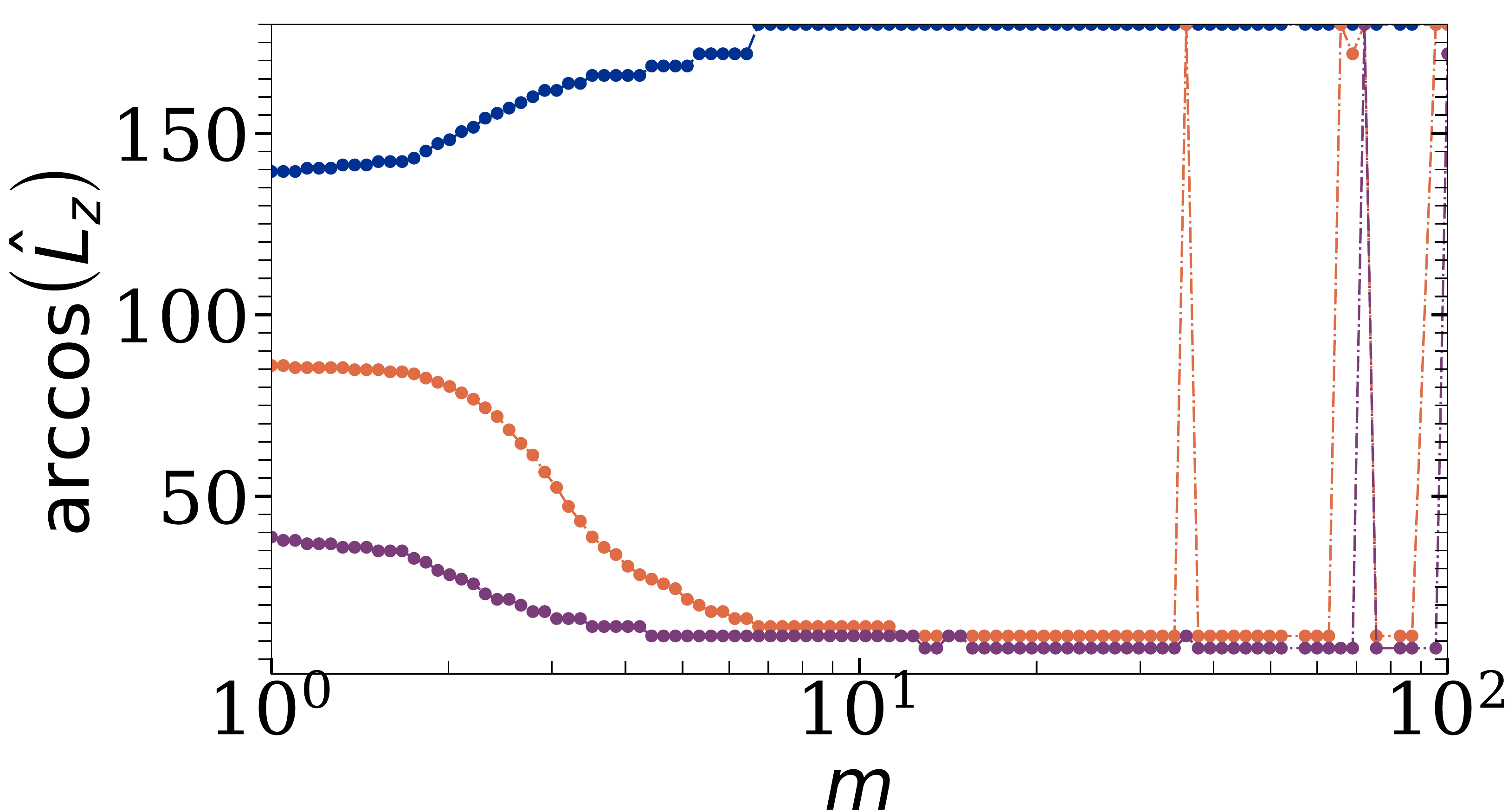}
    \end{subfigure}\hfil
    \begin{subfigure}[b]{0.33\textwidth}
      \centering
      \includegraphics[width=1.\textwidth]{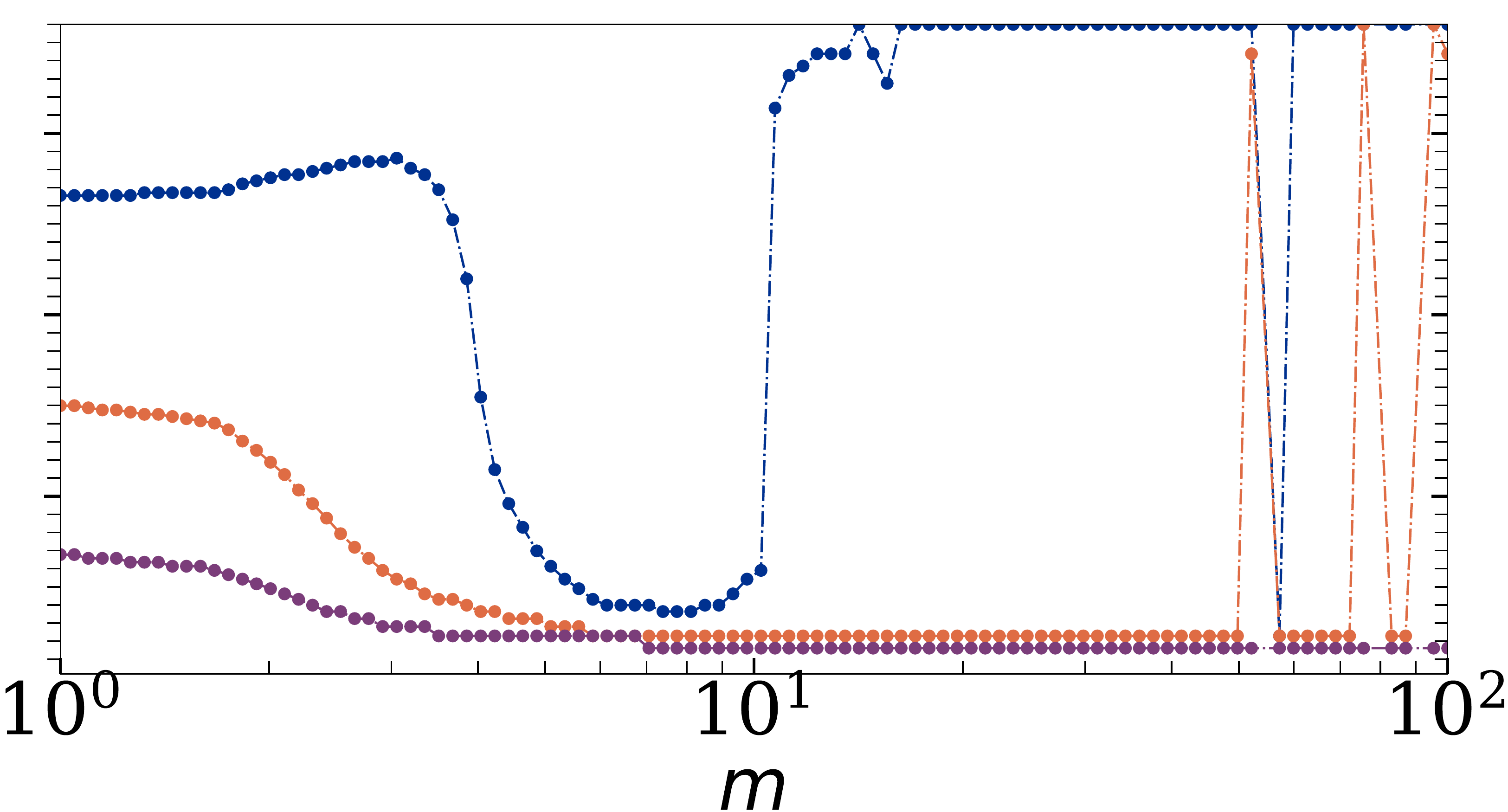}
    \end{subfigure}\hfil
    \begin{subfigure}[b]{0.33\textwidth}
      \centering
      \includegraphics[width=1.\textwidth]{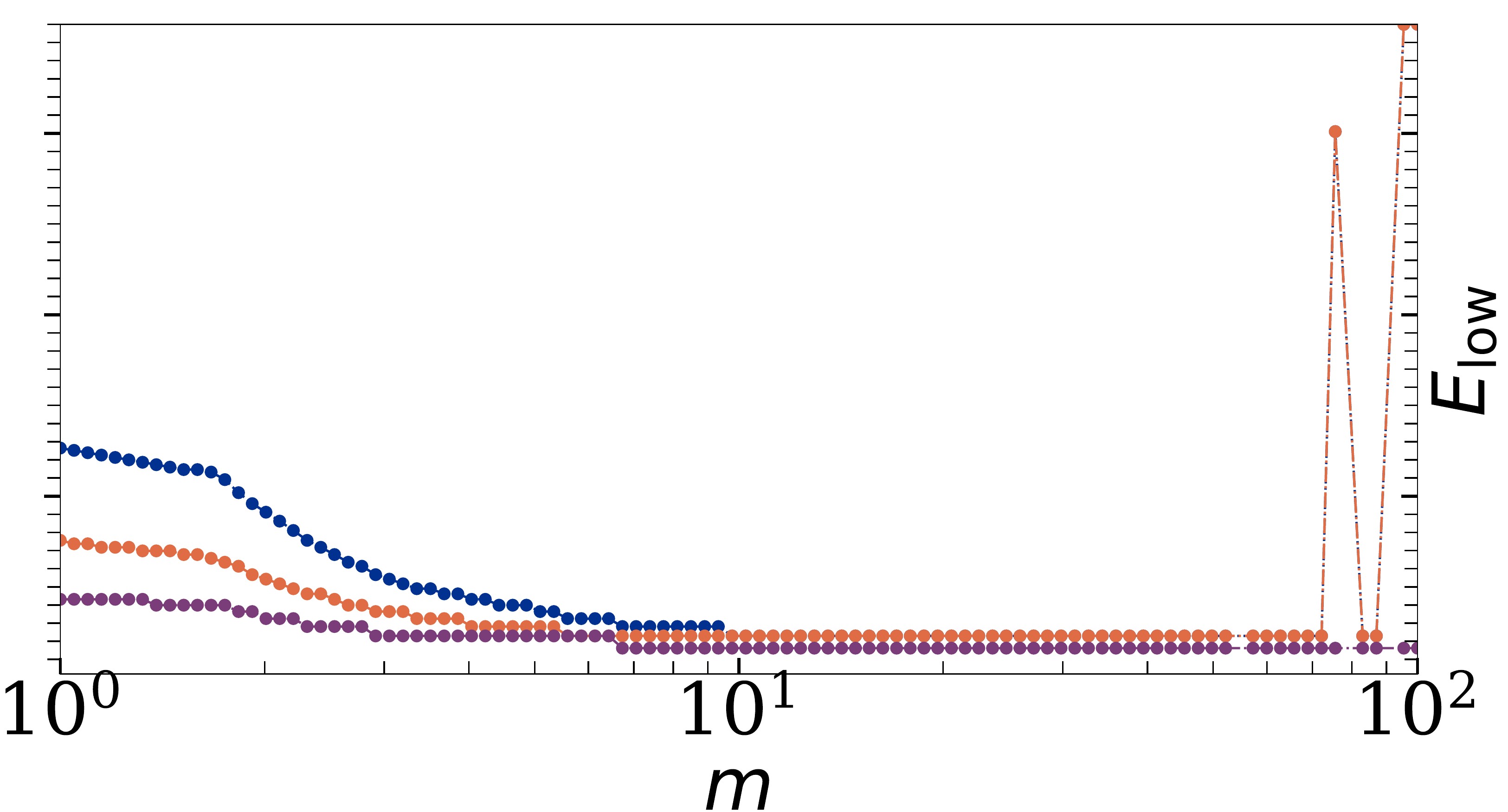}
    \end{subfigure}
\caption{
The ensemble and MCMC-step-averaged cumulative distributions of the inclination angle
$\cos^{-1}(\hat{L}_z)$ (shown in degrees) as a function of mass for systems with $(E_{\rm tot},L_{\rm tot})$ given by Table~\ref{tab:regions}. The lines show the moving average with respect to $m$ where $\lg_{10} (m/m_{\rm min})$ is between $\lg_{10} (m/m_{\rm min})\pm \lg_{10}(2)$. The blue, orange and violet colored dash-dotted lines show the $25\%,50\%\ \rm and\ 75\%$ of the cumulative distributions. High energy cases resemble a spherical distribution with a median inclination close to $90^{\circ}$ except for stars heavier than $m/m_{\rm min}\geq 10$. Heavy stars tend to systematically shift toward an inclination of 0 or $180^\circ$. For moderate and low energies the distribution is more flattened, where the disc thickness decreases with higher $m$. Note that for low $L_{\rm tot}$ (left panels) there is a flattened counter-rotating disc component implying that the $25\%$ and $75\%$ cumulative levels approach $0$ and $180^\circ$, especially for high masses and for low $E_{\rm tot}$ (see also Figure~\ref{fig:mSelectedTimeEnsembleAver}). Also note the the discontinuous variations for large $m$ are due to low number statistics. For $(E_{\rm med},L_{\rm med})$ and $(E_{\rm low},L_{\rm med})$, the trough shows that the counter-rotating fraction is does not reach $25\%$ for intermediate masses.
}
\label{fig:commulativeInclinationInMasses}
\end{figure*}

\begin{figure}
	\includegraphics[width=1.\columnwidth]{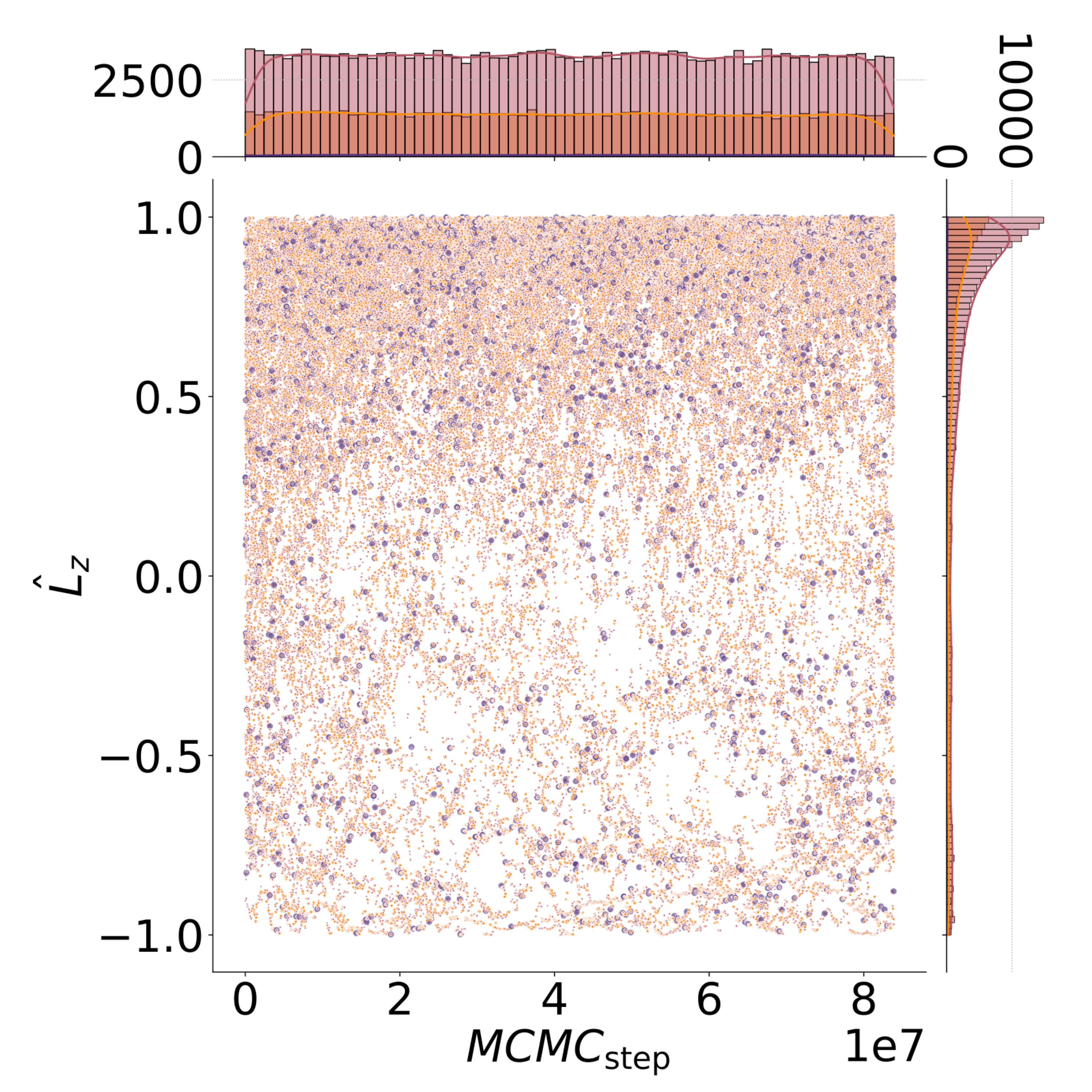}
    \caption{The distribution of the orbital inclination cosine $\hat{L}_{z}$ of the accepted steps for a particle with mass $m=45.81m_{\min}$, semimajor axis $a=12.95a_{\min}$ and eccentricity $e=0.24$ in accepted MCMC steps for an ensemble of 100 different simulation with high energy $E_{\rm high}$ and moderate total angular momentum $L_{\rm med}$. The color coding scheme matches Figures~\ref{fig:ksTest},~\ref{fig:mSelectedTimeEnsembleAver},~\ref{fig:commulativeInclinationInMasses}, and larger symbols denote larger interacting partner mass. The histograms on the sides show the marginal distributions in the MCMC steps (top) and in $\hat{L}_z$ (right). The former resembles a uniform distribution as expected where the average number of accepted steps decreases for increasing partner-mass, and the latter clearly indicates a systematic excess at $\hat{L}_z=1$ representing a disc structure. 
    }
    \label{fig:heavyAcceptedStepsEnsemDistribution}
\end{figure}

\begin{figure*}
\centering
\begin{subfigure}[b]{.5\textwidth}
      \centering
      \includegraphics[width=1.\textwidth]{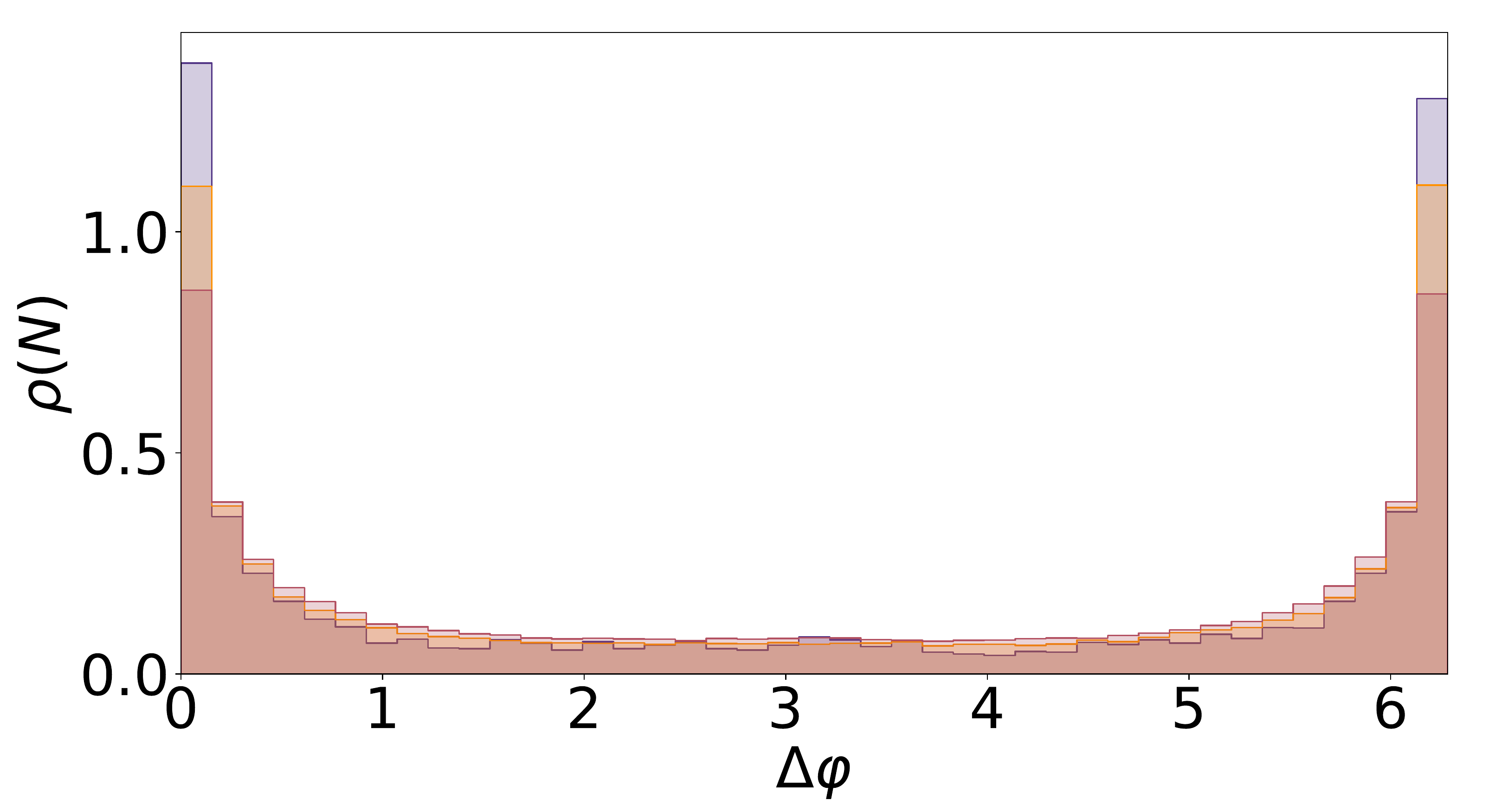}
    \end{subfigure}\hfil
    \begin{subfigure}[b]{.5\textwidth}
      \centering
      \includegraphics[width=1.\textwidth]{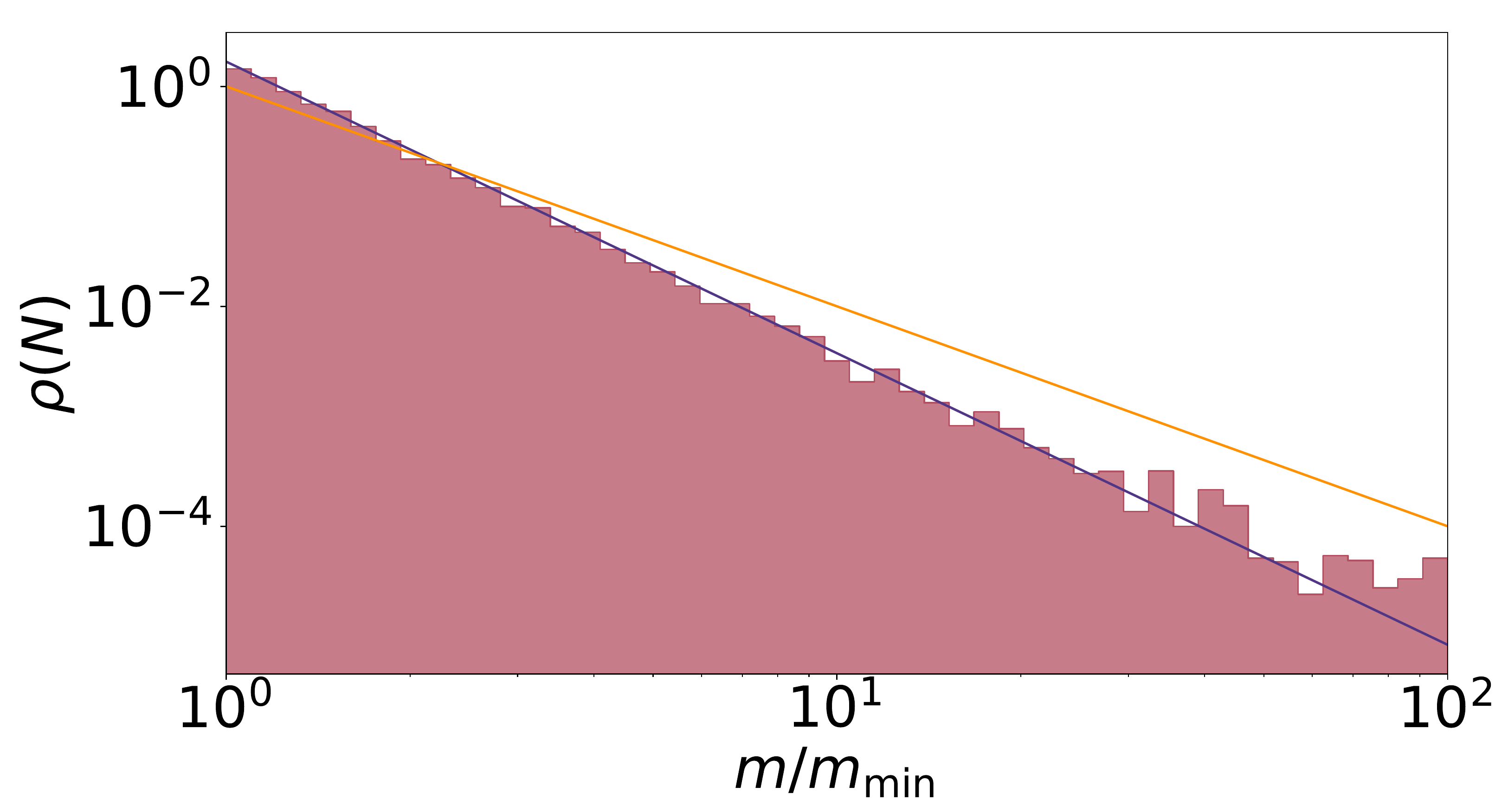}
    \end{subfigure}\hfil
\caption{The distribution of the rotation angle $\Delta\varphi$ (left) and the companion mass $m/m_{\rm min}$ (right) for the accepted steps of a particle with mass of $m/m_{\rm min}=45.81$, semimajor axis of $a/a_{\rm min}=12.95$ and eccentricity of $e=0.24$ in an ensemble of 100 different simulations with $(E_{\rm tot},L_{\rm tot})=(E_{\rm high},L_{\rm med})$ (see Tab.~\ref{tab:regions}). For the rotation angle, we used the same coloring scheme as for Figures~\ref{fig:ksTest},~\ref{fig:mSelectedTimeEnsembleAver},~\ref{fig:commulativeInclinationInMasses}. The figure shows that the distribution of light particles tends to populate the region around $\pi$ more, which corresponds to large rotation angles. While with increasing mass small rotation angles are favored which corresponds to the higher peaks of moderate and large masses particles. The double peaked distribution of $\Delta\varphi$ is a consequence of the fact that $-\Delta\varphi=2\pi-\Delta\varphi$. The distribution of companion masses $\propto m^{-2.65}$ (fitted with violet) deviates slightly from the initial one of $\propto m^{-2}$ (orange). Interactions with light particles are over represented compared to those with heavy particles. This is the natural consequence of the fact that interaction of heavy particles gets rejected on average more often since that corresponds to a higher $\Delta E$ energy change on average. 
    }
\label{fig:heavyParticlePartnerStatistics}
\end{figure*}

\begin{figure*}
\centering
\begin{subfigure}[b]{.33\textwidth}
      \centering
      \includegraphics[width=1.\textwidth]{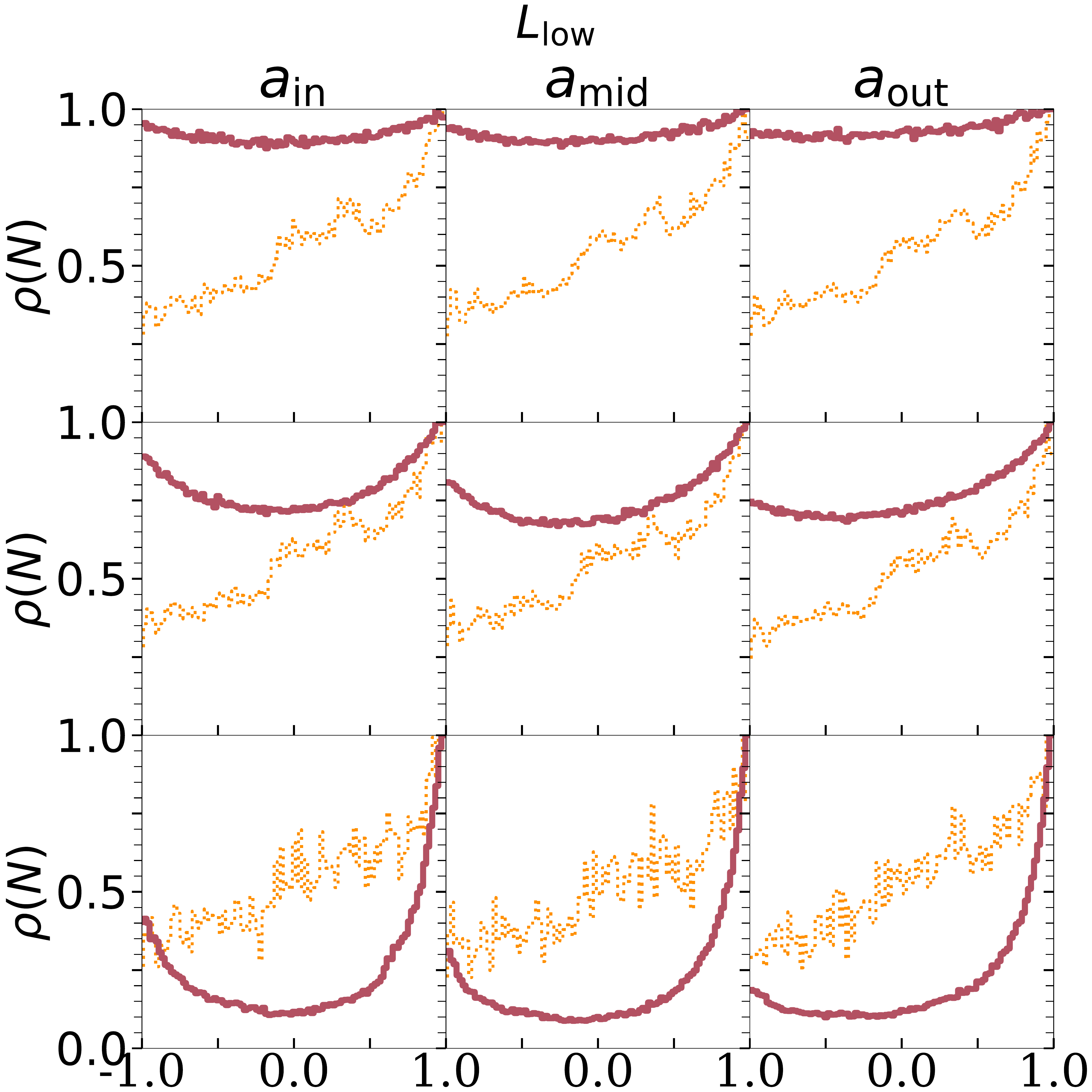}
    \end{subfigure}\hfil
    \begin{subfigure}[b]{.33\textwidth}
      \centering
      \includegraphics[width=1.\textwidth]{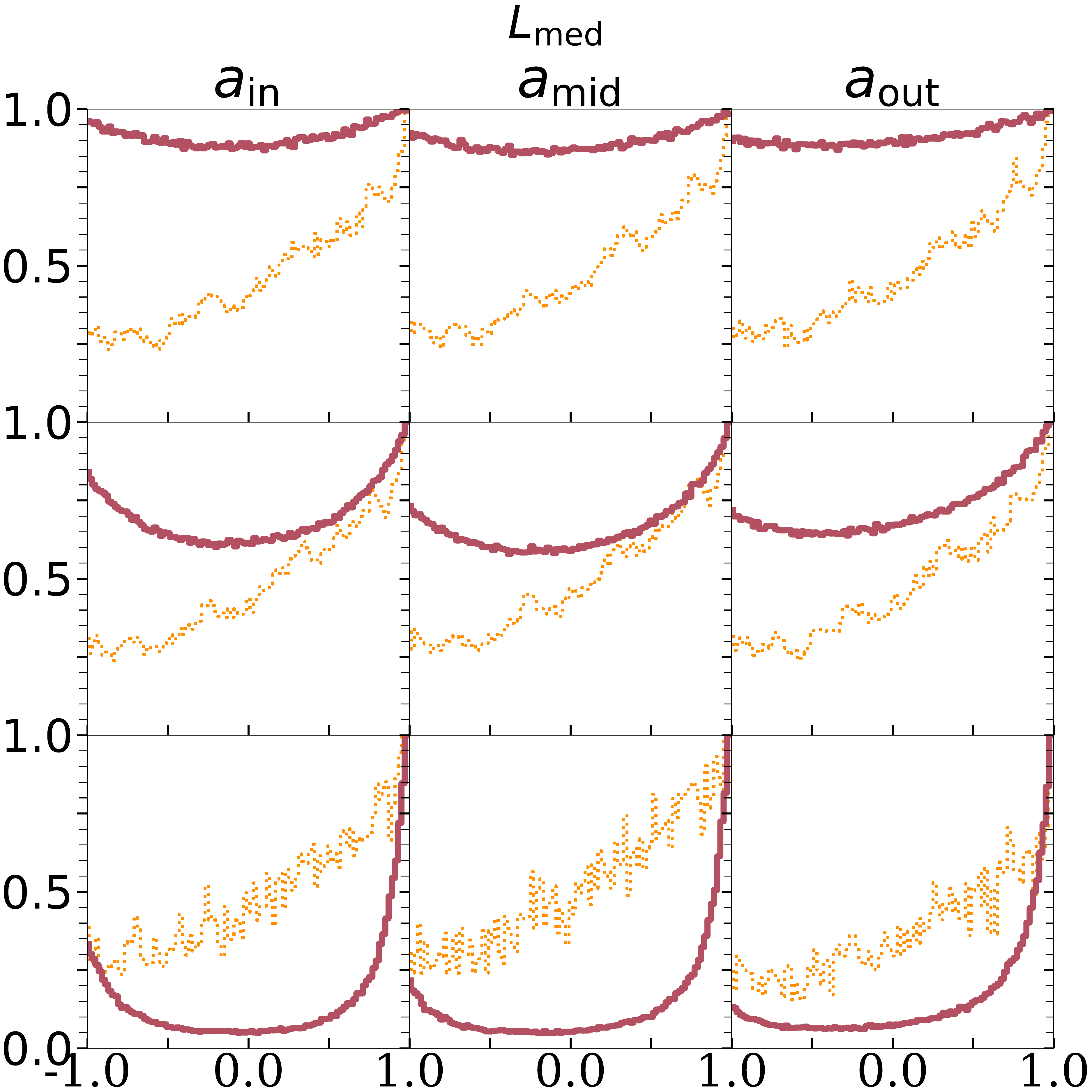}
    \end{subfigure}\hfil
    \begin{subfigure}[b]{.33\textwidth}
      \centering
      \includegraphics[width=1.\textwidth]{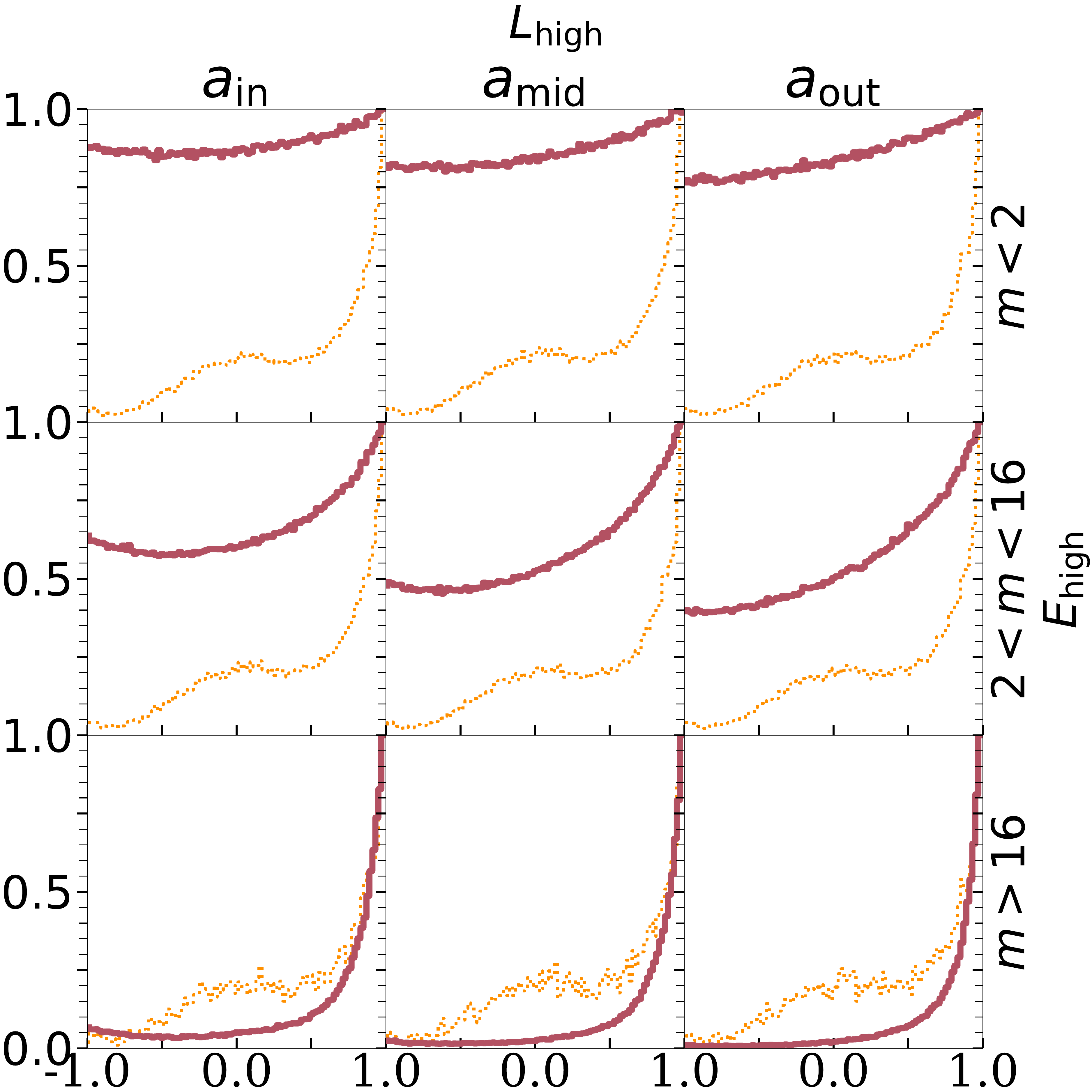}
    \end{subfigure}

    \medskip
    \begin{subfigure}[b]{0.33\textwidth}
      \centering
      \includegraphics[width=1.\textwidth]{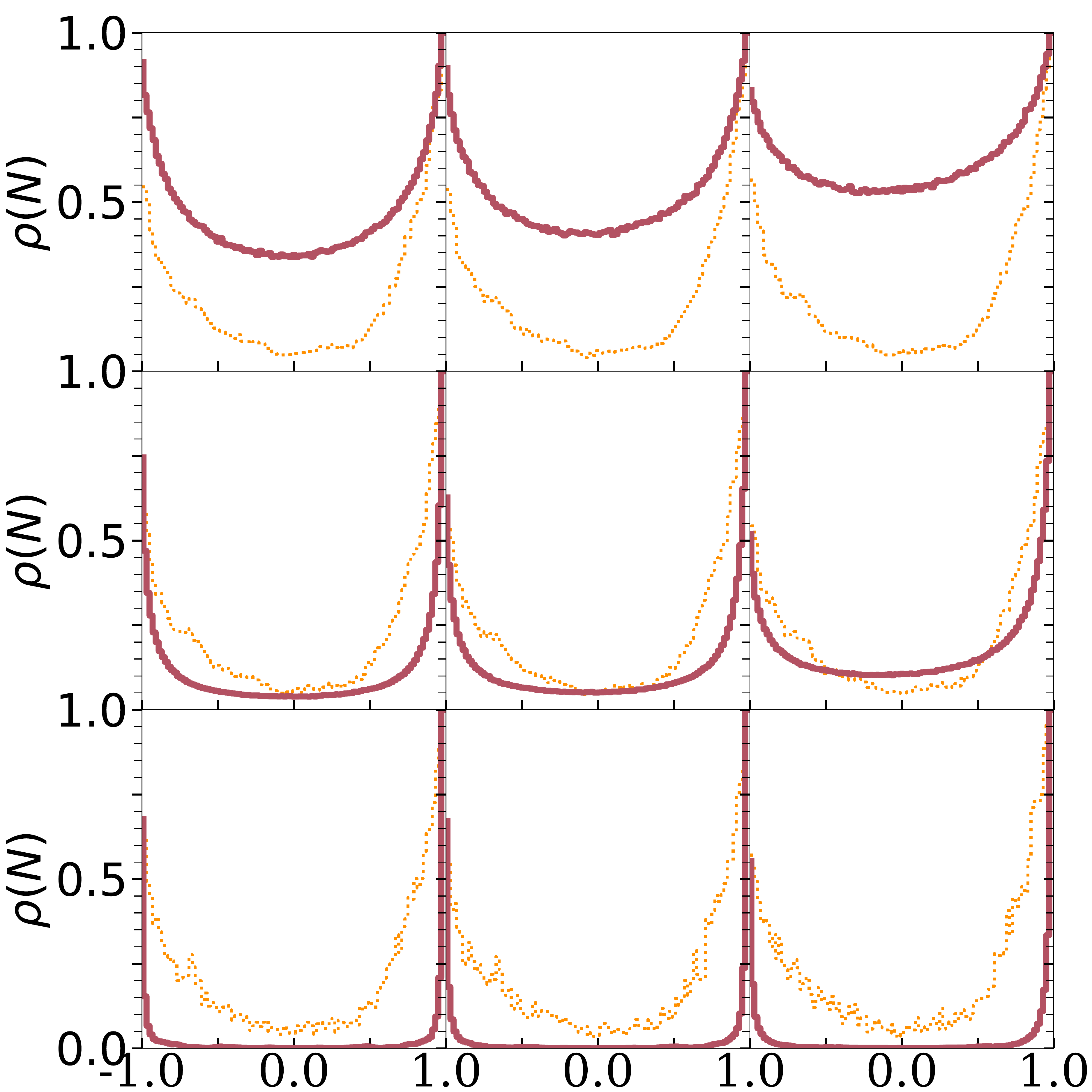}
    \end{subfigure}\hfil
    \begin{subfigure}[b]{0.33\textwidth}
      \centering
      \includegraphics[width=1.\textwidth]{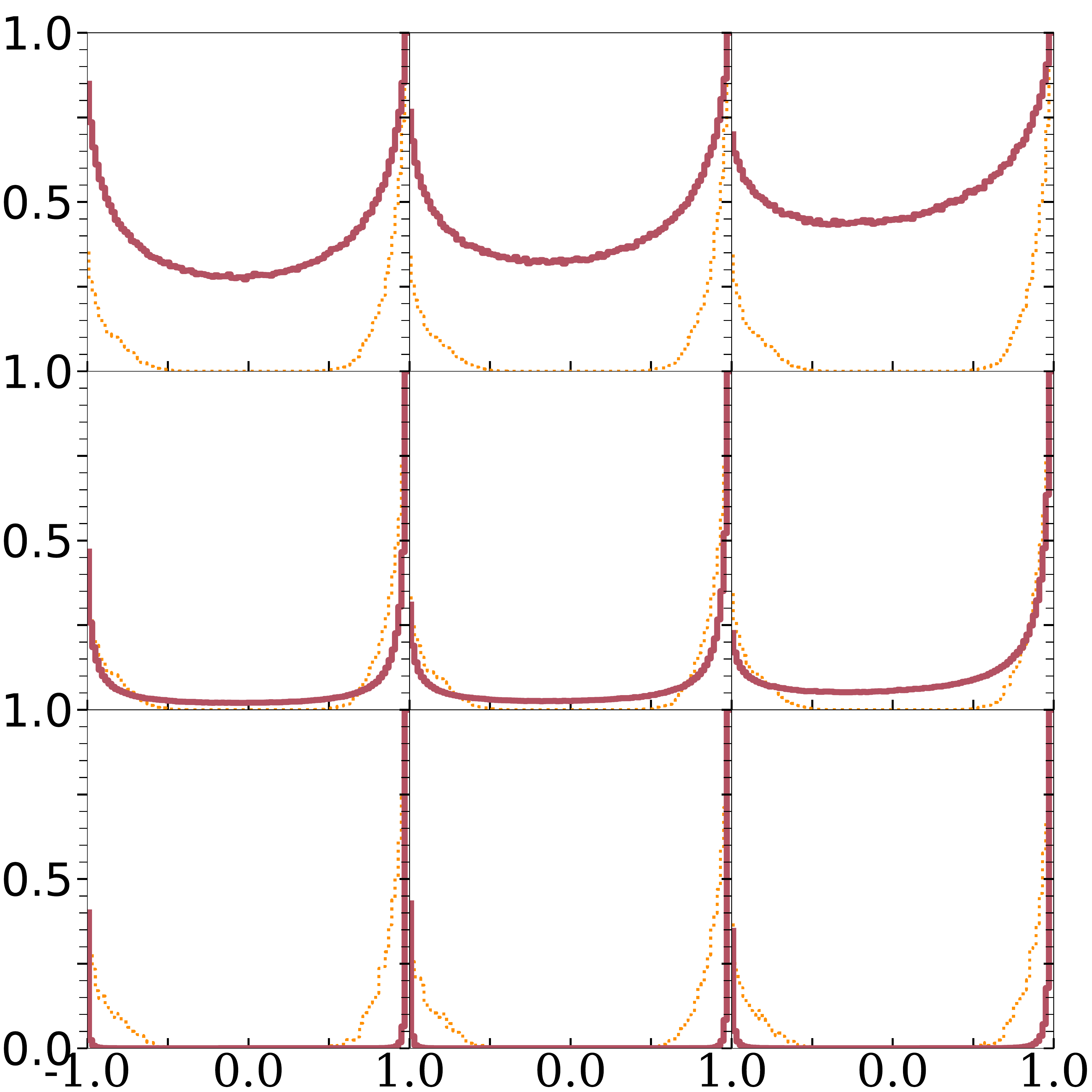}
    \end{subfigure}\hfil
    \begin{subfigure}[b]{0.33\textwidth}
      \centering
      \includegraphics[width=1.\textwidth]{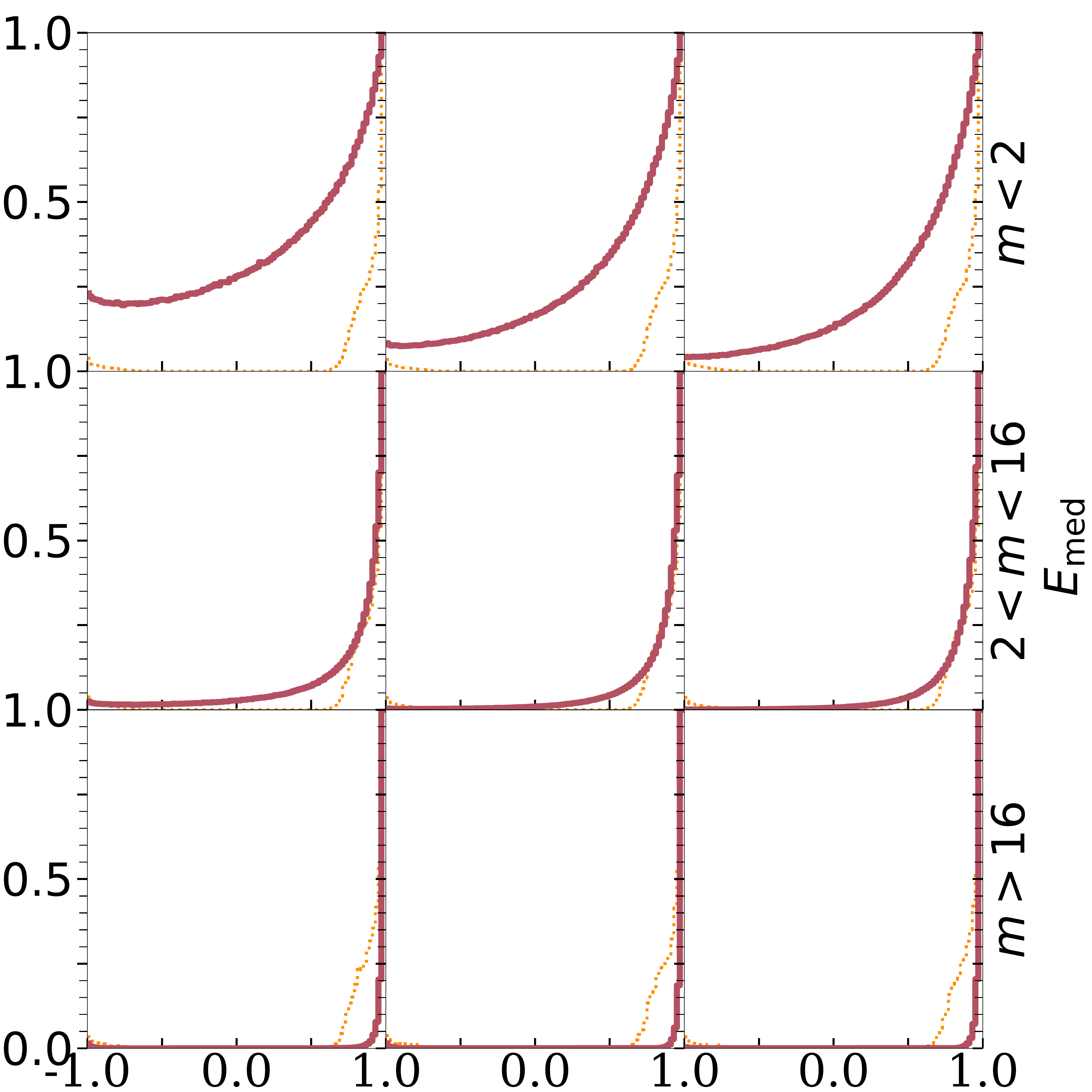}
    \end{subfigure}
    
    \medskip
    \begin{subfigure}[b]{0.33\textwidth}
      \centering
      \includegraphics[width=1.\textwidth]{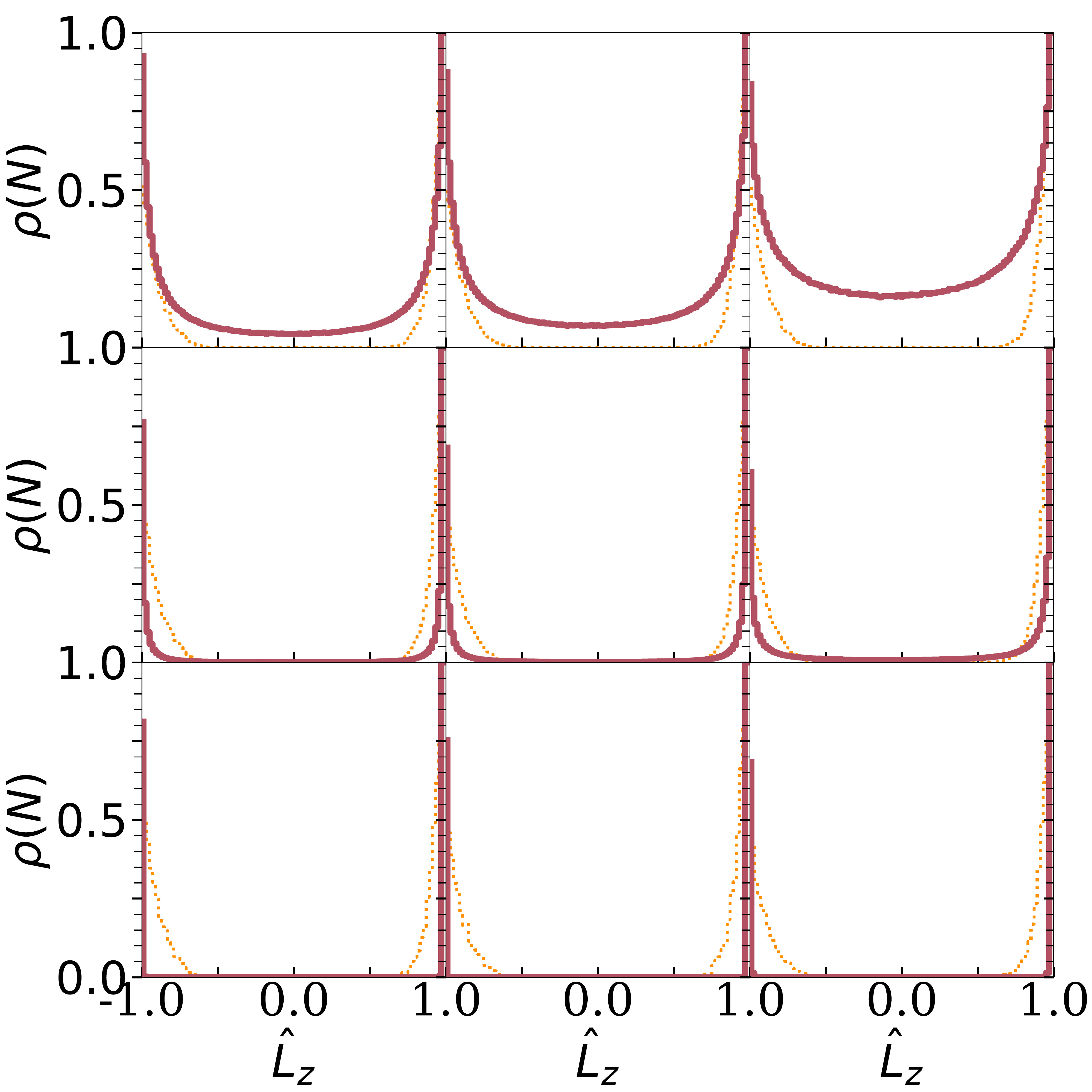}
    \end{subfigure}\hfil
    \begin{subfigure}[b]{0.33\textwidth}
      \centering
      \includegraphics[width=1.\textwidth]{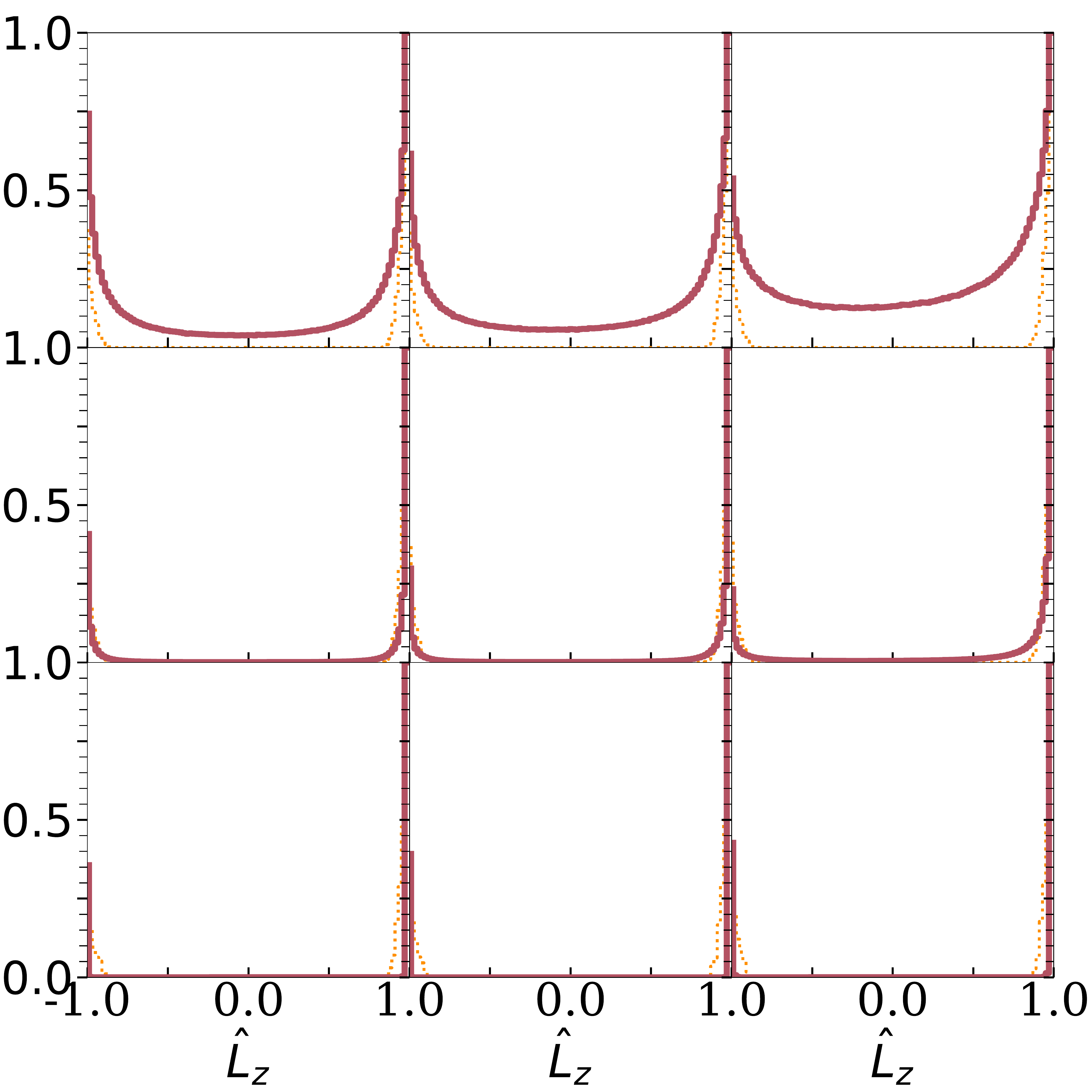}
    \end{subfigure}\hfil
    \begin{subfigure}[b]{0.33\textwidth}
      \centering
      \includegraphics[width=1.\textwidth]{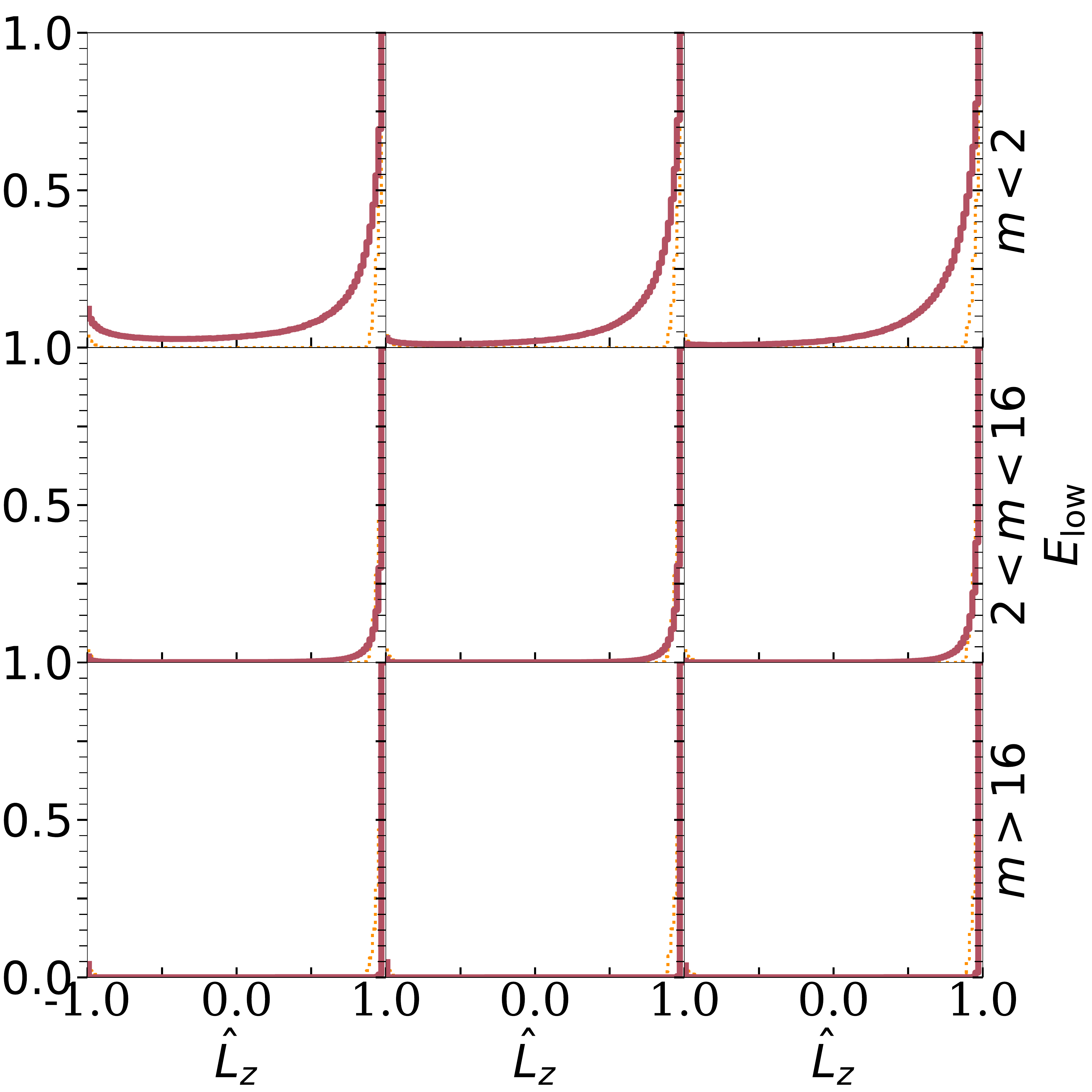}
    \end{subfigure}
\caption{
    Similar to Figure~\ref{fig:mSelectedTimeEnsembleAver} showing the distribution of the cosine of the orbital inclinations, $\hat{L}_z$, but showing both the mass and semimajor axis dependence. Orange dotted lines represent the ensemble-averaged initial conditions, the burgundy solid lines show the stacked final distributions. There are $3\times 3$ large blocks of panels with $(E_{\rm tot}, L_{\rm tot})$ given in Table~\ref{tab:regions}, and each block consists of $3\times 3$ subpanels with different $a$ (inner, middle and outer regions on a logarithmic scale shown by columns from left to right) and $m$ (low, intermediate, high for rows from top to bottom), see text for definitions.
    The figure shows that different regions in $a$ weakly affect the distributions with respect to variations due to $m$, the distribution is qualitatively the same for different semimajor axis. Systematic changes are most prominent for $L_{\rm high}$ or $E_{\rm high}$ showing a larger spherical component in the innermost region.
    }
\label{fig:aSelectedmSelectedTimeEnsembleAverV1}
\end{figure*}

\section{Results}
\label{sec:results}
We generate a sample of $100$ systems with different random realisations of the initial conditions specified in Section~\ref{sec:methods} for the 9 $(E_{\rm tot},L_{\rm tot})$ pair listed in Table~\ref{tab:regions} for the fiducial eccentricity distribution truncated at $e=0.3$. In addition, for the thermal and superthermal distributions we generate $100$ systems for two pairs: $(E_{\rm tot}, L_{\rm tot})=(-0.58,0.88)$ and $(-0.58,0.38)$, respectively with a maximum eccentricity of $e=0.95$. We run \textsc{Nring-MCMC} for $8.39\times 10^7$ steps for all of these $13\times 100$ systems to approach the microcanonical equilibrium. Here we examine the equilibrium geometry represented by the distribution of angular momentum vectors for objects with different masses, semimajor axes, and eccentricities.

\subsection{Mass dependence}
The $3\times 3$ panels of Figure~\ref{fig:mSelectedTimeEnsembleAver} show the equilibrium distribution of the $z$ component of the normalized angular momentum unit vectors, i.e. the cosine of the orbital inclination, for the selected $3\times3$ pairs of $(E_{\rm tot},L_{\rm tot})$ of Table~\ref{tab:regions}. Here and in all other figures below, we stack the final $3.36\times 10^6$ MCMC steps 
of all 100 realisations for a fixed $(E_{\rm tot},L_{\rm tot})$ to minimise the statistical fluctuations and maximise the resolution. The solid lines represent the equilibrium distributions for different mass groups, while the faint dash-dotted show the initial distribution (approximately identical for different mass groups) for reference. We distinguish 3 mass groups (light, intermediate, heavy) as $1\leq m/m_{\rm min}<2$, $2\leq m/m_{\rm min}<16$, $16\leq m/m_{\rm min}\leq100$, which may respectively resemble typical main sequence stars, B-stars, and O-stars or stellar BHs for example. The inclination distributions are normalized with  the maximum values for each histogram, so that $dN/d {\hat{L}_z}=1$ at $\hat{L}_z=1$ by construction. The shape of the initial distributions are qualitatively similar for all the different mass populations, but they strongly depend on the $E_{\rm tot}-L_{\rm tot}$ parameters. We find that the mean variation of samples (not shown) is less than $8.5\%$ with respect to the stacked equilibrium distribution shown in the figure. 

Figure~\ref{fig:mSelectedTimeEnsembleAver} shows that the equatorial regions of the angular momentum sphere are depopulated by VRR as $E_{\rm tot}$ decreases. (Note that $E_{\rm tot}$ decreases from top to bottom rows of panels and $L_{\rm tot}$ increases from left to right columns of panels as labelled and as in Table~\ref{tab:regions}.) 
The equilibrium distributions
show the coexistence
of spherical and disc components
for all $E_{\rm tot},L_{\rm tot}$ pairs. 
We decompose the systems crudely into an isotropic and an anisotropic component, labelled ``sphere'' and ``disc'' by
\begin{equation}\label{eq:fsphere}
    f_{\rm sphere} 
    = \frac{2 \min \rho(\hat{L}_z) }{\int_{-1}^{1} \rho(\hat{L}_z) \,d\hat{L}_z }\,,\quad
    f_{\rm disc} =  1- f_{\rm sphere} 
\end{equation}
where $\rho(\hat{L}_z)=dN/d\hat{L}_z$ is calculated using the stacked histograms for a given $(E_{\rm tot},L_{\rm tot})$. This decomposition assigns all of the anisotropy to the ``disc'' component, thus corresponding to an upper limit of the disc fraction and a lower limit for the spherical fraction. The ``thickness'' of the disc may be specified by the width of the peak of the $\hat{L}_z$ distribution near 1 (corotating component with the total angular momentum vector) and -1 (counter-rotating component), elaborated in more detail in Figure~\ref{fig:commulativeInclinationInMasses} below. We calculate $f_{\rm sphere}$ and $f_{\rm disc}$ fraction for stars in the three different mass bins, separately.

The different mass groups show the following trends.
\begin{enumerate}
    \item Light particles (burgundy solid lines) resemble a nearly isotropic distribution for $E_{\rm high}$ with $f_{\rm sphere} = 96\%$ of the objects in the isotropic distribution. For $E_{\rm med}$ and $E_{\rm low}$, the distribution has a more significant disc component with a significant increase for $L_{\rm high}$ cases; the isotropic fraction is $f_{\rm sphere} = (78,63)\%$  and $(46,36)\%$, respectively. The thickness of the disc of light particles is clearly larger than for the heavier mass components.
    \item Intermediate mass particles (orange solid lines) are also close to a spherical distribution  for $E_{\rm high}$ with $f_{\rm sphere} = 86\%$. For $E_{\rm med}$ the isotropic fraction decreasing with a heavy drop for $L_{\rm high}$. $f_{\rm sphere} = (40,30)\%$ is for $L_{\rm low}$, $L_{\rm med}$ and $L_{\rm high}$ respectively. For $E_{\rm low}$ $f_{\rm sphere}$ is around $6\%$. In those cases, these stars clearly settle into a flattened disc-like structure. 
    \item Heavy particles (violet solid lines) settle in a disc even for $E_{\rm high}$, here $f_{\rm sphere}=33\%$. The distribution for $E_{\rm med}$ and $E_{\rm low}$ clearly resembles a thin disc. 
\end{enumerate}

Note that the initial distributions (dotted lines) are statistically identical for the different mass stars in each panel, respectively. Thus, the simulations show evidence of anisotropic mass-segeration due to VRR in which the heavier components settle into a disc-like equilibrium  for all $(E_{\rm tot},L_{\rm tot})$.

Figure~\ref{fig:cosThetaMassMaxBinNormed} shows the equilibrium distribution function in the two-dimensional plane spanned by $\hat{L}_z$ (i.e. the cosine inclination) and the particle mass $m$ for the same $3\times 3$ values of $(E_{\rm tot}, L_{\rm tot})$ as in the panels of Figure~\ref{fig:mSelectedTimeEnsembleAver}. We stack again the 100 different realisations of the systems (ensemble-average) and the last $3.35 \times 10^6$ MCMC steps and normalize the 2D histograms with the highest values in each logarithmic mass bin, such that the density attains a maximum of unity (typically at $\hat{L}_z=1$) for each $m$. Thus we show the equilibrium distribution function $f_{\rm eq}(m,\hat{L}_z)/\max_{\hat{L}_z}(f_{\rm eq}(m,\hat{L}_z))$. This representation provides detailed information of anisotropic mass segregation as a function of mass. High energy simulations show that the angular momentum vector distribution is predominantly spherical for particles with masses $1\leq m/m_{\rm min}\leq 12$, but it is highly anisotropic for higher masses. For higher masses the distribution is restricted to a small region around $\hat{L}_z=1$, implying a thin disc configuration in physical space, where the disc thickness reduces with increasing $m$ especially for $E_{\rm high}$ and $E_{\rm med}$. The distribution function shows anisotropy (i.e. variations in the $\hat{L}_z$ direction) for the counter-rotating component for $L_{\rm low}$ and $L_{\rm med}$, but not for $L_{\rm high}$.

Figure~\ref{fig:commulativeInclinationInMasses} shows the cumulative distribution of the inclination angle $\arccos\left(\hat{L}_z\right)$ as a function of mass evaluated as a moving average with $\lg_{10} (m/m_{\rm min})$ between $\lg_{10}(m/m_{\rm min})\pm \lg_{10}(2)$. Blue, orange and violet dash-dotted lines show the $25\%,50\%,75\%$ levels of the cumulative distribution. We stack the realisations and the  last $3.36\times 10^6$ MCMC steps as before. For $E_{\rm high}$ the distributions are widely distributed in inclinations, they are close to spherical with the (25\%, 50\% 75\%) levels near $60^\circ, 90^\circ, 120^\circ$ for $m/m_{\rm min}\leq 10$, $L_{\rm low}$ with a systematic shift towards lower inclinations for high masses for $L_{\rm med}$, $L_{\rm high}$. For smaller $E_{\rm tot}$ the equilibrium configuration tends to be more disc-like where the $(25\% - 75\%)$ levels either move close to small inclinations towards $0^\circ$ creating a unidirectional disc (i.e. for $(E_{\rm low}, L_{\rm high})$) or close to $180^\circ$ to create a counter-rotating disc (i.e. for  $(E_{\rm low}, L_{\rm low})$). In both cases, the disc thickness decreases with increasing mass. Further, note that the fraction of counter-rotating stars decreases to below $25\%$ for intermediate mass stars for $(E_{\rm med},L_{\rm med})$ and $(E_{\rm low},L_{\rm med})$. The discontinuities at high masses are due to small number statistics.

The fact that the higher mass particles tend to settle to discs is the natural consequence of the system evolving towards its maximum entropy state. As massive particles decrease their mutual inclination with the rest of the cluster on average a large number of light particles are able to get perturbed on orbits with higher mutual inclination which resembles the isotropic distribution in the limiting case, which by definition maximises the system entropy. In order to conserve the total VRR energy and angular momentum, the system cannot become completely isotropic. In our simulations any proposed step with a heavy partner tends on average to decrease the mutual inclination of the heavy particle with respect to the cluster's angular momentum.

Figure~\ref{fig:heavyAcceptedStepsEnsemDistribution} shows the evolution of the cosine orbital inclination of a particular heavy particle of mass $m/m_{\min} = 45.81$, semimajor axis $a/a_{\min}=12.95$ and eccentricity $e=0.24$ in all the accepted MCMC steps for an ensemble of 100 simulations for $(E_{\rm high},L_{\rm med})$. This figure uses the same coloring scheme as Figures~\ref{fig:ksTest},~\ref{fig:mSelectedTimeEnsembleAver},~\ref{fig:commulativeInclinationInMasses} and additionally encodes the mass (light, intermediate, heavy) of the interaction partner with the marker size. Light particles are plotted with smaller markers with an increasing marker size for heavier objects. Both the $2D$ and the marginal distributions support our findings that heavy particles tend to settle to a disc. The marginal distribution shows a significant excess at $\hat{L}_z=1$. This represents orbits which on average reside in a thin, counter-rotating disc like structure in physical space.

However note that the tendency for heavy objects to form discs may depend on the assumed level of initial deviation from the exactly isotropic distribution, i.e. $E_{\rm tot}=-0.03$ instead of 0. The expected deviation from isotropy in an idealized, one component system is proportional to $\propto 1/\sqrt{N}$, where $N$ is the total number of particles. Thus, for $N\sim 10^6$, spherical systems may in principle exist even closer to isotropy with $E_{\rm tot}\sim -0.001$, where forming a disc population of heavy objects would be more difficult due to the conservation of the total VRR energy. This expectation is confirmed by recent direct $N$-body numerical simulations of \citet{Panamarev_Kocsis2022}, where an almost exactly isotropic massive spherical system drives the disc into a spherical configuration including its massive stars. This is not seen in our models with $N=4096$, where shot noise leads to a larger deviations from isotropy.

\subsection{Semimajor axis dependence}
Figure~\ref{fig:aSelectedmSelectedTimeEnsembleAverV1} shows the distribution of the angular momentum vector directions as in Figure~\ref{fig:mSelectedTimeEnsembleAver}, but here we group the particles not only by mass but also by their semimajor axes. There are now $3\times 3$ blocks of panels with $3\times 3$ values of $(E_{\rm tot},L_{\rm tot})$ and there are $3\times 3$ subpanels with different $(a,m)$ within each block as labelled. The equilibrium distributions are shown by burgundy solid lines and the initial distribution is shown with orange dotted lines for reference. The $3$ mass groups (light, intermediate, heavy) are the same as before ($1\leq m/m_{\rm min}<2$, $2\leq m/m_{\rm min}<16$, $16\leq m/m_{\rm min}\leq100$) and are shown in different rows of subpanels, and the $3$ semimajor axis groups (inner, middle, and outer) are chosen uniformly on a logarithmic scale as ($1\leq a/a_{\rm min}<4.64$, $4.64\leq a/a_{\rm min}<21.5$, $21.5\leq a/a_{\rm min}\leq 100$) and shown in different subcolumns. The $L_{\rm tot}$ parameter increases from left to right while the $E_{\rm tot}$ decreases from top to bottom in the $3\times 3$ panels as in Table~\ref{tab:regions}. The stacking and normalization are the same as in Figure~\ref{fig:mSelectedTimeEnsembleAver}. During VRR, the level of anisotropic mass segregation is weakly inhomogeneous in semimajor axis. We find that that the spherical fraction $f_{\rm sphere}$ (Eq.~\ref{eq:fsphere}) is systematically higher in the innermost region particularly for the most spherical initial condition $E_{\rm high}$ for all $L_{\rm tot}$ or for systems with a high net rotation $L_{\rm high}$ for all $E_{\rm tot}$.\footnote{Note that the conclusions may be affected by small number statistics for high mass stars $(m>16\,m_{\rm min})$ in the innermost region $(a_{\rm in})$ and by the slow convergence of the MCMC method.} For $(E_{\rm med},L_{\rm low})$ and $(E_{\rm med},L_{\rm med})$, the middle region has the smallest $f_{\rm sphere}$.

\subsection{Eccentricity dependence}
The equilibrium distributions are qualitatively very similar for the thermal and super-thermal eccentricity distribution simulations to the results presented above for which the distribution was truncated at $e\leq 0.3$. This is not surprising as the VRR energy (particularly $\mathcal{J}_{ij\ell}$ in Eq.~\ref{eq:HVRRij}) is weakly sensitive to $e$ for $e\leq 0.9$, see \citet{Kocsis2015}.

\section{Conclusions and Discussion}
\label{sec:conclusion}

In this paper we generalised the work of \citet{Szolgyen2018} to determine the range of possible statistical equilibria of orbital inclinations during VRR for a variety of astrophysically motivated initial conditions see Section~\ref{sec:initialConditions}. Since the equilibria depend on the initial conditions only through the conserved quantities, the total VRR energy and total angular momentum, $(E_{\rm tot}, L_{\rm tot})$, we explored the outcome on a $3\times 3$ grid in this space for low, intermediate, and high values (Table~\ref{tab:regions} and Equations~\ref{eq:Etotnorm} and \ref{eq:Ltotnorm}). We considered nuclear star clusters with multiple mass, semimajor axis, and eccentricity components with powerlaw distributions for each. For the eccentricity distribution we mostly focused on the case limited to $e\leq 0.3$ as observed for the young stars in the Galactic centre, but also examined two other cases with a thermal and superthermal distribution and found the outcome to be independent of eccentricity. The initial distributions for the orbital inclinations were chosen to be the same for objects with different $m$, $a$, and $e$, so the systematic dependence of orbital inclinations in the equilibrium sample must come from VRR. We constructed a large sample of initial distributions and evolved them using \textsc{Nring-MCMC} to obtain the microcanonical ensemble for fixed $(E_{\rm tot}, L_{\rm tot})$. The MCMC method reached convergence at the $8\%$ level or better (Figure \ref{fig:ksTest}). The resulting equilibrium distributions show systematic differences for different $E_{\rm tot}$, $L_{\rm tot}$, $m$, and $a$,

Our findings show that anistropic mass segregation is a general ubiquitous outcome of VRR beyond the isolated cases found previously \citep{Szolgyen2018, Szolgyen+2019,Tiongco+2021,MagnanFouvry2021}, see Figures~\ref{fig:mSelectedTimeEnsembleAver}, \ref{fig:cosThetaMassMaxBinNormed}, \ref{fig:commulativeInclinationInMasses} above. Light particles generally settle to a more spherical spatial configuration, while the heavy particles segregate to a more disc-like configuration  than the initial configuration (Figure~\ref{fig:cosThetaMassMaxBinNormed}). In doing so, the heavier objects do not merely retain the initial level of anisotropy but even amplify it, while lighter objects become more spherically distributed. The overall level of anisotropy is set by $E_{\rm tot}$ and $L_{\rm tot}$. Small $E_{\rm tot}$ leads to a disc-like structure of both heavy and light objects, but in which the thickness of the disc is smaller for the heavier objects (Figure~\ref{fig:commulativeInclinationInMasses}). High $E_{\rm tot}$ leads to the coexistence of spherical and disc-like distributions. In this case a large fraction of objects are spherically distributed with the exception of the heaviest objects which settle into a thick disc. The disc is counter-rotating for low $L_{\rm tot}$ and mostly unidirectional for high $L_{\rm tot}$. 

We find weak systematic trends with semimajor axis, where the level of anisotropy is slightly larger for low $L_{\rm tot}$ and slightly smaller for high $L_{\rm tot}$ in the inner regions with respect to that in the outer regions (Figure~\ref{fig:aSelectedmSelectedTimeEnsembleAverV1}). In other words, the disc is slightly thinner in the inner region for low $L_{\rm tot}$ but thicker in the inner region for large $L_{\rm tot}$. However, we note that while the systematic trends with mass are more pronounced and are expected to be robust, the trends with respect to semimajor axis may be affected by the large variation of the VRR timescale and the timescales of other effects such as scalar resonant relaxation and two body relaxation for a vast range of semimajor axes in the nuclear star cluster.

The cases explored in the space of $(E_{\rm tot},L_{\rm tot})$ are useful to understand the qualitative outcome for arbitrary initial conditions. For example, to obtain the equilibrium outcome for the mixture of two systems with an initial spherical and a disc component or with two discs, it is sufficient to calculate the total normalized $(E_{\rm tot},L_{\rm tot})$. This is similar to asking the outcome for mixing water and ice; the total thermal energy per particle of the mixture specifies its temperature and hence the outcome to be either water or ice, or the mixture of the two. Similarly, $(E_{\rm tot},L_{\rm tot})$ are preserved during VRR, and given its value for the mixture, one can simply interpolate the results shown in this paper to obtain the qualitative behaviour of the equilibrium configuration. Interestingly, in this context, we find that heavier objects form an ordered flattened structure even in the presence of an initially nearly spherical disordered configuration. In this analogy, the ``freezing point'' of the angular momentum distribution is effectively higher for the heavier objects, implying that they may settle to the ordered disc state in thermal equilibrium with lower mass objects being in the disordered spherical phase. Furthermore, there is an inevitable angular momentum shot noise anisotropy for a finite number of objects, even if initially nearly-spherically distributed ($4096$ in our simulations), and this anisotropy is absorbed by the distribution of highest mass stars to form an ordered disc within the disordered spherical ambient medium. 

These findings may also be interesting with regards to the statistical physics of long range interacting systems \citep{Campa2014} and tantalizing connections to condensed matter physics. It has been shown that the leading-order quadrupolar approximation of VRR is equivalent to the Maier Saupe model of liquid crystals \citep{Kocsis2015,Roupas2017}. Indeed, orbit-averaging over the apsidal precession time results in axisymmetric mass annuli which resemble the axisymmetric molecules of liquid crystals. In nearly coplanar cases, the VRR interaction resembles that of vortex crystals, and the N-vector model of spin systems \citep{Kocsis2011,Kocsis2015,Touma_Tremaine2014}. This correspondance leads to a qualitative similarity between the statistical physics of these systems. Interestingly, VRR admits the curious possibility of negative absolute temperature equilibria given that the VRR energy is bounded from above \citep{Kocsis2011,Roupas2017,Takacs2018}. Further, given the non-additive nature of the energy of subsystems, different ensembles (e.g. canonical and microcanonical) are inequivalent for VRR \citep{Roupas2017,Roupas2020}. Both VRR and liquid crystals exhibit a phase transition between a disordered isotropic phase and an ordered disc-like nematic phase \citep{Roupas2017,Takacs2018,Roupas2020} which may be lopsided \citep{Touma_Tremaine2019,Tremaine2020a,Tremaine2020b,Zderic+2020}. The phase transition is known to be first order for the single-component model (i.e. all objects have the same fixed $m$, $a$, and $e$) under certain conditions, namely if the total angular momentum of the star cluster is below a critical value for VRR, and if the external magnetic field is small for liquid crystals. While phase separation is generally unexpected for non-additive systems \citep{Campa2014}, the findings of this paper show in contrast that flattened structures are ubiquitous for the high mass components of the system, such as stellar mass black holes, and may coexist with a spherical distribution of low mass objects. 

An exploration of the allowed diversity of equilibrium configurations may be relevant in a variety of contexts beyond explaining the geometry of the discs in the Galactic centre. If massive stellar objects settle into flattened disc-like configurations, the resulting equilibrium distributions should contain a large number of stellar mass black holes. Such a population of black holes (i) may influence the stellar orbits affecting precision tests of general relativity \citep{Merritt2010,Meyer2012,Gravity+2020}, (ii) could regulate the accretion flow in active galactic nuclei potentially forming IMBHs \citep{Kocsis2011b,McKernan+2012}, (iii) could lead to X-ray flares as they traverse through gas \citep{Bartos2013}, (iv) lead to a distinct class of sources of gravitational waves for LIGO/VIRGO/KAGRA observations \citep{Bartos2017,Stone2017,Tagawa2020} and pulsar timing array observations \citep{Kocsis2012}, and (v) can produce extreme mass ratio inspirals observable by LISA \citep{Kocsis2011b,Gair2004,AmaroSeoane2007,Babak2017}.

\section*{Acknowledgements}

We are grateful to Mária Kolozsvári for help with logistics and administration related to the research. This work was supported by the European Research Council (ERC) under the European Union’s Horizon 2020 research and innovation programme under grant agreement No 638435 (GalNUC) and by the Hungarian National Research, Development, and Innovation Office grant NKFIH KH125675. The calculations were carried out on the Hungarian National Information Infrastructure Development NIIF HPC cluster at the University of Szeged, Hungary.

\section*{Data availability}
Data available on request. Due to the size of the underlying data set the permanent storage on a public server is not feasible. However on dedicated request we are glad to share and explain the data set in detail.


\bibliographystyle{mnras}
\bibliography{mathe_et_al}

\begin{thebibliography}{}
\makeatletter
\relax
\def\mn@urlcharsother{\let\do\@makeother \do\$\do\&\do\#\do\^\do\_\do\%\do\~}
\def\mn@doi{\begingroup\mn@urlcharsother \@ifnextchar [ {\mn@doi@}
  {\mn@doi@[]}}
\def\mn@doi@[#1]#2{\def\@tempa{#1}\ifx\@tempa\@empty \href
  {http://dx.doi.org/#2} {doi:#2}\else \href {http://dx.doi.org/#2} {#1}\fi
  \endgroup}
\def\mn@eprint#1#2{\mn@eprint@#1:#2::\@nil}
\def\mn@eprint@arXiv#1{\href {http://arxiv.org/abs/#1} {{\tt arXiv:#1}}}
\def\mn@eprint@dblp#1{\href {http://dblp.uni-trier.de/rec/bibtex/#1.xml}
  {dblp:#1}}
\def\mn@eprint@#1:#2:#3:#4\@nil{\def\@tempa {#1}\def\@tempb {#2}\def\@tempc
  {#3}\ifx \@tempc \@empty \let \@tempc \@tempb \let \@tempb \@tempa \fi \ifx
  \@tempb \@empty \def\@tempb {arXiv}\fi \@ifundefined
  {mn@eprint@\@tempb}{\@tempb:\@tempc}{\expandafter \expandafter \csname
  mn@eprint@\@tempb\endcsname \expandafter{\@tempc}}}

\bibitem[\protect\citeauthoryear{{Ali} et~al.,}{{Ali} et~al.}{2020}]{Ali2020}
{Ali} B.,  et~al., 2020, \mn@doi [\apj] {10.3847/1538-4357/ab93ae}, \href
  {https://ui.adsabs.harvard.edu/abs/2020ApJ...896..100A} {896, 100}

\bibitem[\protect\citeauthoryear{{Amaro-Seoane}, {Gair}, {Freitag}, {Miller},
  {Mandel}, {Cutler}  \& {Babak}}{{Amaro-Seoane}
  et~al.}{2007}]{AmaroSeoane2007}
{Amaro-Seoane} P.,  {Gair} J.~R.,  {Freitag} M.,  {Miller} M.~C.,  {Mandel} I.,
   {Cutler} C.~J.,   {Babak} S.,  2007, \mn@doi [Class. Quantum Gravity]
  {10.1088/0264-9381/24/17/R01}, \href
  {http://adsabs.harvard.edu/abs/2007CQGra..24R.113A} {24, R113}

\bibitem[\protect\citeauthoryear{{Antonini}}{{Antonini}}{2013}]{Antonini2013}
{Antonini} F.,  2013, \mn@doi [\apj] {10.1088/0004-637X/763/1/62}, \href
  {http://adsabs.harvard.edu/abs/2013ApJ...763...62A} {763, 62}

\bibitem[\protect\citeauthoryear{{Antonini}}{{Antonini}}{2014}]{Antonini2014}
{Antonini} F.,  2014, \mn@doi [\apj] {10.1088/0004-637X/794/2/106}, \href
  {http://adsabs.harvard.edu/abs/2014ApJ...794..106A} {794, 106}

\bibitem[\protect\citeauthoryear{{Antonini}, {Capuzzo-Dolcetta},
  {Mastrobuono-Battisti}  \& {Merritt}}{{Antonini} et~al.}{2012}]{Antonini2012}
{Antonini} F.,  {Capuzzo-Dolcetta} R.,  {Mastrobuono-Battisti} A.,   {Merritt}
  D.,  2012, \mn@doi [\apj] {10.1088/0004-637X/750/2/111}, \href
  {http://adsabs.harvard.edu/abs/2012ApJ...750..111A} {750, 111}

\bibitem[\protect\citeauthoryear{{Antonini}, {Barausse}  \& {Silk}}{{Antonini}
  et~al.}{2015}]{Antonini2015}
{Antonini} F.,  {Barausse} E.,   {Silk} J.,  2015, \mn@doi [\apj]
  {10.1088/0004-637X/812/1/72}, \href
  {http://adsabs.harvard.edu/abs/2015ApJ...812...72A} {812, 72}

\bibitem[\protect\citeauthoryear{Arca~Sedda}{Arca~Sedda}{2019}]{arca_sedda_2019}
Arca~Sedda M.,  2019, \mn@doi [Proc. Int. Astron. Union.]
  {10.1017/S1743921319007324}, 14, 51–55

\bibitem[\protect\citeauthoryear{{Arca-Sedda}, {Capuzzo-Dolcetta}, {Antonini}
  \& {Seth}}{{Arca-Sedda} et~al.}{2015}]{ArcaSedda2015}
{Arca-Sedda} M.,  {Capuzzo-Dolcetta} R.,  {Antonini} F.,   {Seth} A.,  2015,
  \mn@doi [\apj] {10.1088/0004-637X/806/2/220}, \href
  {http://adsabs.harvard.edu/abs/2015ApJ...806..220A} {806, 220}

\bibitem[\protect\citeauthoryear{{Arca-Sedda}, {Kocsis}  \&
  {Brandt}}{{Arca-Sedda} et~al.}{2018}]{ArcaSedda_Kocsis2018}
{Arca-Sedda} M.,  {Kocsis} B.,   {Brandt} T.~D.,  2018, \mn@doi [\mnras]
  {10.1093/mnras/sty1454}, \href
  {https://ui.adsabs.harvard.edu/abs/2018MNRAS.479..900A} {479, 900}

\bibitem[\protect\citeauthoryear{{Babak} et~al.,}{{Babak}
  et~al.}{2017}]{Babak2017}
{Babak} S.,  et~al., 2017, \mn@doi [\prd] {10.1103/PhysRevD.95.103012}, \href
  {http://adsabs.harvard.edu/abs/2017PhRvD..95j3012B} {95, 103012}

\bibitem[\protect\citeauthoryear{{Bahcall} \& {Wolf}}{{Bahcall} \&
  {Wolf}}{1977}]{Bachall1977}
{Bahcall} J.~N.,  {Wolf} R.~A.,  1977, \mn@doi [\apj] {10.1086/155534}, 216,
  883

\bibitem[\protect\citeauthoryear{{Bar-Or} \& {Fouvry}}{{Bar-Or} \&
  {Fouvry}}{2018}]{Bar-Or_Fouvry2018}
{Bar-Or} B.,  {Fouvry} J.-B.,  2018, \mn@doi [\apjl]
  {10.3847/2041-8213/aac88e}, \href
  {https://ui.adsabs.harvard.edu/abs/2018ApJ...860L..23B} {860, L23}

\bibitem[\protect\citeauthoryear{{Bartko} et~al.,}{{Bartko}
  et~al.}{2009}]{Bartko2009}
{Bartko} H.,  et~al., 2009, \mn@doi [\apj] {10.1088/0004-637X/697/2/1741},
  \href {https://ui.adsabs.harvard.edu/abs/2009ApJ...697.1741B} {697, 1741}

\bibitem[\protect\citeauthoryear{{Bartko} et~al.}{{Bartko}
  et~al.}{2010}]{Bartko2010}
{Bartko} H.,  et~al., 2010, \mn@doi [\apj] {10.1088/0004-637X/708/1/834}, 708,
  834

\bibitem[\protect\citeauthoryear{{Bartos}, {Haiman}, {Kocsis}  \&
  {M{\'a}rka}}{{Bartos} et~al.}{2013}]{Bartos2013}
{Bartos} I.,  {Haiman} Z.,  {Kocsis} B.,   {M{\'a}rka} S.,  2013, \mn@doi
  [Phys. Rev. Lett.] {10.1103/PhysRevLett.110.221102}, \href
  {http://adsabs.harvard.edu/abs/2013PhRvL.110v1102B} {110, 221102}

\bibitem[\protect\citeauthoryear{{Bartos}, {Kocsis}, {Haiman}  \&
  {M{\'a}rka}}{{Bartos} et~al.}{2017}]{Bartos2017}
{Bartos} I.,  {Kocsis} B.,  {Haiman} Z.,   {M{\'a}rka} S.,  2017, \mn@doi
  [\apj] {10.3847/1538-4357/835/2/165}, \href
  {http://adsabs.harvard.edu/abs/2017ApJ...835..165B} {835, 165}

\bibitem[\protect\citeauthoryear{Bianchini, Varri, Bertin  \& Zocchi}{Bianchini
  et~al.}{2013}]{Bianchini2013}
Bianchini P.,  Varri A.~L.,  Bertin G.,   Zocchi A.,  2013, ApJ, 772, 67

\bibitem[\protect\citeauthoryear{Boberg, Vesperini, Friel, Tiongco  \&
  Varri}{Boberg et~al.}{2017}]{Boberg2017}
Boberg O.~M.,  Vesperini E.,  Friel E.~D.,  Tiongco M.~A.,   Varri A.~L.,
  2017, ApJ, 841, 114

\bibitem[\protect\citeauthoryear{Breen, Heggie  \& Varri}{Breen
  et~al.}{2017}]{Breen2017}
Breen P.~G.,  Heggie D.~C.,   Varri A.~L.,  2017, \mn@doi [MNRAS]
  {10.1093/mnras/stx1750}, 471, 2778

\bibitem[\protect\citeauthoryear{Campa, Dauxois, Fanelli  \& Ruffo}{Campa
  et~al.}{2014}]{Campa2014}
Campa A.,  Dauxois T.,  Fanelli D.,   Ruffo S.,  2014, Physics of Long-Range
  Interacting Systems.
Oxford University Press, Oxford,
  \mn@doi{10.1093/acprof:oso/9780199581931.001.0001}, \url
  {http://www.oxfordscholarship.com/10.1093/acprof:oso/9780199581931.001.0001/acprof-9780199581931}

\bibitem[\protect\citeauthoryear{{Cuadra}, {Armitage}  \& {Alexander}}{{Cuadra}
  et~al.}{2008}]{Cuadra+2008}
{Cuadra} J.,  {Armitage} P.~J.,   {Alexander} R.~D.,  2008, \mn@doi [\mnras]
  {10.1111/j.1745-3933.2008.00500.x}, \href
  {https://ui.adsabs.harvard.edu/abs/2008MNRAS.388L..64C} {388, L64}

\bibitem[\protect\citeauthoryear{Eilon, Kupi  \& Alexander}{Eilon
  et~al.}{2009}]{Eilon2009}
Eilon E.,  Kupi G.,   Alexander T.,  2009, \mn@doi [ApJ]
  {10.1088/0004-637X/698/1/641}, 698, 641

\bibitem[\protect\citeauthoryear{Einsel \& Spurzem}{Einsel \&
  Spurzem}{1999}]{Einsel1999}
Einsel C.,  Spurzem R.,  1999, \mn@doi [MNRAS]
  {10.1046/j.1365-8711.1999.02083.x}, 302, 81

\bibitem[\protect\citeauthoryear{Ferraro et~al.,}{Ferraro
  et~al.}{2018}]{Ferraro2018}
Ferraro F.~R.,  et~al., 2018, ApJ, 860, 50

\bibitem[\protect\citeauthoryear{{Fouvry} \& {Bar-Or}}{{Fouvry} \&
  {Bar-Or}}{2018}]{Fouvry_Bar-Or2018}
{Fouvry} J.-B.,  {Bar-Or} B.,  2018, \mn@doi [\mnras] {10.1093/mnras/sty2571},
  \href {https://ui.adsabs.harvard.edu/abs/2018MNRAS.481.4566F} {481, 4566}

\bibitem[\protect\citeauthoryear{{Fouvry}, {Bar-Or}  \& {Chavanis}}{{Fouvry}
  et~al.}{2019}]{Fouvry2019}
{Fouvry} J.-B.,  {Bar-Or} B.,   {Chavanis} P.-H.,  2019, \mn@doi [\apj]
  {10.3847/1538-4357/ab2f78}, \href
  {https://ui.adsabs.harvard.edu/abs/2019ApJ...883..161F} {883, 161}

\bibitem[\protect\citeauthoryear{{Gair}, {Barack}, {Creighton}, {Cutler},
  {Larson}, {Phinney}  \& {Vallisneri}}{{Gair} et~al.}{2004}]{Gair2004}
{Gair} J.~R.,  {Barack} L.,  {Creighton} T.,  {Cutler} C.,  {Larson} S.~L.,
  {Phinney} E.~S.,   {Vallisneri} M.,  2004, \mn@doi [Class. Quantum Gravity]
  {10.1088/0264-9381/21/20/003}, \href
  {http://adsabs.harvard.edu/abs/2004CQGra..21S1595G} {21, S1595}

\bibitem[\protect\citeauthoryear{{Gallego-Cano}, {Sch{\"o}del},
  {Nogueras-Lara}, {Dong}, {Shahzamanian}, {Fritz}, {Gallego-Calvente}  \&
  {Neumayer}}{{Gallego-Cano} et~al.}{2020}]{Gallego-Cano+2020}
{Gallego-Cano} E.,  {Sch{\"o}del} R.,  {Nogueras-Lara} F.,  {Dong} H.,
  {Shahzamanian} B.,  {Fritz} T.~K.,  {Gallego-Calvente} A.~T.,   {Neumayer}
  N.,  2020, \mn@doi [\aap] {10.1051/0004-6361/201935303}, \href
  {https://ui.adsabs.harvard.edu/abs/2020A&A...634A..71G} {634, A71}

\bibitem[\protect\citeauthoryear{{Genzel} et~al.}{{Genzel}
  et~al.}{2010}]{Genzel2010}
{Genzel} R.,  et~al., 2010, \mn@doi [Rev. Mod. Phys.]
  {10.1103/RevModPhys.82.3121}, 82, 3121

\bibitem[\protect\citeauthoryear{{Gillessen}, {Eisenhauer}, {Trippe},
  {Alexander}, {Genzel}, {Martins}  \& {Ott}}{{Gillessen}
  et~al.}{2009}]{Gillessen+2009}
{Gillessen} S.,  {Eisenhauer} F.,  {Trippe} S.,  {Alexander} T.,  {Genzel} R.,
  {Martins} F.,   {Ott} T.,  2009, \mn@doi [\apj]
  {10.1088/0004-637X/692/2/1075}, \href
  {https://ui.adsabs.harvard.edu/abs/2009ApJ...692.1075G} {692, 1075}

\bibitem[\protect\citeauthoryear{Gnedin et~al.}{Gnedin
  et~al.}{2014}]{Gnedin2014}
Gnedin O.~Y.,  et~al., 2014, \apj, 785, 71

\bibitem[\protect\citeauthoryear{{Gravity Collaboration} et~al.,}{{Gravity
  Collaboration} et~al.}{2020}]{Gravity+2020}
{Gravity Collaboration} et~al., 2020, \mn@doi [\aap]
  {10.1051/0004-6361/202037813}, \href
  {https://ui.adsabs.harvard.edu/abs/2020A&A...636L...5G} {636, L5}

\bibitem[\protect\citeauthoryear{{Gruzinov}, {Levin}  \& {Zhu}}{{Gruzinov}
  et~al.}{2020}]{Gruzinov+2020}
{Gruzinov} A.,  {Levin} Y.,   {Zhu} J.,  2020, \mn@doi [\apj]
  {10.3847/1538-4357/abbfaa}, \href
  {https://ui.adsabs.harvard.edu/abs/2020ApJ...905...11G} {905, 11}

\bibitem[\protect\citeauthoryear{Hopman \& Alexander}{Hopman \&
  Alexander}{2006}]{Hopman2006}
Hopman C.,  Alexander T.,  2006, \mn@doi [ApJ] {10.1086/504400}, 645, 1152

\bibitem[\protect\citeauthoryear{Jeffreson et~al.,}{Jeffreson
  et~al.}{2017}]{Jeffreson2017}
Jeffreson S. M.~R.,  et~al., 2017, \mn@doi [MNRAS] {10.1093/mnras/stx1152},
  469, 4740

\bibitem[\protect\citeauthoryear{Kamann et~al.,}{Kamann
  et~al.}{2018}]{Kamann2018}
Kamann S.,  et~al., 2018, \mn@doi [MNRAS] {10.1093/mnras/stx2719}, 473, 5591

\bibitem[\protect\citeauthoryear{Kocsis \& Levin}{Kocsis \&
  Levin}{2012}]{Kocsis2012}
Kocsis B.,  Levin J.,  2012, \mn@doi [Phys. Rev. D]
  {10.1103/PhysRevD.85.123005}, 85, 123005

\bibitem[\protect\citeauthoryear{Kocsis \& Tremaine}{Kocsis \&
  Tremaine}{2011}]{Kocsis2011}
Kocsis B.,  Tremaine S.,  2011, \mn@doi [MNRAS]
  {10.1111/j.1365-2966.2010.17897.x}, 412, 187

\bibitem[\protect\citeauthoryear{Kocsis \& Tremaine}{Kocsis \&
  Tremaine}{2015}]{Kocsis2015}
Kocsis B.,  Tremaine S.,  2015, \mn@doi [MNRAS] {10.1093/mnras/stv057}, 448,
  3265

\bibitem[\protect\citeauthoryear{{Kocsis}, {Yunes}  \& {Loeb}}{{Kocsis}
  et~al.}{2011}]{Kocsis2011b}
{Kocsis} B.,  {Yunes} N.,   {Loeb} A.,  2011, \mn@doi [\prd]
  {10.1103/PhysRevD.84.024032}, \href
  {http://adsabs.harvard.edu/abs/2011PhRvD..84b4032K} {84, 024032}

\bibitem[\protect\citeauthoryear{Kormendy \& Ho}{Kormendy \&
  Ho}{2013}]{Kormendy2013}
Kormendy J.,  Ho L.~C.,  2013, \mn@doi [Annu. Rev. Astron. Astrophys.]
  {10.1146/annurev-astro-082708-101811}, 51, 511

\bibitem[\protect\citeauthoryear{Lanzoni et~al.,}{Lanzoni
  et~al.}{2018}]{Lanzoni2018}
Lanzoni B.,  et~al., 2018, ApJ, 861, 16

\bibitem[\protect\citeauthoryear{{Loose}, {Kruegel}  \& {Tutukov}}{{Loose}
  et~al.}{1982}]{Loose+1982}
{Loose} H.~H.,  {Kruegel} E.,   {Tutukov} A.,  1982, \aap, \href
  {https://ui.adsabs.harvard.edu/abs/1982A&A...105..342L} {105, 342}

\bibitem[\protect\citeauthoryear{{Lu}, {Ghez}, {Hornstein}, {Morris}, {Becklin}
   \& {Matthews}}{{Lu} et~al.}{2009}]{Lu2009}
{Lu} J.~R.,  {Ghez} A.~M.,  {Hornstein} S.~D.,  {Morris} M.~R.,  {Becklin}
  E.~E.,   {Matthews} K.,  2009, \mn@doi [\apj] {10.1088/0004-637X/690/2/1463},
  \href {http://adsabs.harvard.edu/abs/2009ApJ...690.1463L} {690, 1463}

\bibitem[\protect\citeauthoryear{{Lu}, {Do}, {Ghez}, {Morris}, {Yelda}  \&
  {Matthews}}{{Lu} et~al.}{2013}]{Lu+2013}
{Lu} J.~R.,  {Do} T.,  {Ghez} A.~M.,  {Morris} M.~R.,  {Yelda} S.,   {Matthews}
  K.,  2013, \mn@doi [\apj] {10.1088/0004-637X/764/2/155}, \href
  {https://ui.adsabs.harvard.edu/abs/2013ApJ...764..155L} {764, 155}

\bibitem[\protect\citeauthoryear{{Magnan}, {Fouvry}, {Pichon}  \&
  {Chavanis}}{{Magnan} et~al.}{2021}]{MagnanFouvry2021}
{Magnan} N.,  {Fouvry} J.-B.,  {Pichon} C.,   {Chavanis} P.-H.,  2021, arXiv
  e-prints, \href {https://ui.adsabs.harvard.edu/abs/2021arXiv211109011M} {p.
  arXiv:2111.09011}

\bibitem[\protect\citeauthoryear{{Mapelli}, {Hayfield}, {Mayer}  \&
  {Wadsley}}{{Mapelli} et~al.}{2012}]{Mapelli+2012}
{Mapelli} M.,  {Hayfield} T.,  {Mayer} L.,   {Wadsley} J.,  2012, \mn@doi
  [\apj] {10.1088/0004-637X/749/2/168}, \href
  {https://ui.adsabs.harvard.edu/abs/2012ApJ...749..168M} {749, 168}

\bibitem[\protect\citeauthoryear{{Mastrobuono-Battisti}, {Perets},
  {Gualandris}, {Neumayer}  \& {Sippel}}{{Mastrobuono-Battisti}
  et~al.}{2019}]{Mastrobuono2019}
{Mastrobuono-Battisti} A.,  {Perets} H.~B.,  {Gualandris} A.,  {Neumayer} N.,
  {Sippel} A.~C.,  2019, \mn@doi [\mnras] {10.1093/mnras/stz3004}, \href
  {https://ui.adsabs.harvard.edu/abs/2019MNRAS.490.5820M} {490, 5820}

\bibitem[\protect\citeauthoryear{{McKernan}, {Ford}, {Lyra}  \&
  {Perets}}{{McKernan} et~al.}{2012}]{McKernan+2012}
{McKernan} B.,  {Ford} K.~E.~S.,  {Lyra} W.,   {Perets} H.~B.,  2012, \mn@doi
  [\mnras] {10.1111/j.1365-2966.2012.21486.x}, \href
  {https://ui.adsabs.harvard.edu/abs/2012MNRAS.425..460M} {425, 460}

\bibitem[\protect\citeauthoryear{{Meiron} \& {Kocsis}}{{Meiron} \&
  {Kocsis}}{2019}]{Meiron2019}
{Meiron} Y.,  {Kocsis} B.,  2019, \mn@doi [\apj] {10.3847/1538-4357/ab1b32},
  \href {https://ui.adsabs.harvard.edu/abs/2019ApJ...878..138M} {878, 138}

\bibitem[\protect\citeauthoryear{{Merritt}, {Alexander}, {Mikkola}  \&
  {Will}}{{Merritt} et~al.}{2010}]{Merritt2010}
{Merritt} D.,  {Alexander} T.,  {Mikkola} S.,   {Will} C.~M.,  2010, \mn@doi
  [\prd] {10.1103/PhysRevD.81.062002}, \href
  {http://adsabs.harvard.edu/abs/2010PhRvD..81f2002M} {81, 062002}

\bibitem[\protect\citeauthoryear{{Meyer} et~al.,}{{Meyer}
  et~al.}{2012}]{Meyer2012}
{Meyer} L.,  et~al., 2012, \mn@doi [Science] {10.1126/science.1225506}, \href
  {http://adsabs.harvard.edu/abs/2012Sci...338...84M} {338, 84}

\bibitem[\protect\citeauthoryear{{Mihos} \& {Hernquist}}{{Mihos} \&
  {Hernquist}}{1994}]{Mihos_Hernquist1994}
{Mihos} J.~C.,  {Hernquist} L.,  1994, \mn@doi [\apjl] {10.1086/187679}, \href
  {https://ui.adsabs.harvard.edu/abs/1994ApJ...437L..47M} {437, L47}

\bibitem[\protect\citeauthoryear{{Milosavljevi{\'c}} \&
  {Merritt}}{{Milosavljevi{\'c}} \& {Merritt}}{2001}]{Milosavljevic2001}
{Milosavljevi{\'c}} M.,  {Merritt} D.,  2001, \mn@doi [\apj] {10.1086/323830},
  \href {http://adsabs.harvard.edu/abs/2001ApJ...563...34M} {563, 34}

\bibitem[\protect\citeauthoryear{{Neumayer}, {Seth}  \& {B{\"o}ker}}{{Neumayer}
  et~al.}{2020}]{Neumayer_Seth_Boker2020}
{Neumayer} N.,  {Seth} A.,   {B{\"o}ker} T.,  2020, \mn@doi [\aapr]
  {10.1007/s00159-020-00125-0}, \href
  {https://ui.adsabs.harvard.edu/abs/2020A&ARv..28....4N} {28, 4}

\bibitem[\protect\citeauthoryear{{O'Leary} et~al.}{{O'Leary}
  et~al.}{2009}]{OLeary2009}
{O'Leary} R.~M.,  et~al., 2009, \mn@doi [\mnras]
  {10.1111/j.1365-2966.2009.14653.x}, 395, 2127

\bibitem[\protect\citeauthoryear{{Panamarev} \& {Kocsis}}{{Panamarev} \&
  {Kocsis}}{2022}]{Panamarev_Kocsis2022}
{Panamarev} T.,  {Kocsis} B.,  2022, arXiv e-prints, \href
  {https://ui.adsabs.harvard.edu/abs/2022arXiv220706398P} {p. arXiv:2207.06398}

\bibitem[\protect\citeauthoryear{{Pei{\ss}ker}, {Eckart}, {Zaja{\v{c}}ek},
  {Ali}  \& {Parsa}}{{Pei{\ss}ker} et~al.}{2020}]{Peissker2020}
{Pei{\ss}ker} F.,  {Eckart} A.,  {Zaja{\v{c}}ek} M.,  {Ali} B.,   {Parsa} M.,
  2020, \mn@doi [\apj] {10.3847/1538-4357/ab9c1c}, \href
  {https://ui.adsabs.harvard.edu/abs/2020ApJ...899...50P} {899, 50}

\bibitem[\protect\citeauthoryear{{Perets} \& {Mastrobuono-Battisti}}{{Perets}
  \& {Mastrobuono-Battisti}}{2014}]{Perets_Mastro2014}
{Perets} H.~B.,  {Mastrobuono-Battisti} A.,  2014, \mn@doi [\apjl]
  {10.1088/2041-8205/784/2/L44}, \href
  {https://ui.adsabs.harvard.edu/abs/2014ApJ...784L..44P} {784, L44}

\bibitem[\protect\citeauthoryear{Rauch \& Tremaine}{Rauch \&
  Tremaine}{1996}]{Rauch1996}
Rauch K.~P.,  Tremaine S.,  1996, \mn@doi [New Astron.]
  {https://doi.org/10.1016/S1384-1076(96)00012-7}, 1, 149

\bibitem[\protect\citeauthoryear{{Roupas}}{{Roupas}}{2020}]{Roupas2020}
{Roupas} Z.,  2020, \mn@doi [J. Phys. A] {10.1088/1751-8121/ab5f7b}, \href
  {https://ui.adsabs.harvard.edu/abs/2020JPhA...53d5002R} {53, 045002}

\bibitem[\protect\citeauthoryear{Roupas, Kocsis  \& Tremaine}{Roupas
  et~al.}{2017}]{Roupas2017}
Roupas Z.,  Kocsis B.,   Tremaine S.,  2017, \mn@doi [ApJ]
  {10.3847/1538-4357/aa7141}, 842, 90

\bibitem[\protect\citeauthoryear{{Samsing} et~al.,}{{Samsing}
  et~al.}{2020}]{Samsing2020}
{Samsing} J.,  et~al., 2020, arXiv e-prints, \href
  {https://ui.adsabs.harvard.edu/abs/2020arXiv201009765S} {p. arXiv:2010.09765}

\bibitem[\protect\citeauthoryear{{Sch{\"o}del}, {Gallego-Cano}, {Dong},
  {Nogueras-Lara}, {Gallego-Calvente}, {Amaro-Seoane}  \&
  {Baumgardt}}{{Sch{\"o}del} et~al.}{2018}]{Schodel+2018}
{Sch{\"o}del} R.,  {Gallego-Cano} E.,  {Dong} H.,  {Nogueras-Lara} F.,
  {Gallego-Calvente} A.~T.,  {Amaro-Seoane} P.,   {Baumgardt} H.,  2018,
  \mn@doi [\aap] {10.1051/0004-6361/201730452}, \href
  {https://ui.adsabs.harvard.edu/abs/2018A&A...609A..27S} {609, A27}

\bibitem[\protect\citeauthoryear{{Seth}, {Blum}, {Bastian}, {Caldwell}  \&
  {Debattista}}{{Seth} et~al.}{2008}]{Seth2008}
{Seth} A.~C.,  {Blum} R.~D.,  {Bastian} N.,  {Caldwell} N.,   {Debattista}
  V.~P.,  2008, \mn@doi [\apj] {10.1086/591935}, \href
  {http://adsabs.harvard.edu/abs/2008ApJ...687..997S} {687, 997}

\bibitem[\protect\citeauthoryear{{Stone}, {Metzger}  \& {Haiman}}{{Stone}
  et~al.}{2017}]{Stone2017}
{Stone} N.~C.,  {Metzger} B.~D.,   {Haiman} Z.,  2017, \mn@doi [\mnras]
  {10.1093/mnras/stw2260}, \href
  {http://adsabs.harvard.edu/abs/2017\mnras.464..946S} {464, 946}

\bibitem[\protect\citeauthoryear{Sz\"olgy\'en \& Kocsis}{Sz\"olgy\'en \&
  Kocsis}{2018}]{Szolgyen2018}
Sz\"olgy\'en A.,  Kocsis B.,  2018, \mn@doi [Phys. Rev. Lett.]
  {10.1103/PhysRevLett.121.101101}, 121, 101101

\bibitem[\protect\citeauthoryear{{Sz{\"o}lgy{\'e}n}, {Meiron}  \&
  {Kocsis}}{{Sz{\"o}lgy{\'e}n} et~al.}{2019}]{Szolgyen+2019}
{Sz{\"o}lgy{\'e}n} {\'A}.,  {Meiron} Y.,   {Kocsis} B.,  2019, \mn@doi [\apj]
  {10.3847/1538-4357/ab50bb}, \href
  {https://ui.adsabs.harvard.edu/abs/2019ApJ...887..123S} {887, 123}

\bibitem[\protect\citeauthoryear{{Sz{\"o}lgy{\'e}n}, {M{\'a}th{\'e}}  \&
  {Kocsis}}{{Sz{\"o}lgy{\'e}n} et~al.}{2021}]{Szolgyen+2021}
{Sz{\"o}lgy{\'e}n} {\'A}.,  {M{\'a}th{\'e}} G.,   {Kocsis} B.,  2021, \mn@doi
  [\apj] {10.3847/1538-4357/ac13ab}, \href
  {https://ui.adsabs.harvard.edu/abs/2021ApJ...919..140S} {919, 140}

\bibitem[\protect\citeauthoryear{{Tagawa}, {Haiman}  \& {Kocsis}}{{Tagawa}
  et~al.}{2020}]{Tagawa2020}
{Tagawa} H.,  {Haiman} Z.,   {Kocsis} B.,  2020, \mn@doi [\apj]
  {10.3847/1538-4357/ab9b8c}, \href
  {https://ui.adsabs.harvard.edu/abs/2020ApJ...898...25T} {898, 25}

\bibitem[\protect\citeauthoryear{Tak\'acs \& Kocsis}{Tak\'acs \&
  Kocsis}{2018}]{Takacs2018}
Tak\'acs A.,  Kocsis B.,  2018, \mn@doi [ApJ] {10.3847/1538-4357/aab268}, 856,
  113

\bibitem[\protect\citeauthoryear{{Tiongco}, {Vesperini}  \& {Varri}}{{Tiongco}
  et~al.}{2017}]{Tiongco2017}
{Tiongco} M.~A.,  {Vesperini} E.,   {Varri} A.~L.,  2017, \mn@doi [\mnras]
  {10.1093/mnras/stx853}, \href
  {http://adsabs.harvard.edu/abs/2017MNRAS.469..683T} {469, 683}

\bibitem[\protect\citeauthoryear{Tiongco, Vesperini  \& Varri}{Tiongco
  et~al.}{2018}]{Tiongco2018}
Tiongco M.~A.,  Vesperini E.,   Varri A.~L.,  2018, \mn@doi [MNRAS: Letters]
  {10.1093/mnrasl/sly009}, 475, L86

\bibitem[\protect\citeauthoryear{{Tiongco}, {Collier}  \& {Varri}}{{Tiongco}
  et~al.}{2021}]{Tiongco+2021}
{Tiongco} M.,  {Collier} A.,   {Varri} A.~L.,  2021, \mn@doi [\mnras]
  {10.1093/mnras/stab1968}, \href
  {https://ui.adsabs.harvard.edu/abs/2021MNRAS.506.4488T} {506, 4488}

\bibitem[\protect\citeauthoryear{{Touma} \& {Tremaine}}{{Touma} \&
  {Tremaine}}{2014}]{Touma_Tremaine2014}
{Touma} J.,  {Tremaine} S.,  2014, arXiv e-prints, \href
  {https://ui.adsabs.harvard.edu/abs/2014arXiv1401.5534T} {p. arXiv:1401.5534}

\bibitem[\protect\citeauthoryear{{Touma}, {Tremaine}  \& {Kazandjian}}{{Touma}
  et~al.}{2019}]{Touma_Tremaine2019}
{Touma} J.,  {Tremaine} S.,   {Kazandjian} M.,  2019, \mn@doi [\prl]
  {10.1103/PhysRevLett.123.021103}, \href
  {https://ui.adsabs.harvard.edu/abs/2019PhRvL.123b1103T} {123, 021103}

\bibitem[\protect\citeauthoryear{{Tremaine}}{{Tremaine}}{2020a}]{Tremaine2020a}
{Tremaine} S.,  2020a, \mn@doi [\mnras] {10.1093/mnras/stz3181}, \href
  {https://ui.adsabs.harvard.edu/abs/2020MNRAS.491.1941T} {491, 1941}

\bibitem[\protect\citeauthoryear{{Tremaine}}{{Tremaine}}{2020b}]{Tremaine2020b}
{Tremaine} S.,  2020b, \mn@doi [\mnras] {10.1093/mnras/staa420}, \href
  {https://ui.adsabs.harvard.edu/abs/2020MNRAS.493.2632T} {493, 2632}

\bibitem[\protect\citeauthoryear{{Tremaine}, {Ostriker}  \&
  {Spitzer}}{{Tremaine} et~al.}{1975}]{Tremaine1975}
{Tremaine} S.~D.,  {Ostriker} J.~P.,   {Spitzer} Jr. L.,  1975, \mn@doi [\apj]
  {10.1086/153422}, \href {http://adsabs.harvard.edu/abs/1975ApJ...196..407T}
  {196, 407}

\bibitem[\protect\citeauthoryear{{Tsatsi}, {Mastrobuono-Battisti}, {van de
  Ven}, {Perets}, {Bianchini}  \& {Neumayer}}{{Tsatsi}
  et~al.}{2017}]{Tsatsi2017}
{Tsatsi} A.,  {Mastrobuono-Battisti} A.,  {van de Ven} G.,  {Perets} H.~B.,
  {Bianchini} P.,   {Neumayer} N.,  2017, \mn@doi [\mnras]
  {10.1093/mnras/stw2593}, \href
  {https://ui.adsabs.harvard.edu/abs/2017MNRAS.464.3720T} {464, 3720}

\bibitem[\protect\citeauthoryear{Valtonen \& Karttunen}{Valtonen \&
  Karttunen}{2006}]{Valtonen2006}
Valtonen M.,  Karttunen H.,  2006, The Three-Body Problem.
Cambridge University Press, \mn@doi{10.1017/CBO9780511616006}

\bibitem[\protect\citeauthoryear{{Yelda}, {Ghez}, {Lu}, {Do}, {Meyer}, {Morris}
   \& {Matthews}}{{Yelda} et~al.}{2014}]{Yelda2014}
{Yelda} S.,  {Ghez} A.~M.,  {Lu} J.~R.,  {Do} T.,  {Meyer} L.,  {Morris} M.~R.,
    {Matthews} K.,  2014, \mn@doi [\apj] {10.1088/0004-637X/783/2/131}, \href
  {https://ui.adsabs.harvard.edu/abs/2014ApJ...783..131Y} {783, 131}

\bibitem[\protect\citeauthoryear{{Zderic}, {Collier}, {Tiongco}  \&
  {Madigan}}{{Zderic} et~al.}{2020}]{Zderic+2020}
{Zderic} A.,  {Collier} A.,  {Tiongco} M.,   {Madigan} A.-M.,  2020, \mn@doi
  [\apjl] {10.3847/2041-8213/ab91a0}, \href
  {https://ui.adsabs.harvard.edu/abs/2020ApJ...895L..27Z} {895, L27}

\bibitem[\protect\citeauthoryear{{von Fellenberg} et~al.,}{{von Fellenberg}
  et~al.}{2022}]{vonFellenberg2022}
{von Fellenberg} S.~D.,  et~al., 2022, \mn@doi [\apjl]
  {10.3847/2041-8213/ac68ef}, \href
  {https://ui.adsabs.harvard.edu/abs/2022ApJ...932L...6V} {932, L6}

\makeatother
\end{thebibliography}





\bsp	
\label{lastpage}
\end{document}